\documentclass[fleqn,usenatbib]{mnras}

\usepackage{graphicx}
\usepackage{amsmath,footmisc}
\usepackage{amsfonts,amssymb}
\usepackage{bm}
\usepackage{placeins}
\usepackage{gensymb}
\usepackage{cleveref}
\usepackage{preamble}
\usepackage{orcidlink}

\newcommand\FSDFidModel[2]{{\texttt{FSD-{#1}-{#2}}}}
\newcommand\FSDtModel[2]{{\texttt{FSD-{#1}-t{#2}}}}

\newcommand\FSD{{\texttt{FSD}}}
\newcommand\PSD{{\texttt{PSD}}}
\newcommand\PSDlow{{\texttt{PSD-R1}}}
\newcommand\PSDhigh{{\texttt{PSD-R4}}}
\newcommand\FSDcomp{{\texttt{FSD-Comp}}}
\newcommand\FSDsol{{\texttt{FSD-Sol}}}
\newcommand\FSDhigh{{\texttt{FSD-E50}}}
\newcommand\FSDlow{{\texttt{FSD-E7}}}
\newcommand\FSDcomphigh{{\texttt{FSD-Comp-E50}}}
\newcommand\FSDcomplow{{\texttt{FSD-Comp-E7}}}
\newcommand\FSDsolhigh{{\texttt{FSD-Sol-E50}}}
\newcommand\FSDsollow{{\texttt{FSD-Sol-E7}}}

\title[Comparative Study of Isothermal Turbulence]{A Comparative Study of Isothermal Turbulence Statistics: Fourier Space Driving vs. Point Source Driving}

\author[T. Desire et al.]{
Tejahni Desire\orcidlink{0009-0008-3202-2972},$^{1}$
Chang-Goo Kim\orcidlink{0000-0003-2896-3725},$^{1}$\thanks{E-mail: cgkim@astro.princeton.edu}
Rajsekhar Mohapatra\orcidlink{0000-0002-1600-7552}$^{1}$
\\
$^{1}$Department of Astrophysical Sciences, Princeton University, Princeton NJ 08544, USA\\
}

\date{Accepted XXX. Received YYY; in original form ZZZ}
\pubyear{\the\year{}}
\begin{document}

\label{firstpage}
\pagerange{\pageref{firstpage}--\pageref{lastpage}}
\maketitle

\begin{abstract}
The turbulence driving parameter ($b \equiv \sigma_{\rho/\langle{\rho}\rangle}/\mathcal{M}$; the ratio of the density to velocity fluctuations) is widely used to infer the dominant mode of energy injection in the interstellar turbulence.
Numerical simulations of turbulence using Fourier Space Driving (FSD) establish a mapping from $b\approx1/3$ for purely solenoidal to $b\approx1$ for purely compressive driving.
We test the robustness of this calibration by comparing FSD against Point Source Driving (PSD), which stochastically injects radial momentum at random locations mimicking supernovae.
Using isothermal hydrodynamic simulations in a periodic box with AthenaK, we run a suite of carefully curated simulations to match Mach numbers between the two driving methods and compare morphology, probability density functions, and power spectra of density and velocity.
Despite injecting purely compressive motions, the PSD models yield $b=0.33$–$0.49$, values that the FSD calibration would associate with more solenoidal driving.
With mass-weighted mean Mach number (excluding high velocity bubble interior), $b_M=0.74$–$0.79$ still does not recover the expected $b\approx1$ for volume-filling, pure compressive driving.
More broadly, the PSD models show density and velocity statistics closer to solenoidal and compressive FSD models, respectively, and exhibit unique features, including non-Gaussian velocity tails and a positive density–Mach number correlation at high densities.
Within the FSD framework itself, varying the forcing correlation time changes $b$ by a factor of $>$3 for compressive driving.
These results demonstrate that $b$ is degenerate with both the spatial locality and the temporal correlation of the driving, limiting its utility as a standalone diagnostic of the energy injection mode.
\end{abstract}

\section{Introduction} \label{sec:intro}

The gaseous interstellar medium (ISM) is a high Reynolds number fluid, typically exceeding $\mathrm{Re} \sim 10^3 - 10^9$ depending on the phase and scale \citep[e.g.,][]{2011piim.book.....D,2013SSRv..178..163B}. As the negligible viscosity compared to inertial forces renders the flow highly turbulent, the ISM turbulence is thus ubiquitous, spanning a vast range of spatial scales from the kiloparsec scales of galactic disks down to the astronomical unit scales of protostellar cores \citep{2004ARA&A..42..211E}.
This multi-scale chaotic motion provides critical support against gravitational collapse on global scales while simultaneously promoting localized density enhancements, playing key roles in regulating star formation rates at different scales \citep[e.g.,][for reviews]{2007ARA&A..45..565M,2014prpl.conf...77P}.

Unlike terrestrial fluids often modelled as incompressible, the ISM is a highly compressible fluid where the flow velocity frequently exceeds the local sound speed. In this regime, velocity fluctuations introduced by turbulence also generate profound density fluctuations through shock compressions and rarefactions \citep{1992pavi.book.....S}. Consequently, classical incompressible turbulence theories, such as the \citet{1941DoSSR..30..301K} theory, are not directly applicable to the ISM without modification. Instead, a statistical description that incorporates both density and velocity fluctuations is warranted \citep[e.g.,][]{2020ApJ...904..160F}. In compressible turbulence, for example, the energy cascade is better described by density-weighted variables, such as $\rho^{1/3}v$ \citep[e.g.,][]{1955IAUS....2..121L, 2007ApJ...665..416K}.

The simplest statistical measures in compressible turbulence are the global averages of the fluctuations of the density and velocity fields. Namely, one can calculate the standard deviation of the linear density field (normalized by the mean density), $\sigma_{\rho/\langle\rho\rangle}$, and the turbulent rms Mach number, $\mathcal{M}$. The ratio of the two defines the so-called turbulence driving parameter \citep{1997ApJ...474..730P,2008ApJ...688L..79F}:
\begin{equation}\label{eq:b}
    b \equiv \frac{\sigma_{\rho/\langle\rho\rangle}}{\mathcal{M}}.
\end{equation}
Physically, $b$ serves as a measure of the level of density fluctuation introduced given velocity fluctuation, which is sensitive to the turbulence driving mechanism. From extensive numerical simulations, \citet{2010A&A...512A..81F} show that purely solenoidal (divergence-free) driving typically yields $b \approx 1/3$, while purely compressive (curl-free) driving yields $b \approx 1$.

If the density field follows a log-normal probability density function (PDF), which is often the case in solenoidal driving \citep[e.g.,][]{1994ApJ...423..681V,1998PhRvE..58.4501P,2001ApJ...546..980O}, the linear density variance relates to the logarithmic density variance $\sigma_s^2 = \ln(1 + \sigma_{\rho/\langle\rho\rangle}^2)$. Substituting \autoref{eq:b} into this identity yields the $\sigma_s$--$\mathcal{M}$ relation \citep[][see \citealt{2012ApJ...755L..19B} for the relation for column density]{1997MNRAS.288..145P,2008ApJ...688L..79F}:
\begin{equation}
    \sigma_s^2 = \ln(1 + b^2\mathcal{M}^2).
\end{equation}
Modifications to the $b$ parameter and $\sigma_s$-$\mathcal{M}$ relation in a magnetized medium are also proposed \citep{2001ApJ...546..980O,2008ApJ...682L..97L,2011ApJ...727L..21P,2012MNRAS.423.2680M}.

The $b$-parameter has quickly transitioned from a theoretical construct to a widely adopted observational diagnostic. By measuring the density variance (inferred from column density maps; e.g., \citealt{2010MNRAS.405L..56B}) and velocity variance (inferred from spectral line widths or centroid velocity dispersions), this theoretical connection provides a diagnostic tool to constrain the nature of turbulence injection in various astrophysical environments. This diagnostic framework has been applied across a diverse range of scales, from individual molecular clouds and the diffuse ISM within the Milky Way \citep{2010A&A...513A..67B,2013ApJ...779...50G,2013A&A...549A..53K,2016ApJ...832..143F,2021ApJ...908..186M} to the ISM of nearby external galaxies \citep{2022MNRAS.509.2180S,2023MNRAS.526..982G}. The $\sigma_s-\mathcal{M}$ relation is also used to estimate the Mach number \citep[e.g.,][]{2023MNRAS.524.2379H,2025ApJ...994...80L}.

Despite the utility of the $b$-parameter and $\sigma_s$-$\mathcal{M}$ relation, a methodological gap exists in the literature. The canonical relations between $b$ and the forcing mode were derived primarily using idealized simulations where turbulence is driven by a ``Fourier Space Driving'' (FSD) algorithm \citep{2008ApJ...688L..79F,2010A&A...512A..81F}. In FSD, energy is injected at specific wavenumbers in spectral space, which in turn is transformed back to the real space, thereby affecting the velocity field globally (i.e., volume-filling stochastic driving). However, the main energy sources of the ISM turbulence are often discrete, spatially and temporally \emph{localized} events in real space, most notably via expanding bubbles driven by stellar feedback, including supernovae (SNe), stellar winds, H\,{\sc ii} regions, and protostellar outflows \citep[e.g.,][]{2004RvMP...76..125M,2007ARA&A..45..565M}. This ``Point Source Driving'' (PSD) can lead to fundamentally different turbulence statistics, while the energy injection itself is purely compressive.

To our knowledge, there has been no direct confirmation that the $b$-parameter characterization derived from FSD simulations holds validity in simulations utilizing localized, real-space energy injection. Validating this link is crucial for ensuring that observational interpretations calibrated on FSD models are accurate.

To address this uncertainty, we conduct a direct comparative study using isothermal hydrodynamic simulations in a periodic box with \texttt{AthenaK}, described in \autoref{sec:methods}. We compare PSD, which injects radial momentum at random locations mimicking SNe (\autoref{subsec:psd}), against FSD with purely solenoidal and compressive modes (\autoref{subsec:fsd}). In \autoref{sec:psd_fsd}, we compare the two methods through morphology of density field and local compressive fraction of the velocity field (\autoref{subsec:morphology}), global statistics including the $b$-parameter (\autoref{subsec:global_stat}), probability density functions of the density and velocity fields (\autoref{subsec:pdfs}), and power spectra with Helmholtz decomposition (\autoref{subsec:spectra}). We additionally examine whether the sensitivity of major statistical properties in \FSD\ itself by varying the forcing correlation time $\tcorr$ in \autoref{sec:tcorr}. Our principal findings and their implications for interpreting $b$ as an observational diagnostic of the energy injection mode are presented in \autoref{sec:summary}.

\section{Methods and Models} \label{sec:methods}
\subsection{Governing Equations and Numerical Setup} \label{subsec:eqs}

We solve the isothermal hydrodynamics equations with forcing. The governing equations are
\begin{align}
    \frac{\partial \rho}{\partial t} +\nabla \cdot (\rho \bm{v}) &= 0,\label{eq:hydro1}\\
    \frac{\partial (\rho \bm{ \bm{v}})}{\partial t} + \nabla \cdot (\rho \bm{v} \bm{v} +P \mathbf{I}) &=  \bm{f}, 
    \label{eq:hydro2}
\end{align}
for density $\rho$, gas velocity $\bm{v}$ and pressure $P=\rho c_s^2$ with a constant sound speed $c_s$.
$\bm{f}$ represents a forcing term to drive turbulence, for which we use either an acceleration field generated in Fourier space or momentum injection with a radial velocity field representing SNe. We describe each driving method in further detail in \autoref{subsec:psd} and \autoref{subsec:fsd}.

We utilize \texttt{AthenaK} \citep{2024arXiv240916053S}, a performance-portable, GPU accelerated version of \texttt{Athena++} \citep{2020ApJS..249....4S}.
We use a piecewise linear method for spatial reconstruction, HLLE Riemann solver for the computation of fluxes, and the 3rd order Runge-Kutta for time integration \citep{2024arXiv240916053S}. To ensure the stability of the simulations containing strong rarefaction and high Mach number shocks, we use the first order flux correction \citep{2009ApJ...691.1092L,2024arXiv240916053S}, a fallback scheme to replace the fluxes of the problematic cells (i.e., cells with negative density) with those calculated using the first order reconstruction (piecewise constant) and local Lax-Friedrich Riemann solver.

For the interpretation of our results in the context of SN driven turbulence in the diffuse ISM, we adopt the length, time, and mass units
\begin{equation}
    l_0 = 1\kpc,\quad
    t_0 = 100\Myr,\quad
    m_0 = 10^7\Msun.
\end{equation}
We use a periodic cube with a side length of $L=0.5l_0=500\pc$.
The adopted initial density $\rho_0=m_0/l_0^3$ and sound speed $c_s=l_0/t_0$ of the simulations correspond to typical conditions of the warm neutral/ionized medium with the hydrogen number density of $n_{\rm H,0}=\rho_{0}/(1.4 m_H)=0.29 \ {\rm cm^{-3}}$ and the sound speed of $9.8\kms$. The initial thermal pressure is $P_{\rm th,0}=\rho_0 c_s^2
= 4.7\times 10^{3}k_B\Kel\pcc$, where $k_B$ is the Boltzmann constant. These values are also similar to the solar neighborhood ISM condition \citep{Jenkins2011}.

\subsection{Point Source Driving}  \label{subsec:psd}

We first consider a Point Source Driving (PSD) method that resembles SNe in the ISM. Although we choose the parameters relevant to the momentum injection by radiative SN remnants \citep{2015ApJ...802...99K, 2020ApJ...905...35K}, this can be interpreted as turbulence driving by expanding bubbles from stochastic, localized point sources in real space, as opposed to driving by global force field realizations in Fourier space.

Coupling the radial momentum to the existing gas with non-uniform density and velocity can be subtle and complicated \citep[e.g.,][]{2025OJAp....8E..44H}. In this paper, we take the simplest approach by injecting constant mass and radial momentum densities within a homogenised spherical region with a fixed radius $r_{\rm inj}$. We first calculate the average mass and momentum densities within the injection volume, $\rho_{\rm avg}$ and $\mathbf{p}_{\rm avg}$, respectively. We inject a total radial momentum of $\mathcal{P}_{\rm inj}$ and mass of $m_{\rm inj}$, yielding the injection mass and momentum densities $\rho_{\rm inj}=m_{\rm inj}/V_{\rm inj}$ and $\mathbf{p}_{\rm inj}=\mathcal{P}_{\rm inj}/{V_{\rm inj}}\rhat$, respectively, for the volume of the injection region $V_{\rm inj}$.
The new mass and momentum densities within the injection region then become $\rho_{\rm new}=\rho_{\rm avg} +\rho_{\rm inj}$ and $\mathbf{p}_{\rm new}=\mathbf{p}_{\rm avg} + \mathbf{p}_{\rm inj}$.

Throughout the paper, we use $r_{\rm inj}=40 \pc$, $m_{\rm inj}=10 \Msun$, and $\mathcal{P}_{\rm inj}=3.0\times 10^{5} \Msun\kms$.
The choice of the injection radius and radial momentum is motivated by the shell formation radius and terminal momentum of radiative SN remnants in a uniform medium with density similar to the adopted initial density of $n_{\rm H,0}=0.3\pcc$ \citep{2015ApJ...802...99K}.
On average, we expect to couple the momentum $P_{\rm inj}$ with the total mass within the injection volume $\rho_{0}V_{\rm inj}+m_{\rm inj}$, yielding a typical value of the injection velocity of $v_{\rm inj}=112\kms$, but it can in principle be as high as $\mathcal{P}_{\rm inj}/m_{\rm inj}=3\times10^4\kms$ if the density within the injection volume is very small. To avoid frequent failures introduced by excessively strong shocks and corresponding short time steps, we apply a velocity ceiling such that $v_{\rm inj}\le v_{\rm max}=500\kms$. We then apply a density floor, for which we adopt $\rho_{\rm floor}=10^{-5}\rho_0$. We check the effects of the velocity ceiling and density floor and confirm that they do not affect our conclusions (\autoref{subsec:vel_ciel}).

From the star formation rate surface density in the solar vicinity, $\Sigma_{\rm SFR,\odot} = 3\times10^{-3}\Msun\kpc^{-2}\yr^{-1}$ \citep{Fuchs2009,2023A&A...669A..10Z}, we have a reference SN event rate within the simulation volume as $\dot{N}_\odot \equiv (\Sigma_{\rm SFR,\odot}/m_*)L^2 = 7.5\times10^{-6}(L/500\pc)^2\yr^{-1}$, where $m_*=100\Msun$ is the mass of new stars per SN estimated using the Kroupa IMF \citep{2001MNRAS.322..231K} and a STARBURST99 population synthesis model \citep{1999ApJS..123....3L}.
The kinetic energy of each event is roughly $E_{\rm kin,SN}\approx \mathcal{P}_{\rm inj}^2/2\rho_{0}V_{\rm inj}=3.3\times10^{50} \erg$. The associated kinetic energy injection rate of the PSD method is then approximately
\begin{equation}
\begin{split}
    \dot{E}_{\rm PSD} &\approx E_{\rm kin,SN}\dot{N}_{\rm SN} = 7.9\times10^{37} (\dot{N}_{\rm SN}/\dot{N}_\odot)\ergs \\
    &= 13(\dot{N}_{\rm SN}/\dot{N}_\odot) E_0/t_0
\end{split}
\end{equation}
where the code energy unit is $E_0= m_0(l_0/t_0)^2$.

For a given SN rate $\dot{N}_{\rm SN}$, we draw the number of events from a Poisson distribution with a mean of $\dot{N}_{\rm SN} dt$ where $dt$ is the simulation time step, and choose the event location randomly within the simulation domain.

\subsection{Fourier Space Driving} \label{subsec:fsd}

In the majority of turbulence simulations, the turbulence forcing is generated in the Fourier space with a prescribed power spectrum \citep[e.g.,][]{2010A&A...512A..81F}. We refer to this method as Fourier Space Driving (FSD).
We implement it in AthenaK similar to \citet[][see also \citealt{2025arXiv251100229M}]{2022ascl.soft04001F}.

Let the acceleration field be $\bm{a} = \bm{f}/\rho$
and its Fourier transform be $\bm{\Tilde{a}}(\bm{k})$.
\begin{enumerate}
    \item First, we implement the spectral weighting in $k$-space as a parabolic function of the spherical wavenumber from $k_{\rm min}$ to $k_{\rm max}$, with a peak at $k_{\rm peak}$:
    $$
    W(k) =
    \begin{cases}
        \left|1 - \dfrac{4\,(k-k_{\rm peak})^2}{(k_{\rm max}-k_{\rm min})^2}\right|, & k_{\rm min} \le k \le k_{\rm max},\\
        0, & \text{otherwise},
    \end{cases}
    $$
    with
    $$
    \big|\bm{\Tilde{a}}^{(1)}(\bm{k})\big| \propto \sqrt{W(k)}\,\frac{k_{\rm peak}}{k}
    $$
    for the 3D driving case. In practice, each Fourier component is assigned independent Gaussian random coefficients with this spectral weighting.

    \item Then we apply the solenoidal/compressive decomposition with fraction $F_{\rm sol}$ in solenoidal modes:
    \begin{subequations}
    \begin{align}
        \bm{\Tilde{a}}^{(2)}(\bm{k}) &=
        (1-F_{\rm sol})\,\bm{\Tilde{a}}_{\rm comp}(\bm{k})
        + F_{\rm sol}\,\bm{\Tilde{a}}_{\rm sol}(\bm{k})\text{, where}\\
        \bm{\Tilde{a}}_{\rm comp}(\bm{k}) &=
        \frac{\big(\bm{k}\cdot\bm{\Tilde{a}}^{(1)}(\bm{k})\big)\,\bm{k}}{k^2},\\
        \bm{\Tilde{a}}_{\rm sol}(\bm{k}) &=
        \bm{\Tilde{a}}^{(1)}(\bm{k}) - \bm{\Tilde{a}}_{\rm comp}(\bm{k}).
    \end{align}
    \end{subequations}

    \item We then construct the real acceleration field in the physical space, remove any net momentum input, and scale to achieve the target energy injection rate $\dot{E}_{\rm FSD}$:
    $$
    \bm{a} = \mathcal{N}_{\rm FSD}\,\bm{a}^{(2)},
    $$
    where $\mathcal{N}_{\rm FSD}$ is determined from
    $$
    m_0 \mathcal{N}_{\rm FSD}^2 + m_1 \mathcal{N}_{\rm FSD} = \dot{E}_{\rm FSD},
    $$
    with
    $$
    m_0 = \int \frac{1}{2}\rho\,|\bm{a}^{(2)}|^2\,\Delta t\,\mathrm{d}V,
    \qquad
    m_1 = \int \rho\,\bm{v}\cdot\bm{a}^{(2)}\,\mathrm{d}V.
    $$

    \item At discrete intervals $\Delta t_{\rm turb}$, we update the acceleration field using an Ornstein-Uhlenbeck process to impose a finite temporal correlation time $\tcorr$:
    $$
    \bm{a}^{(2)}(t+\Delta t_{\rm turb}) =
    \epsilon\,\bm{a}^{(2)}(t)
    + \sqrt{1-\epsilon^2}\,\bm{a}^{(2)\prime}(t),
    $$
    where $\epsilon = e^{-\Delta t_{\rm turb}/\tcorr}$ controls the degree of temporal memory.
    Here $\bm{a}^{(2)\prime}(t)$ is a newly generated real-space acceleration field, computed in a similar manner to $\bm{a}^{(2)}(t)$ from Fourier modes with the same weighting $W(k)$ and $F_{\rm sol}$ but independent random phases.
    The final acceleration field $\bm{a}(t+\Delta t_{\rm turb})$, is then constructed from $\bm{a}^{(2)}(t+\Delta t_{\rm turb})$ to achieve the target energy injection rate $\dot{E}_{\rm FSD}$.
    We fix $\Delta t_{\rm turb}/t_0$ throughout the simulation suite to $0.001$.
\end{enumerate}

\subsection{Models} \label{subsec:models}
For the PSD method, we keep our fiducial parameter choice as described in \autoref{subsec:psd} with two SN rates $\dot{N}_{\rm SN}=\dot{N}_{\odot}$ and $4\dot{N}_{\odot}$, which we refer to as \PSDlow\ and \PSDhigh, respectively.

In this work, our goal is to compare the characteristics of turbulence driven by two driving methods. To achieve this goal, we carefully choose the FSD parameters to make the resulting turbulence as close as possible to the PSD models in multiple aspects. We conduct a large parameter survey with low resolution FSD simulations to settle on the model parameters.
We choose $Lk_{\rm peak}/2\pi=5$ with $Lk_{\rm min}/2\pi=4$ and $Lk_{\rm max}/2\pi=7$ to place the driving scale similar to that of the PSD models.\footnote{Although the PSD models have no single driving scale, we get the integral scale $L_{\rm in}/L\sim 1/5$ using the mean inverse wavenumber weighted by velocity power spectrum. See \autoref{table:main_stats} and \autoref{subsec:spectra} for the definition.}
We only consider two types of forcing: purely solenoidal $F_{\rm sol}=1$ and purely compressive $F_{\rm sol}=0$.
For each forcing, we choose two amplitudes $\dot{E}_{\rm FSD}=7E_0/t_0$ and $50 E_0/t_0$ to match the resulting volume and mass-weighted Mach numbers with those in the PSD models. We note that these numbers are close to the approximate energy injection rate of $\dot{E}_{\rm PSD}\approx 13E_0/t_0$ and $52E_0/t_0$ but not exact as the actual kinetic energy injection rate depends on the coupling mass and other parameter choices.
The model name follows a naming convention \FSDFidModel{mode}{EXX} for {\tt mode=Comp} or {\tt Sol} and $\dot{E}_{\rm FSD}=${\tt XX}$E_0/t_0$.

For the fiducial FSD set (\autoref{sec:psd_fsd}), we use $\tcorr/t_0=0.05$ and 0.025 for the {\tt E7} and {\tt E50} models, respectively, yielding $\tcorr/\teddy\approx 1$ (where $\teddy$ is the eddy turnover time; see \autoref{table:main_stats} and \autoref{subsec:spectra} for the definition). Our standard numerical resolution uses the number of zones $N^3=512^3$.

For the exploration of the effect of $\tcorr$ (\autoref{sec:tcorr}), we run models with $\tcorr/ t_0=0.001, 0.01, 0.05, 0.1,$ and $0.2$ for each solenoidal or compressive driving mode. We fix $\dot{E}_{\rm FSD}=7$ for this model series. The naming convention follows \FSDtModel{mode}{ZZZ} for driving mode of {\tt mode=Comp} or {\tt Sol} and correlation time of $\tcorr/t_0=0.${\tt ZZZ}. This set of models uses the number of zones $N^3=256^3$.

\begin{figure*}
    \centering
    \includegraphics[width=\textwidth]{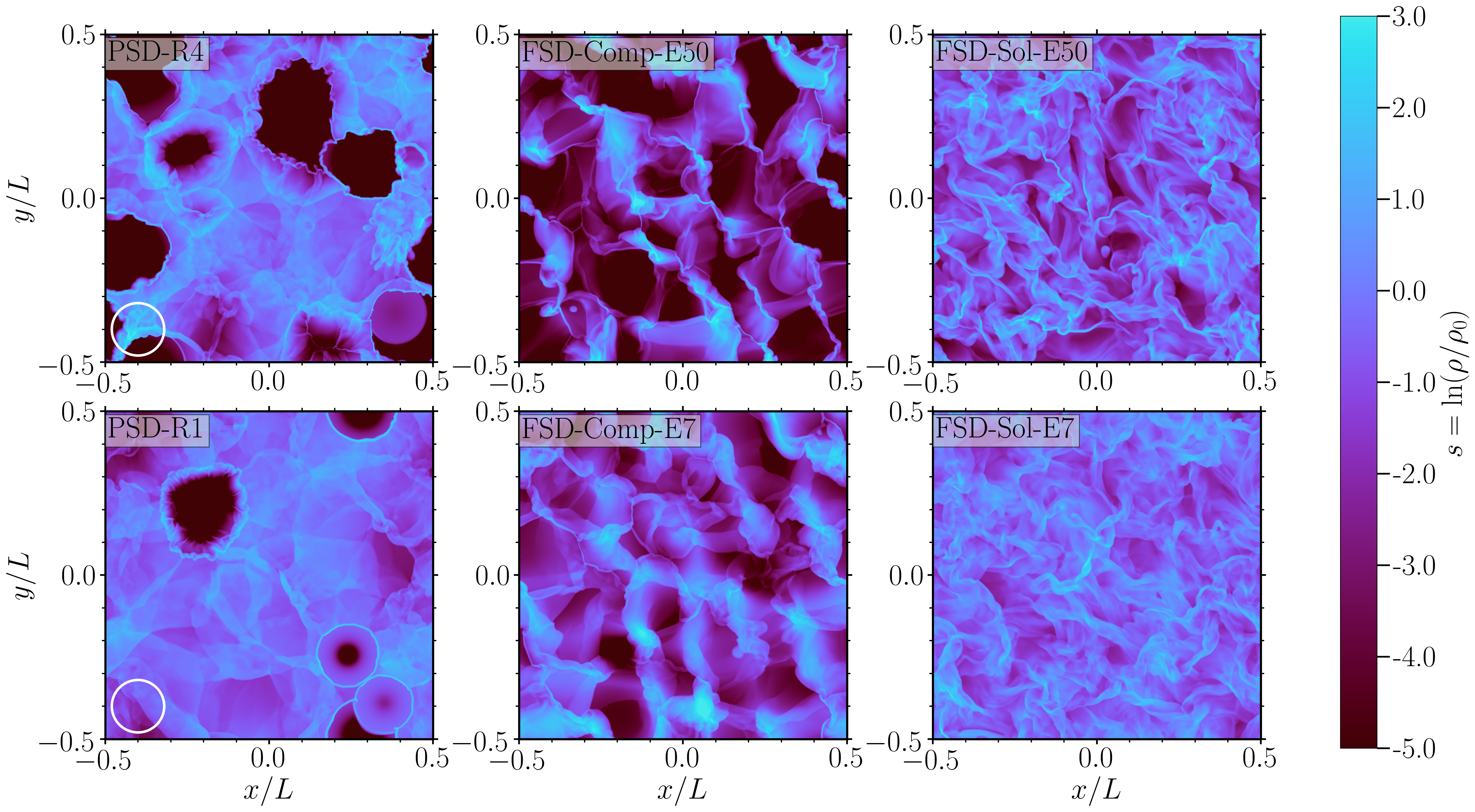}
    \caption{$z=0$ slices through the logarithm of the density $s\equiv \ln (\rho/\rho_0)$ field. From left to right, we show \PSD, \FSDcomp, and \FSDsol, while the models with the higher and lower energy injection rates are shown in the top and bottom rows, respectively. The inset circle in the left column indicates the injection size with a radius of $r_{\rm inj}$.
    \label{fig:slices}}
\end{figure*}

\begin{figure*}
    \centering
    \includegraphics[width=\textwidth]{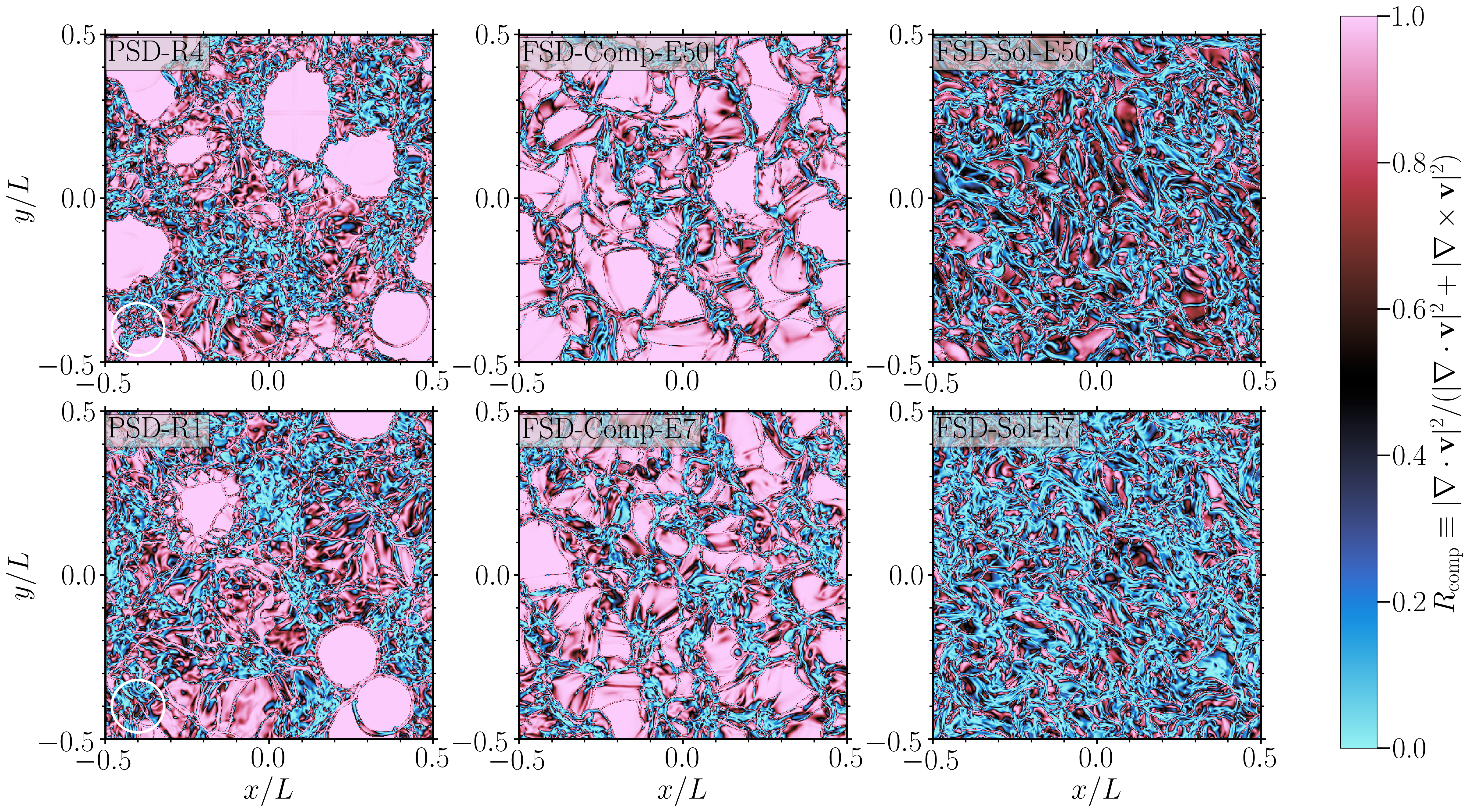}
    \caption{Same as \autoref{fig:slices}, but for the local compressive fraction $\Rcomp \equiv |\nabla\cdot \vel|^2/(|\nabla\cdot \vel|^2+|\nabla\times \vel|^2)$, which measures the fraction of velocity power in compressive modes at each point ($\Rcomp=0$: purely solenoidal; $\Rcomp=1$: purely compressive).
    \label{fig:rcomp_slices}}
\end{figure*}

\section{Comparison of Driving Methods} \label{sec:psd_fsd}

The two driving methods described in \autoref{subsec:psd} and \autoref{subsec:fsd} differ fundamentally in how energy and momentum are deposited: the PSD method injects radial momentum from localized point sources, while the FSD method applies a volume-filling acceleration field constructed in Fourier space. These differences raise the central question of this paper: do the resulting turbulence statistics depend on the driving mechanism, even when the global energy injection rate and resulting velocity dispersion are matched? We address this question by comparing the fiducial \PSD\ and \FSD\ models through morphological inspection (\autoref{subsec:morphology}), global statistics (\autoref{subsec:global_stat}), probability density functions (\autoref{subsec:pdfs}), and power spectra (\autoref{subsec:spectra}).

\subsection{Morphology} \label{subsec:morphology}

\autoref{fig:slices}
shows $z=0$ slices of the logarithmic density $s\equiv \ln (\rho/\rho_0)$ for the higher and lower energy injection rate models in the top and bottom rows, respectively, highlighting
differences in the density structures arising from the driving methods.
The \PSD\ models (left column) display large, circular low-density regions surrounded by high-density shells created by expanding shocks. The shells are thin when they first form, become corrugated as they evolve, and produce high-density filaments where they interact. The \FSDcomp\ models (middle column) also show distinct low-density regions, but without circular shells; their high-density structures appear more node-like, with less prominent filamentary or shell morphology, especially for \FSDcomplow. The \FSDsol\ models (right column) lack the pronounced low-density voids seen in both \PSD\ and \FSDcomp\ models, while their high-density structures are more filamentary.

\autoref{fig:rcomp_slices} shows the local compressive fraction
\begin{equation}\label{eq:Rcomp}
    \Rcomp \equiv \frac{|\nabla\cdot\vel|^2}{|\nabla\cdot\vel|^2+|\nabla\times\vel|^2}
\end{equation}
of the velocity field in the same layout. In compressively driven models (\PSD\ and \FSDcomp), compressive modes ($\Rcomp \approx 1$) are found across the full density range: diverging velocity fields produce high $\Rcomp$ in low-density voids, while converging flows at shock fronts also produce high $\Rcomp$ at high densities. Solenoidal modes, by contrast, arise when converging flows interact with each other and with the inhomogeneous density field, generating vorticity; they are therefore preferentially found in regions outside the low-density voids.
In the \FSDsol\ models, by contrast, $\Rcomp$ is relatively uniform in space, with volume-filling, low-$\Rcomp$ regions.
To quantify the visual impressions, we calculate the volume and mass-weighted mean values of $\Rcomp$ and present them with the Helmholtz decomposed power ratios in \autoref{subsec:spectra}.

\begin{figure}
    \centering
    \includegraphics[width=\linewidth]{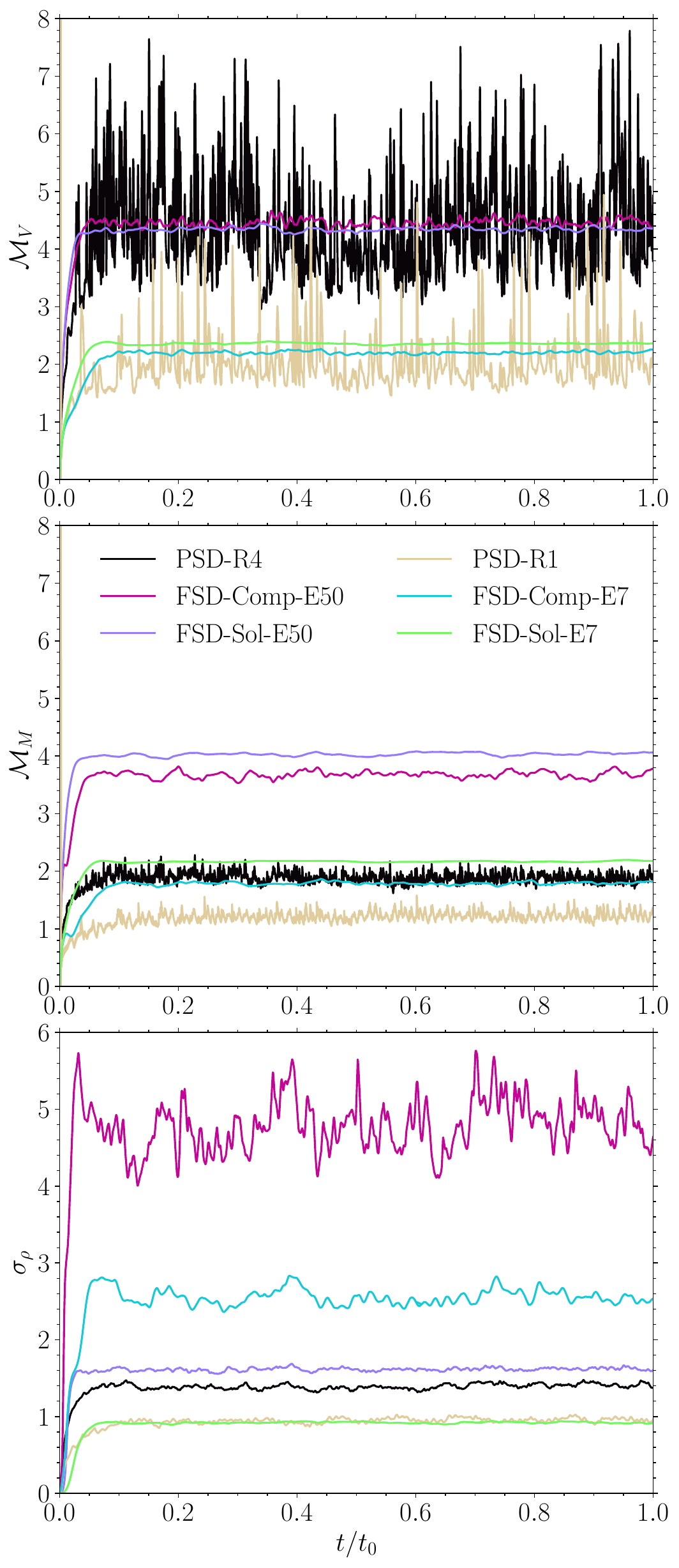}
    \caption{Time evolution of volume-weighted rms Mach number $\mathcal{M}_V$ (top panel), mass-weighted rms Mach number $\mathcal{M}_M$ (middle panel), and standard deviation of density $\sigma_\rho$ (bottom panel). The steady-state average values between $t/t_0=0.4-1$ for these variables are shown in \autoref{table:main_stats}.
    The \FSDhigh\ and \FSDlow\ models roughly match the volume-weighted $\mathcal{M}_V$ of the \PSDhigh\ and \PSDlow\ models, respectively. $\mathcal{M}_M$ of the \PSDhigh\ model is similar to that of \FSDlow.
    \label{fig:mach_dens_hist}}
\end{figure}

\subsection{Global Statistics} \label{subsec:global_stat}
In this subsection, we measure global statistics for density and velocity. We measure the density variance, defined as
$\sigma_\rho^2 \equiv \langle\rho^2\rangle_V - \langle\rho\rangle_V^2$.
The root-mean-square (rms) Mach numbers weighted by volume and mass are defined by $\mathcal{M}_V=\abrackets{v^2}_V^{1/2}/c_s$
and $\mathcal{M}_M=\abrackets{v^2}_M^{1/2}/c_s$, respectively. The angle brackets $\abrackets{q}_V$ denote the average of quantity $q$ over the entire volume, while $\abrackets{q}_M\equiv\abrackets{\rho q}_V/\abrackets{\rho}_V$ denotes the mass-weighted average of $q$. As the mean velocity is nearly zero\footnote{AthenaK adopts a finite volume method that conserves the integrated mass and linear momentum of the system without source terms, keeping $\langle \rho \rangle_V=\rho_0$ and $\langle \rho \mathbf{v}\rangle_V=0$ at machine precision.
But, the turbulence simulations with mass and momentum source terms can lead to non-zero mean velocity and linear momentum.
We find that $|\abrackets{v_i}|<0.1c_s$ and $|\abrackets{\rho v_i}|<10^{-3}\rho_0 c_s$. We note that the PSD method adds mass at a rate $m_{\rm inj} \dot{N}_{\rm SN}$; total added mass over the simulation duration is $\approx 0.006$ and $0.024\rho_0 L^3$ for the \PSDlow\ and \PSDhigh\ models, respectively.}, the rms Mach numbers correspond to the volume and mass-weighted velocity dispersions. We then calculate the ratio of density to velocity dispersion using either volume and mass weighted Mach numbers, $b\equiv \sigma_\rho/\mathcal{M}_V$ and $b_M\equiv \sigma_\rho/\mathcal{M}_M$.\footnote{We note that $b_M$ is a measure of a global $b$-parameter excluding the contribution from low density voids that affects volume-weighted Mach number significantly. This serves as a representative value that is more observationally tractable (\ion{H}{I} or CO) and insensitive to the choice of numerical method of \PSD\ (see \autoref{sec:appendix_psd_param}), rather that a truly mass-weighted $b$-parameter, which would be defined by $\sigma_{\rho,M}/\mathcal{M}_M$ where $\sigma_{\rho,M}^2\equiv \abrackets{\rho^2}_M - \abrackets{\rho}_M^2$.}
Here, we omit the subscript $V$ for $b$ to be consistent with conventional definition in \autoref{eq:b} and simply use $\sigma_{\rho}$ given $\abrackets{\rho}\approx1$.
We summarize the steady-state values averaged over $t/t_0=0.4-1$ in the top five rows of \autoref{table:main_stats}.

\begin{table*}
    \centering
    \caption{Global statistics measured over $t/t_0=0.4-1$. $\mathcal{M}$ and $\mathcal{M}_M$: volume- and mass-weighted rms Mach numbers. $\sigma_\rho$: standard deviations of density. $ b$, $ b_M$: volume- and mass-weighted $b$-parameters. $\abrackets{s}_V$, $\sigma_{s,V}$, $\mathcal{S}_{s,V}$, $\mathcal{K}_{s,V}$ ($\abrackets{s}_M$, $\sigma_{s,M}$, $\mathcal{S}_{s,M}$, $\mathcal{K}_{s,M}$): volume-weighted (mass-weighted) log-density mean, standard deviation, skewness, and kurtosis. $\alphav$, $\alpharho$, $\alphavc$, $\alphavs$: power law slopes of the total velocity, density, compressive velocity, and solenoidal velocity power spectra (power-law fit over $16\le kL/2\pi\le40$ to the median power spectra).
    $L_{\rm in}$, $\teddy$: injection scale and eddy turnover time.
    $\abrackets{\Rcomp}_V$, $\abrackets{\Rcomp}_M$: volume- and mass-weighted mean of local compressive fraction (\autoref{eq:Rcomp}). 
    $\rcomp$: compressive velocity power ratio (\autoref{eq:rcomp}).
    \label{table:main_stats}}
    \begin{tabular}{l c c c c c c}
    \hline
    Quantity &  \PSDhigh & \FSDcomphigh & \FSDsolhigh & \PSDlow & \FSDcomplow & \FSDsollow \\
    \hline
        $\mathcal{M}_V$ & $4.4\pm0.8$ & $4.46\pm0.06$ & $4.34\pm0.03$ & $2.0\pm0.5$ & $2.20\pm0.02$ & $2.36\pm0.01$ \\ 
        $\mathcal{M}_M$ & $1.88\pm0.07$ & $3.69\pm0.05$ & $4.04\pm0.03$ & $1.22\pm0.07$ & $1.79\pm0.02$ & $2.17\pm0.01$ \\ 
        $\sigma_\rho$ & $1.40\pm0.03$ & $4.8\pm0.3$ & $1.62\pm0.02$ & $0.96\pm0.03$ & $2.55\pm0.09$ & $0.92\pm0.01$ \\ 
        $b$ & $0.33\pm0.05$ & $1.08\pm0.06$ & $0.373\pm0.005$ & $0.49\pm0.09$ & $1.16\pm0.04$ & $0.392\pm0.004$ \\ 
        $b_M$ & $0.74\pm0.03$ & $1.3\pm0.09$ & $0.401\pm0.005$ & $0.79\pm0.05$ & $1.42\pm0.06$ & $0.426\pm0.005$ \\ 
    \hline
        $\abrackets{s}_V$ & $-1.5$ & $-2.7$ & $-0.68$ & $-0.67$ & $-1.6$ & $-0.32$ \\ 
        $\sigma_{s,V}$ & $2.9$ & $2.7$ & $1.2$ & $1.9$ & $2.0$ & $0.81$ \\ 
        $\mathcal{S}_{s,V}$ & $-2.0$ & $-0.37$ & $-0.038$ & $-3.2$ & $-0.33$ & $-0.057$ \\ 
        $\mathcal{K}_{s,V}$ & $6.8$ & $3.1$ & $3.0$ & $17.$ & $3.0$ & $3.0$ \\ 
        $\abrackets{s}_M$ & $0.65$ & $2.0$ & $0.67$ & $0.37$ & $1.2$ & $0.32$ \\ 
        $\sigma_{s,M}$ & $0.92$ & $1.7$ & $1.1$ & $0.75$ & $1.4$ & $0.79$ \\ 
        $\mathcal{S}_{s,M}$ & $-0.59$ & $-0.43$ & $-0.091$ & $-0.36$ & $-0.46$ & $-0.090$ \\ 
        $\mathcal{K}_{s,M}$ & $4.4$ & $3.3$ & $2.9$ & $4.3$ & $3.3$ & $3.0$ \\ 
    \hline
        $\alphav$                      & $-1.9$ & $-2.0$ & $-2.2$ & $-1.8$ & $-2.0$ & $-2.1$ \\
        $\alpharho$                    & $-1.1$ & $-1.5$ & $-0.72$ & $-1.3$ & $-2.6$ & $-1.2$ \\
        $\alphavc$                     & $-2.0$ & $-1.9$ & $-1.9$ & $-2.0$ & $-2.0$ & $-1.8$ \\
        $\alphavs$                     & $-1.6$ & $-2.2$ & $-2.3$ & $-1.4$ & $-2.0$ & $-2.2$ \\
        $L_{\rm in}/L$ & $0.23$ & $0.19$ & $0.21$ & $0.22$ & $0.19$ & $0.2$ \\ 
        $\teddy$ & $0.026$ & $0.022$ & $0.024$ & $0.054$ & $0.043$ & $0.041$ \\ 
        $\abrackets{\Rcomp}_V$ & $0.63$ & $0.72$ & $0.41$ & $0.57$ & $0.64$ & $0.35$ \\ 
        $\abrackets{\Rcomp}_M$ & $0.5$ & $0.44$ & $0.34$ & $0.49$ & $0.41$ & $0.32$ \\ 
        $\rcomp$ & $0.76$ & $0.64$ & $0.2$ & $0.81$ & $0.65$ & $0.17$ \\ 
    \hline
    \end{tabular}
\end{table*}

\begin{figure}
    \centering
    \includegraphics[width=\linewidth]{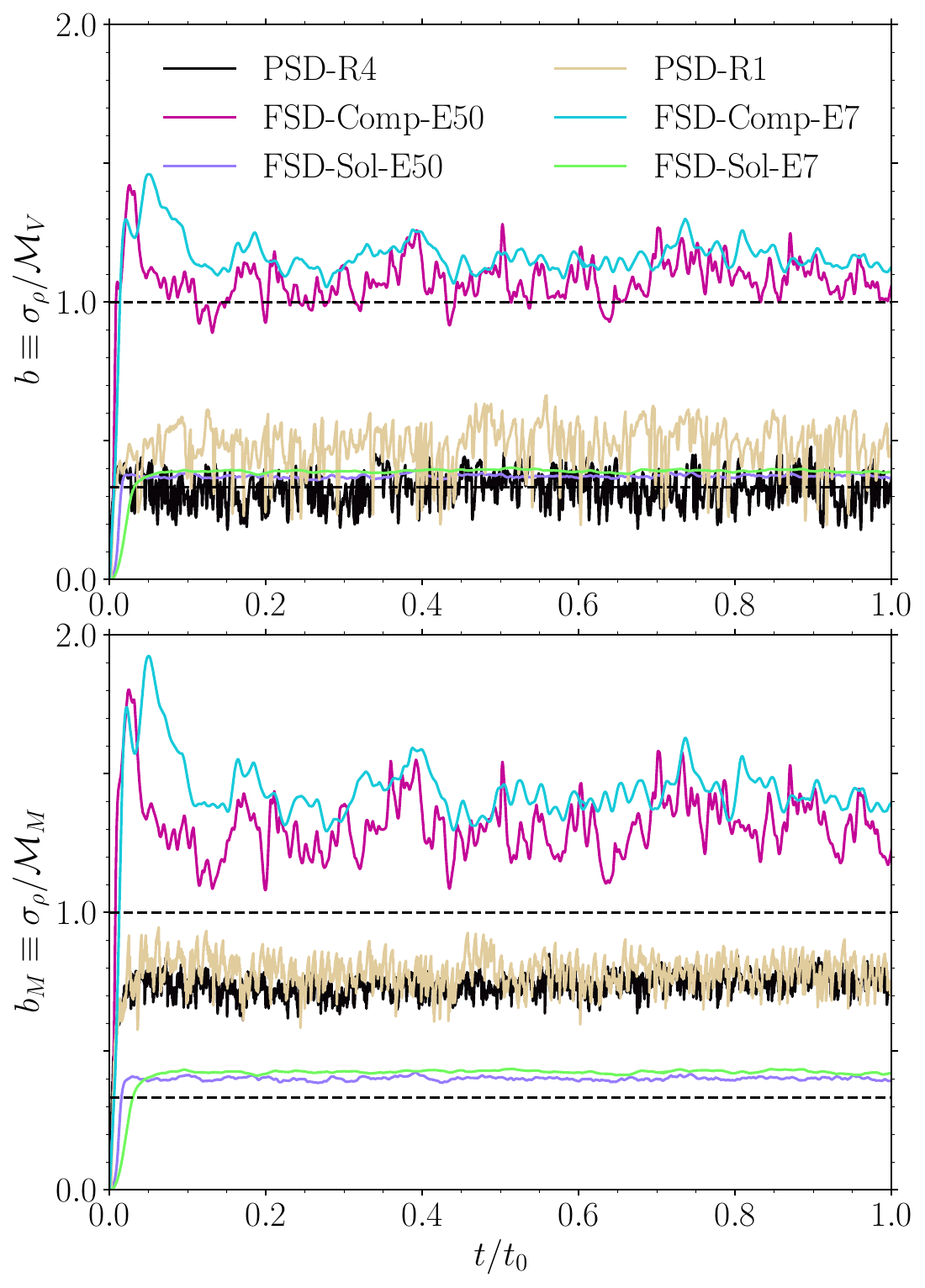}
    \caption{
    Time evolution of $b$-parameters using volume weighted Mach number $b\equiv \sigma_\rho /\mathcal{M}_V$ (top) and mass-weighted Mach number $b_M\equiv \sigma_\rho /\mathcal{M}_M$ (bottom). The steady-state average values between $t/t_0=0.4-1$ for these variables are shown in \autoref{table:main_stats}. The \PSD\ models roughly match the $b$ of the \FSDsol\ models. For $b_M$, which is less sensitive to the density floor applied in the \PSD\ models, the \PSD\ models have values in between the \FSDsol\ and \FSDcomp\ models.
    \label{fig:b_hist}
    }
\end{figure}

\autoref{fig:mach_dens_hist} shows the time evolution of $\mathcal{M}_V$, $\mathcal{M}_M$, and $\sigma_\rho$ for the fiducial model set.
Our choice of $\dot{E}$ leads to the \FSDhigh\ models to match the volume-weighted rms Mach number of the \PSDhigh\ model ($\mathcal{M}_V=4.37$, $4.46$, and $4.34$ respectively), while the \FSDlow\ models have mass-weighted rms Mach numbers ($\mathcal{M}_M=1.79$ and $2.17$) broadly comparable to that of the \PSDhigh\ model ($\mathcal{M}_M=1.88$).

The density fluctuations in the \PSDhigh\ and \PSDlow\ ($\sigma_\rho= 1.40$ and $0.96$, respectively) models are most similar to that in the \FSDsolhigh\ and \FSDsollow\ models ($\sigma_\rho= 1.62$ and $0.92$, respectively), while
both \FSDcomp\ models produce significantly larger $ \sigma_\rho = 4.81$ and $2.55$, respectively, than that in the \PSD\ models. This result is somewhat counterintuitive as the PSD method injects momentum in purely compressive motions.

\autoref{fig:b_hist} plots the driving parameters ($b$ and $b_M$) for the fiducial model set along with the reference lines for $b=1$ and $1/3$ as the predicted values for pure compressive and solenoidal driving, respectively \citep{2010A&A...512A..81F}. For the \FSD\ models, the standard mapping is consistent when using the volume-weighted Mach number. Unexpectedly, however, the \PSD\ models show $b=0.33$ and $0.49$, much closer to those of the \FSDsol\ models ($b\sim 0.37$--$0.39$) than to the \FSDcomp\ models ($b\sim 1.1$--$1.2$), and far below the predicted $b=1$ for purely compressive driving. This is again contrary to the intuition from the fact that PSD injects purely compressive motions to the system.

When using the mass-weighted Mach number, the driving parameters are unchanged for the \FSDsol\ models, while both \FSDcomp\ and \PSD\ models show larger $b_M$ than $b$ as $\mathcal{M}_M$ is smaller than $\mathcal{M}_V$ in these models.
As seen in \autoref{fig:slices}, the low density interiors of expanding bubbles occupy significant volume in the \PSD\ models.
Accounting for this high-velocity, low-density gas requires care because the properties of this region may depend sensitively on the specific momentum injection method as well as is affected by numerical choice like a density floor. At the same time, in observations, gas in a specific density range will be considered depending on gas tracers like HI or CO. We examine how the measured quantities change when a low density mask is applied.
We find that the gas with density lower than $10^{-2}\rho_0$ does not contribute much to the mass-weighted Mach number, while the volume-weighted Mach number is significantly affected by the low density gas. Using this density cut, $\mathcal{M}_V= 4.37\to 3.54$ for model \PSDhigh, while $\sigma_\rho$ is nearly unchanged at $1.40\to1.43$. This moderately increases $b=0.33\to 0.40$, but still far from the predicted $b=1$ for pure compressive driving. These same trends apply for \PSDlow, but with weaker strength: $\mathcal{M}_V=2.04\to 1.77$ yields $b=0.49\to0.55$.

Regardless of different choices of taking averages (volume and mass weighted with and without density cuts), we find that the driving parameters of the \PSD\ models are always smaller than the prediction of $b\approx 1$ for purely compressive driving with the FSD method.

\subsection{Probability Density Functions}\label{subsec:pdfs}

\begin{figure*}
    \centering
    \includegraphics[width=\textwidth]{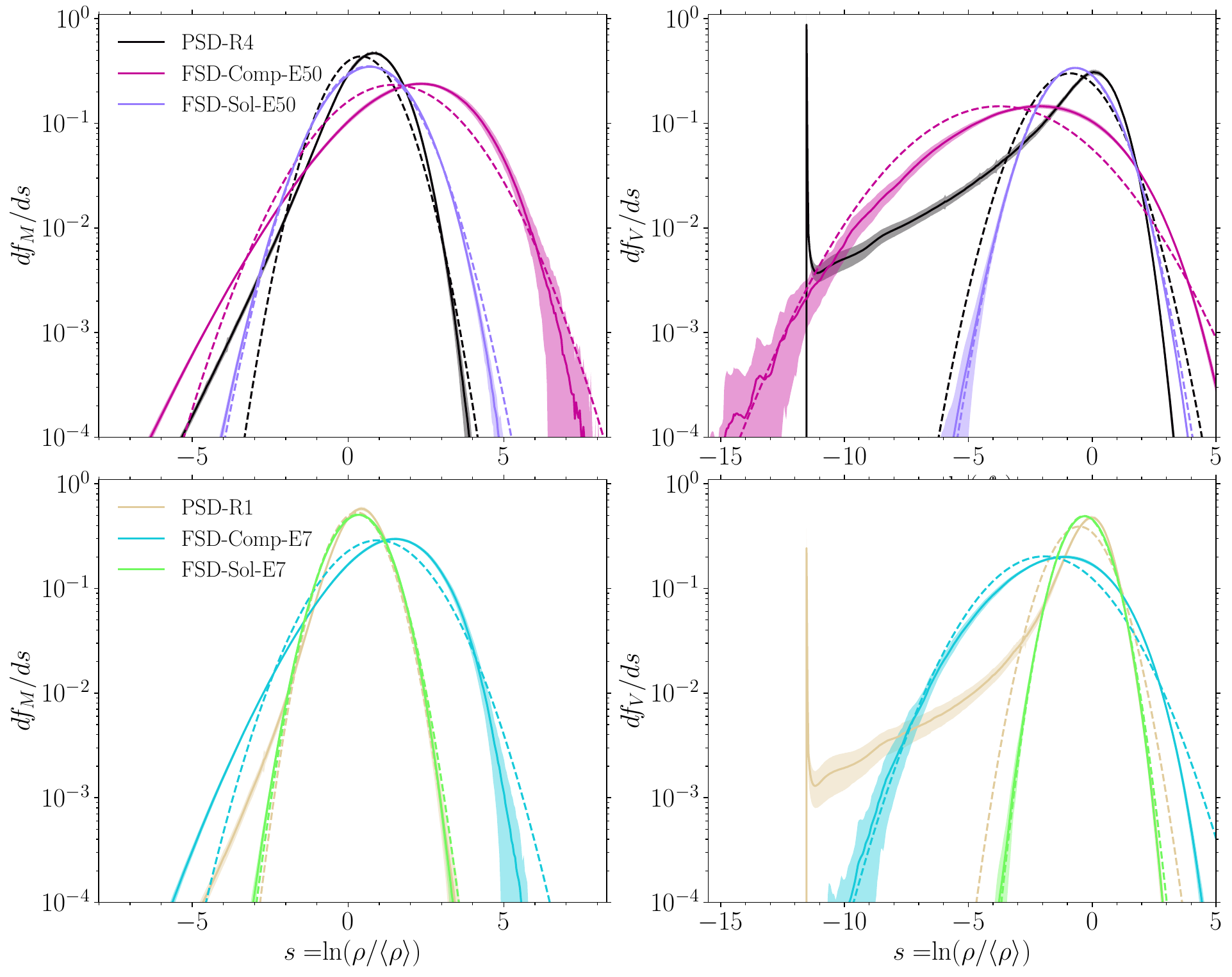}
    \caption{Probability density functions (PDFs) of the logarithmic density $s$ weighted by mass (left column) and volume (right column). Top row collects higher energy injection models (\PSDhigh\ and \FSDhigh), and bottom row collects lower energy injection models (\PSDlow\ and \FSDlow).
    The mean values over $t/t_0=0.4-1$ are shown in solid lines, while the shaded regions depict temporal fluctuations using standard deviations.
    The predicted log-normal PDFs (\autoref{eq:lognorm}) given $\sigma_s$ and $\sigma_{s,M}$ for volume and mass weighted PDFs, respectively,
    in \autoref{table:main_stats} are shown in dashed lines. The $\sigma_s$ value used for the $\PSD$ predicted volume weighted PDF uses a density floor of $\rho>0.01$, shifting the values for $\PSDhigh$ from $2.9\to1.3$, and for $\PSDlow$ from $1.9\to1.0$.
    The density floor applied in the \PSD\ models appear as a spike in the volume-weighted PDFs (right column), which represents tenuous interior of bubbles. The volume fraction of gas at $s<-10$ is still small; 0.042 and 0.012 for \PSDhigh\ and \PSDlow, respectively.
    We discuss the sensitivity of our results to the density floor in \autoref{sec:appendix_psd_param}.
    \label{fig:512_s_pdf_grid}}
\end{figure*}

\begin{figure*}
    \centering
    \includegraphics[width=\textwidth]{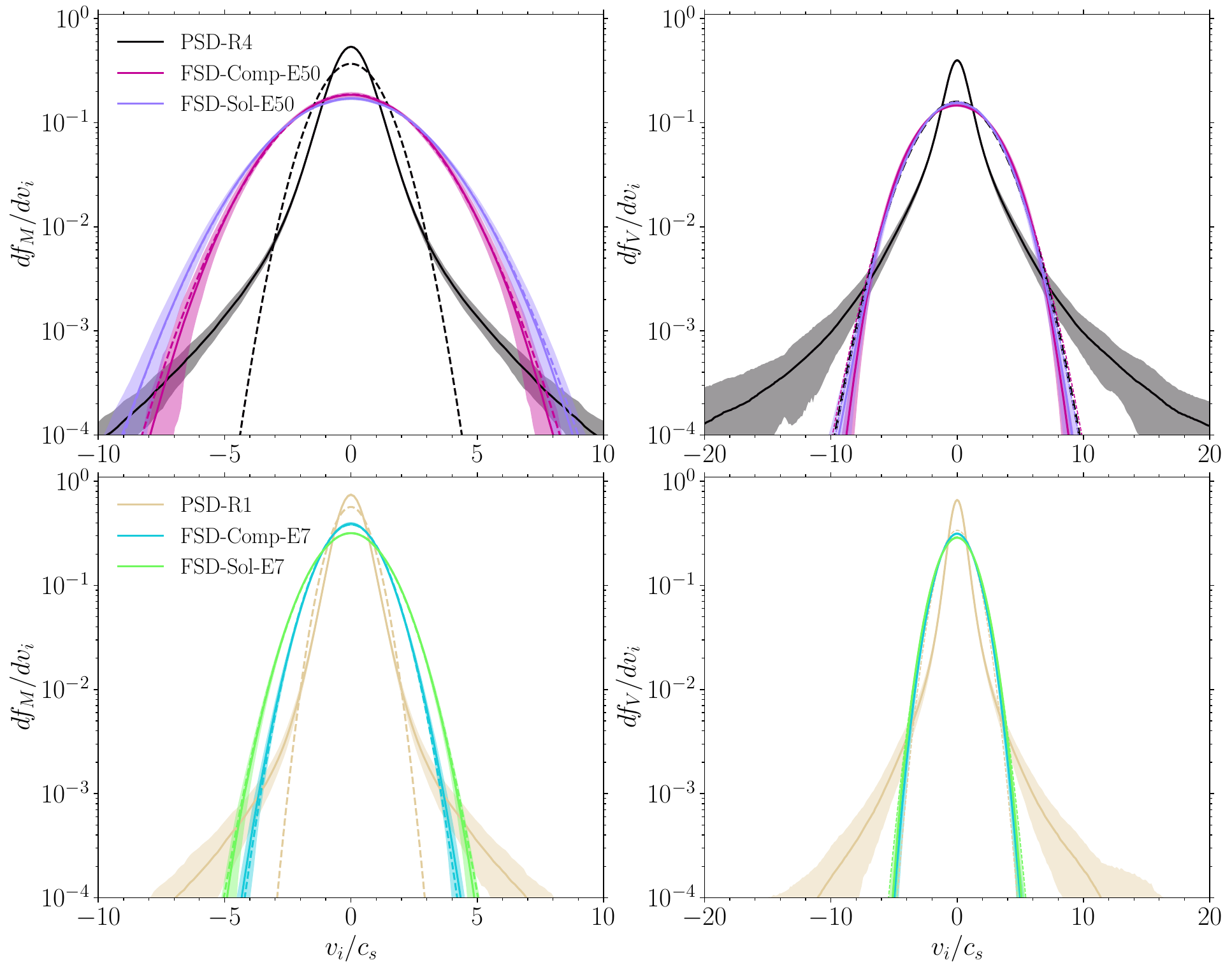}
    \caption{Mass-weighted (left column) and volume-weighted (right column) PDFs of each velocity component $v_i$ ($i=x,y,z$) for models \PSDhigh\ and \FSDhigh\ (top row), and models \PSDlow\ and \FSDlow\ (bottom row). All three components are overlaid for each model, demonstrating statistical isotropy. The PDFs were averaged over 60 Myr in simulation time. Shaded regions indicate $1\sigma$ standard deviations. Dashed lines show Gaussian profiles with zero mean and standard deviations using measured $\mathcal{M}_M$ and $\mathcal{M}_V$ (\autoref{table:main_stats}).  \label{fig:512_vel_pdf_grid}}
\end{figure*}

In the turbulence simulations using the FSD method with pure solenoidal modes, it is well known that the density PDF can be described by a log-normal distribution \citep{1994ApJ...423..681V,2001ApJ...546..980O,2007ARA&A..45..565M,2010A&A...512A..81F}. The PDF of the logarithmic density, $s\equiv \ln( \rho/\abrackets{\rho})$, follows a normal distribution
\begin{align}
    p_{s,V/M}&=\frac{1}{\sqrt{2\pi\sigma_{s,V/M}^2}}\rm exp \left[ -\frac{(s-\langle s \rangle_{V/M})^2}{2\sigma_{s,V/M}^2} \right]. \label{eq:lognorm}
\end{align}
The mass conservation sets a relation between the first and second moments, $\abrackets{s}_{V}=-\sigma_{s,V}^2/2$ and $\abrackets{s}_M=\sigma_{s,M}^2/2$ for the volume and mass weighted PDFs, respectively, with the same standard deviation $\sigma_{s,M}=\sigma_{s,V}$.
\autoref{fig:512_s_pdf_grid} shows both the mass/volume weighted PDFs $p_{s,M/V}=df_{M/V}/ds$ to describe 
the probability that a fractional mass/volume lies in $q$ to $q+dq$.
We show the corresponding log-normal PDF predictions (\autoref{eq:lognorm}) as dashed lines. For the \FSD\ models, the reference log-normal curves use the full-domain $\sigma_{s,V}$ and $\sigma_{s,M}$ listed in \autoref{table:main_stats}. For the \PSD\ models, the low-density interiors of expanding bubbles are dominated by the numerical density floor rather than physical turbulence (see below), so we compute $\sigma_{s,V}$ after excluding cells with $\rho<0.01\rho_0$; this shifts $\sigma_{s,V}$ from $2.9\to1.3$ for \PSDhigh\ and from $1.9\to1.0$ for \PSDlow, providing a more meaningful comparison with the log-normal prediction for the turbulent gas.

In the second half of \autoref{table:main_stats}, we list summary statistics of $s$.
We calculate up to the fourth moments for the PDFs of $s$, i.e., mean $\abrackets{s}_V$, variance $\sigma_{s,V}^2\equiv \abrackets{s^2}_V-\abrackets{s}_V^2$, skewness $\mathcal{S}_{s,V}\equiv \abrackets{(s-\abrackets{s}_V)^3}_V/\sigma_{s,V}^3$, and kurtosis $\mathcal{K}_{s,V}\equiv \abrackets{(s-\abrackets{s}_V)^4}_V/\sigma_{s,V}^4$ for the volume-weighted PDFs. The same quantities are calculated for the mass-weighted PDFs.
The higher-order moments quantify deviations from log-normality. The \FSDsol\ models have $\mathcal{S}_{s,V}\approx0$ and $\mathcal{K}_{s,V}\approx3$, consistent with a Gaussian distribution in $s$ and hence a log-normal density PDF. The \FSDcomp\ models show modest departures ($\mathcal{S}_{s,V}\approx-0.35$, $\mathcal{K}_{s,V}\approx3.1$), indicating a slight asymmetry toward low densities. The \PSD\ models exhibit far stronger departures from log-normality in the volume-weighted statistics: $\mathcal{S}_{s,V}=-2.0$ (\PSDhigh) and $-3.2$ (\PSDlow), with $\mathcal{K}_{s,V}=6.8$ and $17$, respectively, reflecting heavy tails produced by the low-density bubble interiors. Mass weighting suppresses the contribution of this tenuous gas, yielding more moderate values ($\mathcal{S}_{s,M}\approx-0.6$ to $-0.4$, $\mathcal{K}_{s,M}\approx4.3$--$4.4$), though still in excess of the Gaussian reference.
The low probability at high densities compared to the \FSDcomp\ models implies that bubble expansion alone in these isothermal \PSD\ models does not generate very high density gas; additional physical processes, e.g., gravity and thermal instability, may be required to generate equivalently strong converging flows. Even though both \PSD\ and \FSDcomp\ inject energy in purely compressive ways, the resulting density PDFs are sensitive to the exact form of driving.

Having examined the density distributions, we now turn to the velocity statistics. \autoref{fig:512_vel_pdf_grid} presents the mass-weighted (left) and volume-weighted (right) PDFs of each velocity component $v_i$ ($i=x,y,z$) shown together as the velocity field is statistically isotropic. The \FSD\ models exhibit similar, near-Gaussian profiles regardless of driving mode, with only modest departures for compressive forcing. This near-Gaussian character is expected: the acceleration field is constructed as a Gaussian random field in Fourier space (random phases at fixed spectral weights) so that the velocity field inherits much of this structure at scales near the driving range. Non-Gaussianity is somewhat larger for the \FSDcomp\ models, where shocks driven by convergent flows introduce non-Gaussian tails.

\begin{figure*}
    \centering
    \includegraphics[width=\textwidth]{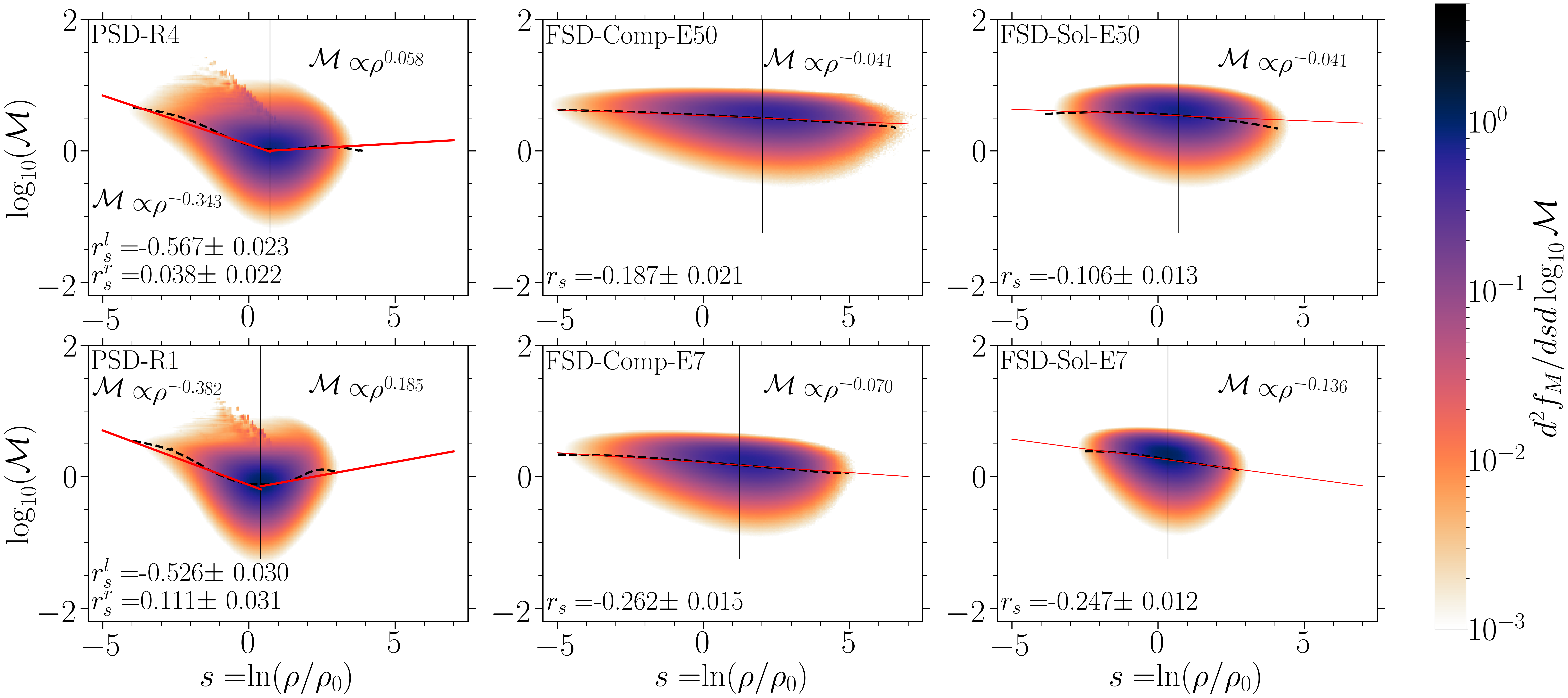}
    \caption{Mass-weighted correlation PDFs between logarithmic density $s$ and Mach number $\log_{10}\mathcal{M}$. From left to right, we show \PSD, \FSDcomp, and \FSDsol, while the models with the higher and lower energy injection rates are shown in the top and bottom rows, respectively. Red lines indicate the linear regression line.  The Spearman correlation coefficient is reported as $r_s$. In the \PSD\ panels, red lines indicate separate linear regression results for low and high density regimes ($s<\langle s\rangle_M$ and $s>\langle s\rangle_M$). The correlation coefficients for each respective regime are reported as $r_s^l$ and $r_s^r$.  Dashed black lines panels indicate $\bar{\mathcal{M}}$ as function of $s$.  \label{fig:512_2D_pdf_grid}}
\end{figure*}

The \PSD\ models, by contrast, display heavier tails extending to large $v$. The most prominent contribution comes from the high-velocity, low-density gas filling the interiors of expanding bubbles, which is why the tails are far more pronounced in the volume-weighted PDFs, consistent with the large separation between $\mathcal{M}_V$ and $\mathcal{M}_M$ noted in \autoref{subsec:global_stat}. However, the mass-weighted PDFs also show departures from Gaussian, indicating that the high-density shell gas swept up by expanding bubbles also carries substantial velocity. As we shall show in \autoref{fig:512_2D_pdf_grid}, the \PSD\ models exhibit an inversion in the $\rho$--$\mathcal{M}$ correlation, with high-density gas showing a weakly positive correlation in contrast to the overall negative trend.
Thus, while the \PSD\ density PDFs resemble the \FSDsol\ models at high densities (\autoref{fig:512_s_pdf_grid}), the velocity distributions are qualitatively distinct: despite matching the rms Mach numbers of the corresponding \FSD\ models, none of the \FSD\ models explored here reproduce the \PSD\ velocity PDFs.

\autoref{fig:512_2D_pdf_grid} shows the mass-weighted joint PDFs of the logarithmic density $s$ and Mach number $\log_{10}\mathcal{M}$, averaged over $t/t_0=0.4$--$1$. The most striking difference between the two driving methods is a sign reversal in the $s$--$\mathcal{M}$ correlation that appears only in the \PSD\ models. The conditional mean Mach number $\bar{\mathcal{M}}(s)$ (dotted black lines) reveals a V-shaped profile centred near the mass-weighted mean density $\abrackets{s}_M$: at low $s$ (rarefied bubble interiors), $\bar{\mathcal{M}}$ decreases steeply with increasing $s$, producing a strong negative correlation; at high $s$ (dense shell gas), $\bar{\mathcal{M}}$ rises with $s$, yielding a weak positive correlation. Separate linear fits to each branch (gray dotted lines) confirm this two-regime structure. Quantitatively, the Spearman coefficient restricted to the high-density regime ($s>\abrackets{s}_M$) is $r_s = -0.045\pm0.027$ for \PSDhigh\ ($\abrackets{s}_M = 0.65$) and $r_s = +0.069\pm0.026$ for \PSDlow\ ($\abrackets{s}_M = 0.37$), consistent with a flat-to-positive correlation in the overdense gas.

The physical origin of this sign reversal is tied to the morphology of \PSD\ driving. In the low-density regime, expanding bubbles evacuate their interiors, generating fast, diverging flows that create the strong negative $s$--$\mathcal{M}$ branch. In the high-density regime, the same bubbles sweep ambient gas into shells that are simultaneously compressed and accelerated outward, so that denser shell gas also moves faster, producing the positive branch. The \FSD\ models, by contrast, show only a monotonically negative $s$--$\mathcal{M}$ correlation at all densities. In \FSD\ turbulence, the densest gas forms at convergence points where opposing flows collide; these stagnation regions are compressed but carry low bulk velocity, so the $s$--$\mathcal{M}$ correlation remains negative throughout. This density-dependent sign reversal constitutes a qualitative signature of \PSD\ driving that is independent of the previously discussed differences in $\sigma_\rho$, $b$, or PDF shape. We show in \autoref{subsec:tcorr_pdfs} that the distinction persists even for \FSDcomp\ models with large $\tcorr$.

\subsection{Power Spectra} \label{subsec:spectra}
We now examine the power spectra of velocity and density, $\PS{v}(k)$ and $\PS{\rho}(k)$.
For a field $q(\bm{x})$ in a periodic domain, the spherical shell-averaged power spectrum is defined as
\begin{equation}
    \PS{q}(k)dk = \int \tilde{q}(\bm{k})\,\tilde{q}^*(\bm{k})4\pi k^2dk,
    \label{eq:ps_def}
\end{equation}
where $\tilde{q}$ is the Fourier transform of $q$ and $4\pi k^2dk$ is the volume of the spherical shell at wavenumber $k$ and $k+dk$.
Each spectrum is characterized by its power-law slope $\alpha_q$ ($\PS{q}(k)\propto k^{\alpha_q}$) fit over the range $16\le kL/2\pi\le40$.
The velocity field is further decomposed via the Helmholtz theorem into compressive and solenoidal components,
with corresponding power spectra $\PS{v,c}(k)$ and $\PS{v,s}(k)$.
We list the power law slopes of power spectra in the third part of \autoref{table:main_stats}.

Using the median velocity power spectra, we additionally list the injection scale $L_{\rm in}/L\equiv 2\pi \int \PS{v}(k) (kL)^{-1} dk/\int \PS{v}(k)dk$ and eddy turnover time $\teddy\equiv L_{\rm in}/(\mathcal{M}_Vc_s)$ \citep[e.g.,][]{2016ApJ...822...11P,2026ApJ...998..270S}.
For the \FSD\ models, the measured injection scales $L_{\rm in}/L\approx0.19$--$0.21$ are consistent with the imposed peak wavenumber $Lk_\mathrm{peak}/2\pi=5$. The resulting eddy turnover times ($\teddy/t_0\approx0.022$--$0.024$ for the high-energy models and $\teddy/t_0\approx0.041$--$0.043$ for the low-energy models) are comparable to the fiducial correlation times ($\tcorr/t_0=0.025$ and $0.05$, respectively), confirming the design choice $\tcorr\approx\teddy$ (\autoref{subsec:fsd}). The \PSD\ models yield similar values ($L_{\rm in}/L\approx0.22$--$0.23$, $\teddy/t_0\approx0.026$--$0.054$), which motivated the \FSD\ parameter choices.

We also present the ratio of compressive to total velocity power
\begin{equation}\label{eq:rcomp}
\rcomp \equiv \frac{\int P_{v,c}(k)dk}{\int(P_{v,c}(k) +  P_{v,s}(k)) dk}
\end{equation}
and the volume and mass weighted mean values of the local compressive fraction (\autoref{eq:Rcomp}), $\abrackets{\Rcomp}_V$ and $\abrackets{\Rcomp}_M$, respectively.
The power ratios of compressively driven models are $\rcomp\approx0.6-0.8$ with slightly higher values in \PSD\ than those of \FSDcomp.
These values are generally consistent with
$\abrackets{\Rcomp}_V\approx 0.5-0.7$, while the trend is reversed between \PSD\ and \FSDcomp. The solenoidally driven models have smaller $\rcomp\approx 0.2$ and $\abrackets{\Rcomp}_V\approx 0.4$. The differences between $\rcomp$ and $\abrackets{\Rcomp}_V$ as well as the reversed trend between \PSD\ and \FSDcomp\ are due to different effective weighting: velocity magnitude for $\rcomp$ and velocity gradient for $\abrackets{\Rcomp}_V$.

In the \PSD\ and \FSDcomp\ models, where purely compressive modes are injected either locally or globally, the mass-weighted mean $\abrackets{\Rcomp}_M\approx0.4-0.5$ is systematically lower than the volume-weighted value. This quantitatively supports visual impressions that solenoidal modes are generated preferentially in denser gas for those models (\autoref{subsec:morphology}). In the \FSDsol\ models, by contrast, $\Rcomp$ is relatively uniform in space, so the mass-weighted mean $\abrackets{\Rcomp}_M\approx0.3$ is only slightly below the volume-weighted value.

\begin{figure*}
    \centering
    \includegraphics[width=\textwidth]{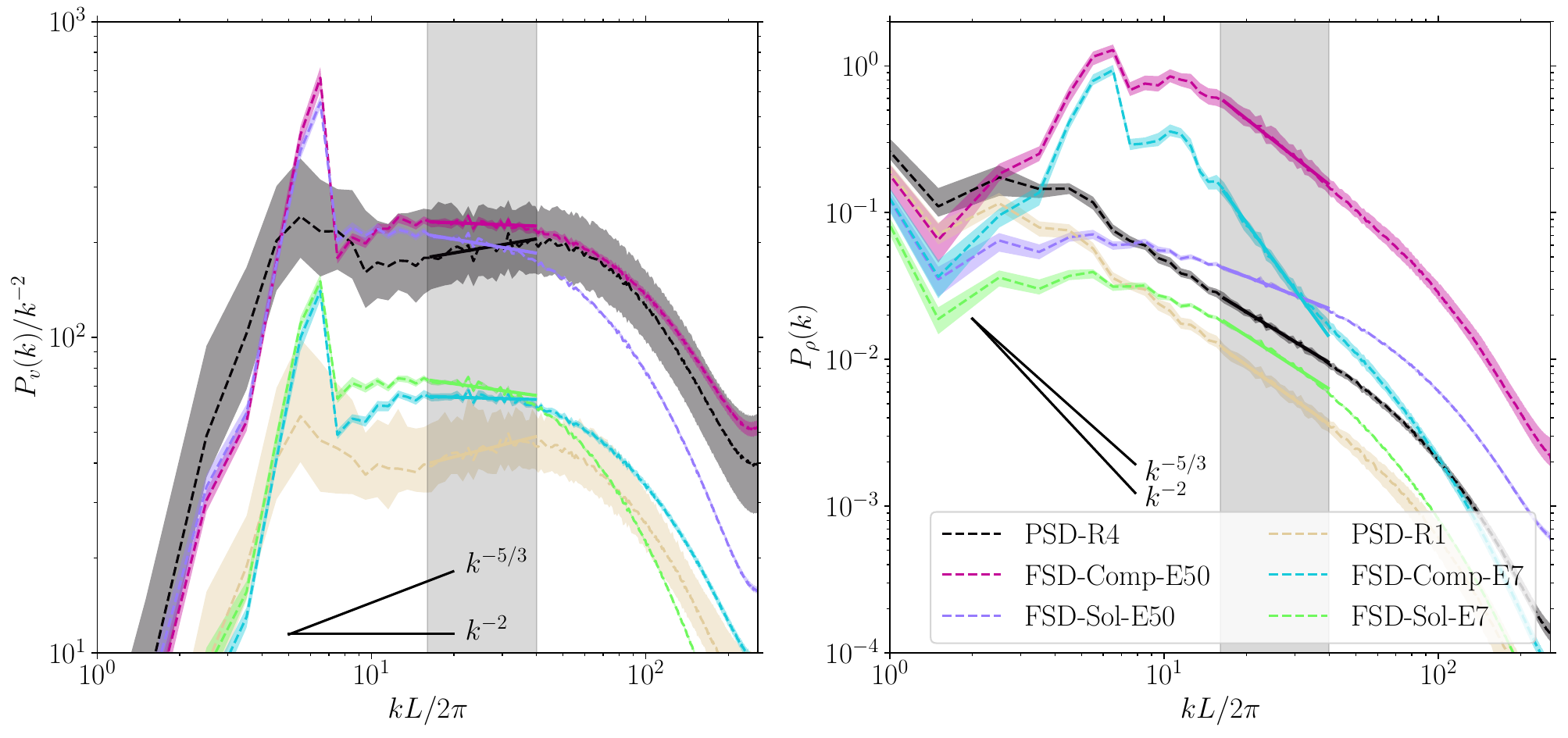}
    \caption{
    Compensated velocity (left) and uncompensated density (right) power spectra, $\PS{v}(k)/k^{-2}$ and $\PS{\rho}(k)$, so that Burgers scaling ($\alpha=-2$) appears as a horizontal line in the left panel. Solid colored lines show best-fit slopes over the fitting range $16\le k L/2\pi\le 40$ (grey shaded band), whose values are reported in \autoref{table:main_stats}. Curves are medians over $t/t_0=0.4-1$; shaded bands indicate the 16th--84th percentile range.
    \label{fig:512_spectra_row}}
\end{figure*}

\begin{figure}
    \centering
    \includegraphics[width=\linewidth]{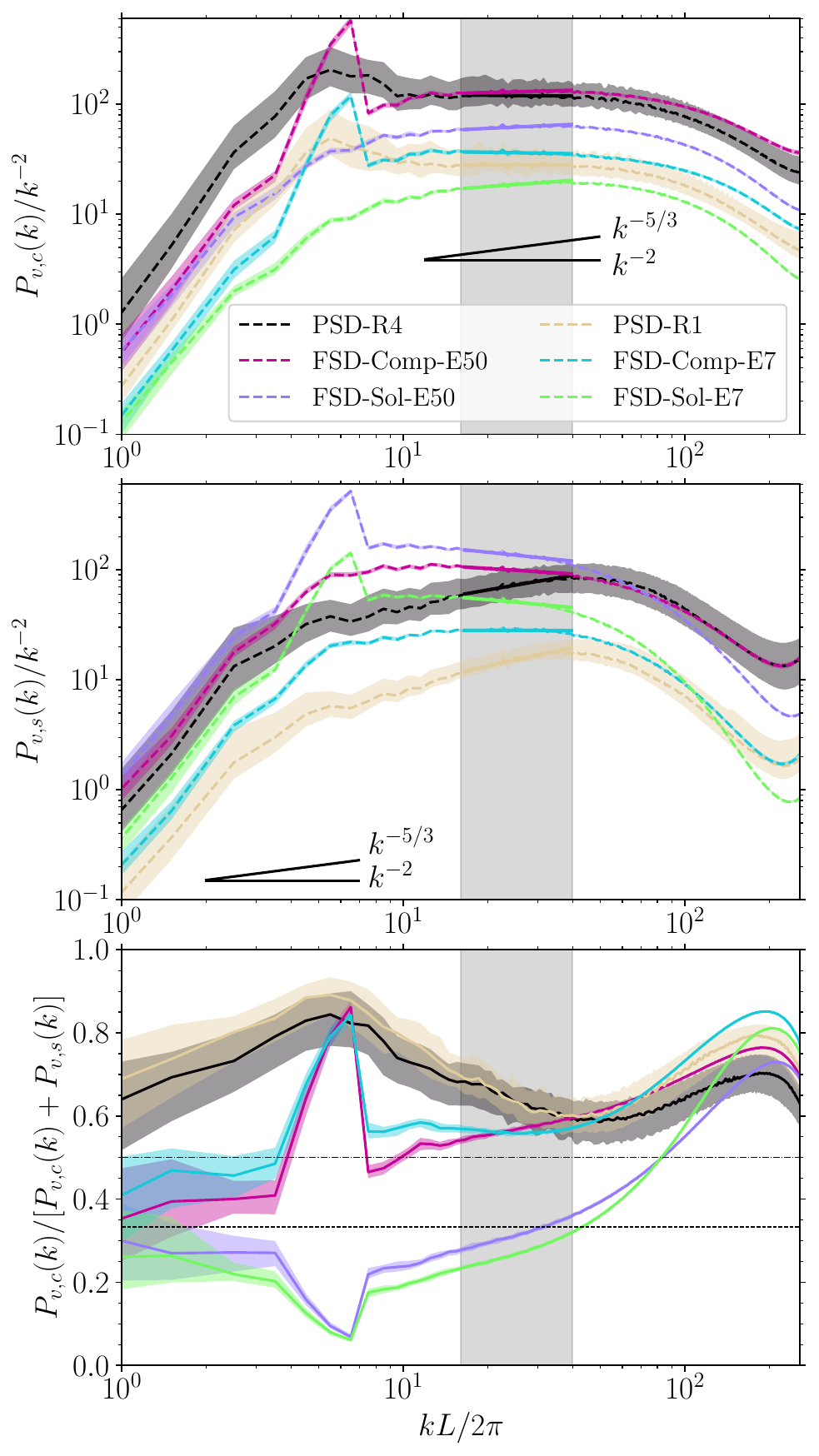}
    \caption{Top: compensated compressive velocity power spectrum $\PS{v,c}(k)/k^{-2}$. Middle: compensated solenoidal velocity power spectrum $\PS{v,s}(k)/k^{-2}$. Bottom: compressive fraction $\PS{v,c}(k)/[\PS{v,c}(k)+\PS{v,s}(k)]$; horizontal lines mark equipartition ($1/2$, dashed) and the natural ratio ($1/3$, dash-dotted). Slopes are fit over $16\le kL/2\pi\le40$ (gray shaded band). Curves are medians over $t/t_0=0.4-1$; shaded bands indicate the 16th--84th percentile range.
    \label{fig:512_helm_grid}}
\end{figure}

In supersonic isothermal turbulence, $\PS{v}$ is expected to follow Burgers scaling ($\alphav=-2$), steeper than Kolmogorov scaling ($\alphav=-5/3$), characteristic of incompressible turbulence. \autoref{fig:512_spectra_row} shows the velocity power spectra compensated by $k^{-2}$ (left), so that Burgers scaling appears as a flat profile, and the uncompensated density power spectra (right).

The velocity power spectra of all models cluster near Burgers scaling ($\alphav \approx -2$), regardless of driving method (\autoref{fig:512_spectra_row}, left).\footnote{We caution that the measured slopes here (almost certainly) may not be converged because (1) our choice of somewhat large $Lk_{\rm peak}/2\pi\sim5$ causes a limited scale separation between driving and (numerical) dissipation, and (2) the resolution requirement to resolve inertial range of turbulence is much more stringent than what we use here \citep{2016PhRvF...1h2403I,2025JFM..1019R...2Y}.} The \PSD\ models have slightly shallower slopes ($-1.9$ and $-1.8$ for \PSDhigh\ and \PSDlow, respectively) compared to the \FSD\ models, which all fall between $-2.0$ and $-2.2$. The \PSD\ models also exhibit broader temporal variance and no sharp peak defining a well-defined driving scale, in contrast to the \FSD\ models which show a clear injection peak at $Lk_\mathrm{peak}/2\pi=5$.

The density spectra (\autoref{fig:512_spectra_row}, right), on the other hand, show far greater variation between models, revealing two trends. First, within each driving method, higher Mach number produces a shallower density slope, consistent with \citet{2005ApJ...630L..45K}: for compressive FSD, \FSDcomplow\ has slope $-2.6$ at $\mathcal{M}_M=1.79$ while \FSDcomphigh\ has slope $-1.5$ at $\mathcal{M}_M=3.69$; for solenoidal FSD, \FSDsollow\ has slope $-1.2$ at $\mathcal{M}_M=2.17$ while \FSDsolhigh\ has a nearly flat spectrum (slope $-0.72$) at $\mathcal{M}_M=4.04$. Second, at similar $\mathcal{M}_M$, solenoidal driving produces shallower density slopes than compressive driving, which is somewhat surprising given that compressive driving more efficiently generates shock-compressed density structures whose Fourier transforms would favor shallower spectra.  Notably, the PSD models align much more closely with the \FSDsol\ models than the \FSDcomp\ ones: \PSDhigh\ (slope $-1.1$, $\mathcal{M}_M=1.88$) and \PSDlow\ (slope $-1.3$, $\mathcal{M}_M=1.22$) sit near \FSDsollow\ rather than \FSDcomplow\ despite having comparable $\mathcal{M}_M$.

The compressive velocity spectra (top panel of \autoref{fig:512_helm_grid}) show that the \FSDcomp\ models have a sharp injection peak at $k_\mathrm{peak}=5$ as imposed, while the \PSD\ models exhibit a broader peak at similar $k$. The compressive slopes are near Burgers scaling ($\alpha\approx-2$) for the \PSD\ and \FSDcomp\ models, while the \FSDsol\ models have somewhat shallower compressive slopes ($-1.88$ and $-1.81$ for \FSDsolhigh\ and \FSDsollow, respectively), likely reflecting the different character of compressive modes generated indirectly through nonlinear coupling rather than by direct compressive forcing.

The solenoidal spectra (middle panel) show that the \PSD\ models have markedly shallower slopes ($\alphavs = -1.6$ and $-1.4$ for \PSDhigh\ and \PSDlow) compared to all \FSD\ models, whose solenoidal slopes ($-2.0$ to $-2.3$) are close to Burgers scaling. $\alphavs\sim -1.5$ in the \PSD\ models is broadly consistent with SN-driven turbulence simulations with cooling and heating both in a periodic box \citep{2002ApJ...576..870P} and a stratified disk \citep{2025ApJ...994..193B, 2026ApJ...997...33C}. 
We note that the measured slopes serve as representative values for comparison between models rather than absolutely converged quantities, given our limited scale separation between driving and dissipation. Especially, the relatively steep slopes of solenoidal modes in our \FSD\ models can be due to a limited inertial range in our simulations; higher resolution simulations show the slopes $\alphavs\sim -1.5$ and $\alphavc\sim-2$ \citep[e.g.,][]{2025NatAs...9.1195B}.

The compressive fraction (bottom panel) most clearly illustrates the difference in driving character. The \FSDcomp\ peak and \FSDsol\ dip at $k_\mathrm{peak}=5$ directly reflect their respective imposed driving modes. The \PSD\ models show a broad peak near $k\sim5$--$6$, broadly consistent with the initial bubble injection radius $r_{\rm inj}\sim40\,\pc$ in a $L=500\,\pc$ box ($k\approx L/2r_{\rm inj}\approx6$), which motivated the choice of $k_\mathrm{peak}=5$ for the \FSD\ models. An experiment with a $2\times$ smaller $r_{\rm inj}$ confirms the peak shifts with injection scale, though not in strict proportion, suggesting it also reflects shell expansion dynamics. Overall, the \PSD\ models maintain a compressive fraction of $\sim$0.65--0.70 across all scales, above both the equipartition value ($1/2$) and the natural ratio ($1/3$).

All models show a rise in the compressive fraction towards small scales ($k\gtrsim40$).  \citet{2010A&A...512A..81F} find that solenoidal and compressive driving converge to $1/3$ and $1/2$, respectively, in an extended $k$ range; our models follow this trend qualitatively but do not fully converge given our limited scale separation. The rise at small scales reflects the preferential (numerical) dissipation of transverse (solenoidal) modes over longitudinal (compressive) modes near the grid scale: in a shock-capturing code, compressive modes are efficiently resolved by the Riemann solver, while vortical structures require finer resolution and are therefore truncated at somewhat larger scales. This is a consequence of grid discretization, and since all models are run with the same code, it does not affect our comparative conclusions. We note that the isothermal equation of state additionally suppresses baroclinic vorticity generation \citep{2011PhRvL.107k4504F,2022MNRAS.514.3139M,2025arXiv250907354B}, so the compressive fractions here should be regarded as upper limits relative to non-isothermal models.

The high compressive fraction of the \PSD\ models presents an apparent inconsistency with the density power spectrum results above, and highlights a key finding of this paper. Although the \PSD\ velocity field is predominantly compressive in energy partition, the resulting density fluctuations behave more like those produced by solenoidal forcing. This suggests that the standard $b$-parameter framework, which links density variance to Mach number and the compressive fraction of the driving, does not straightforwardly extend to the SN-driven turbulence.

\begin{figure*}
    \centering
    \includegraphics[width=\textwidth]{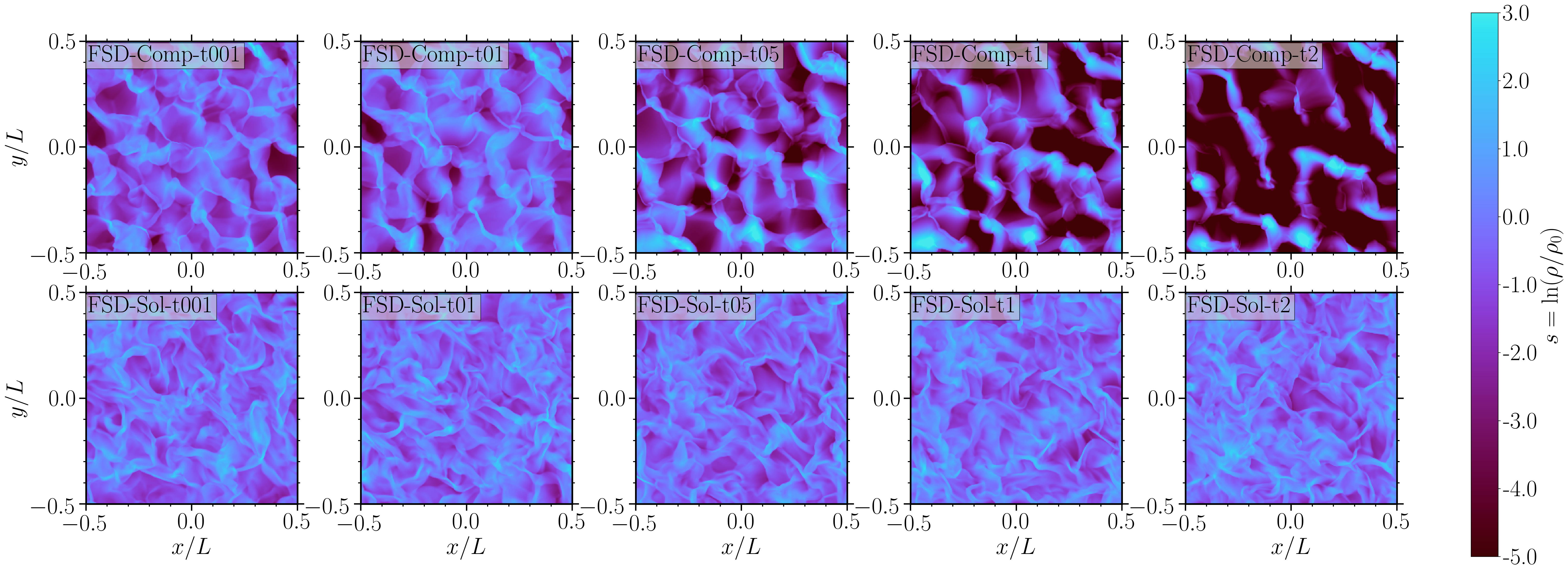}
    \caption{Mid-plane ($z=0$) slices of the logarithmic density field $s$ for the $\tcorr$-varying \FSD\ models at $N=256$. The top row shows the compressive models (\FSDtModel{Comp}{ZZZ}) and the bottom row shows the solenoidal models (\FSDtModel{Sol}{ZZZ}), with $\tcorr/t_0$ increasing from left to right. Longer $\tcorr$ produces progressively deeper and more extended low-density voids in the compressive models, while the solenoidal models show little visible change. \label{fig:tcorr_slices}}
\end{figure*}

\begin{figure*}
    \centering
    \includegraphics[width=\textwidth]{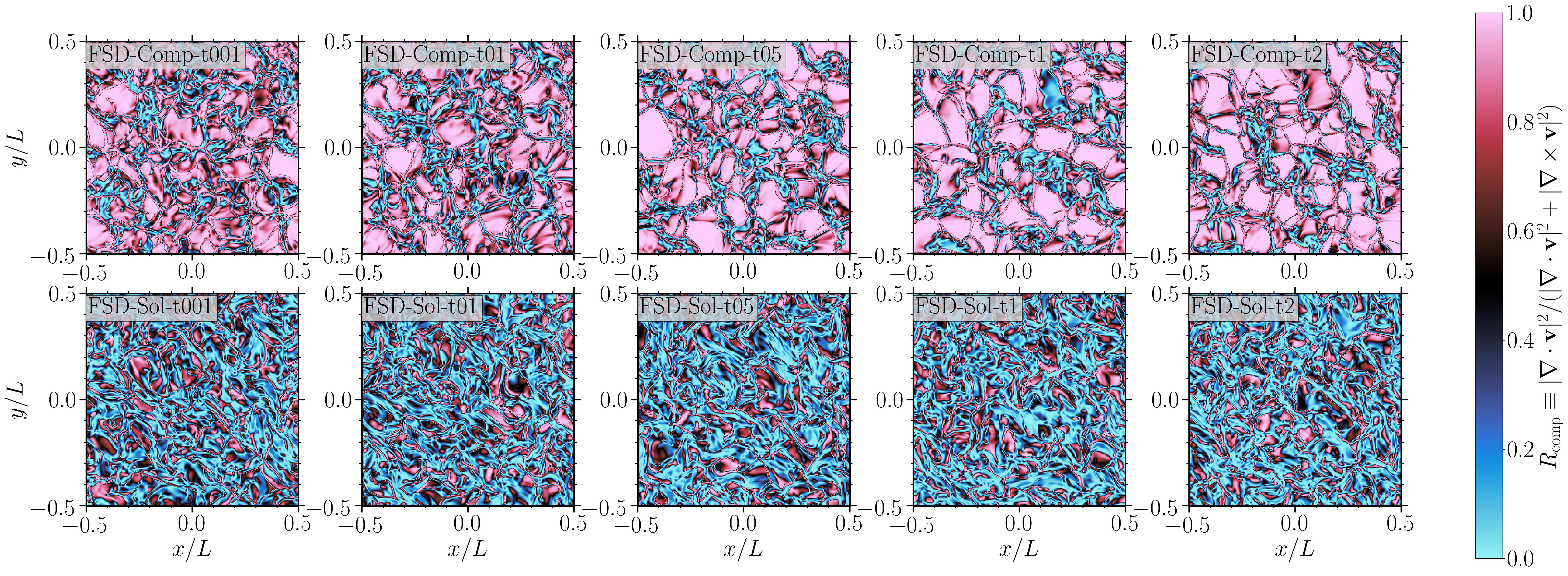}
    \caption{Same as \autoref{fig:tcorr_slices}, but for the local compressive fraction $\Rcomp \equiv |\nabla\cdot \vel|^2/(|\nabla\cdot \vel|^2+|\nabla\times \vel|^2)$. For the compressive models (top row), longer $\tcorr$ produces progressively more extensive regions of high $\Rcomp$, reflecting the buildup of coherent large-scale compressive flows sustained by the longer forcing correlation. The solenoidal models (bottom row) show little variation in $\Rcomp$ across the full range of $\tcorr$. \label{fig:tcorr_divcurl}}
\end{figure*}

\begin{figure}
    \centering
    \includegraphics[width=\linewidth]{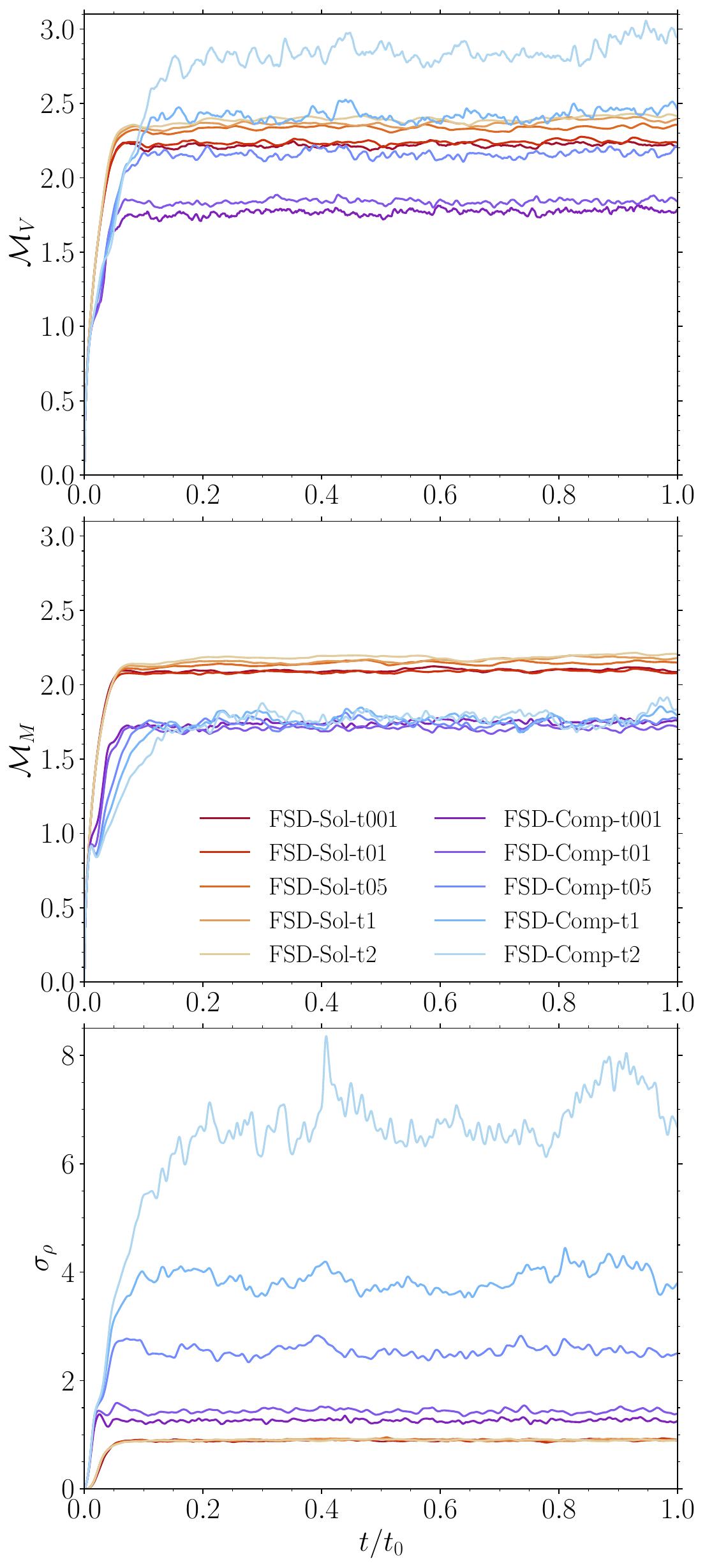}
    \caption{Time evolution of the volume-weighted Mach number $\mathcal{M}_V$ (top), mass-weighted Mach number $\mathcal{M}_M$ (middle), and density standard deviation $\sigma_\rho$ (bottom) for the $\tcorr$-varying \FSD\ models. Time-averaged values are reported in \autoref{table:tcorr_comp_stats} and \autoref{table:tcorr_sol_stats}. Both $\mathcal{M}_V$ and $\sigma_\rho$ increase with $\tcorr$ for the compressive models (\FSDtModel{Comp}{ZZZ}), while $\mathcal{M}_M$ shows a weaker trend, indicating that the highest velocities occur in low-density regions. The solenoidal models (\FSDtModel{Sol}{ZZZ}) display a much weaker dependence on $\tcorr$. The explicit dependence of these time-averaged quantities on $\tcorr$ is shown in \autoref{fig:tcorr_var_dependance}. \label{fig:tcorr_mach_dens_hist}}
\end{figure}

\begin{figure}
    \centering
    \includegraphics[width=\linewidth]{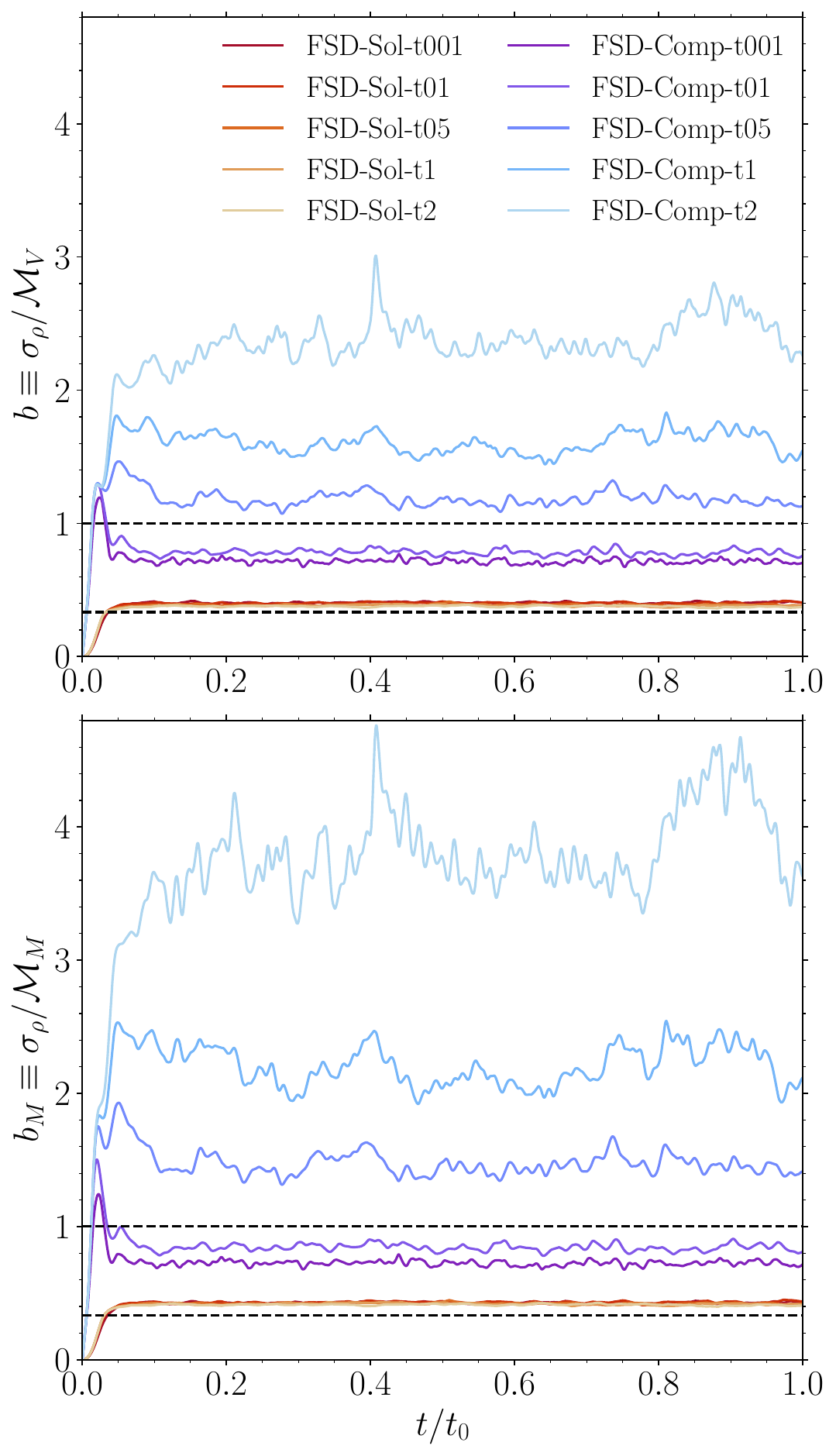}
    \caption{Time evolution of the volume-weighted $b$-parameter (top) and its mass-weighted counterpart $b_M$ (bottom) for the $\tcorr$-varying \FSD\ models. Time-averaged values are reported in \autoref{table:tcorr_comp_stats} and \autoref{table:tcorr_sol_stats}. Both $b$ and $b_M$ increase with $\tcorr$ for the compressive models (\FSDtModel{Comp}{ZZZ}), with the trend being stronger for $b_M$ owing to the weaker $\tcorr$ dependence of $\mathcal{M}_M$ relative to $\mathcal{M}_V$ (see \autoref{fig:tcorr_mach_dens_hist}). \label{fig:tcorr_b_hist}}
\end{figure}

\section{Effect of Forcing Correlation Time in the FSD Models} \label{sec:tcorr}

The comparison in \autoref{sec:psd_fsd} demonstrated that no single \FSD\ model reproduces the turbulence statistics of the \PSD\ simulations. In particular, the $b$-parameter of the \PSD\ models is comparable to that of solenoidal \FSD\ driving, yet a Helmholtz decomposition reveals that the \PSD\ velocity field is predominantly compressive. This fundamental mismatch motivates an exploration of whether additional \FSD\ parameters beyond the driving mode $F_{\rm sol}$ and amplitude $\dot{E}$ can bring the two frameworks into closer agreement. Among these parameters, the forcing correlation time $\tcorr$ deserves particular scrutiny: it is poorly constrained for real ISM turbulence and has no direct analogue in the spatially and temporally stochastic momentum injection of \PSD\ driving \citep{2018ApJ...858L..19G}.

To isolate the effect of $\tcorr$, we run 10 models at resolution $N=256$ with $\dot{E}=7\,E_0/t_0$, varying $\tcorr/t_0=0.001,\,0.01,\,0.05,\,0.1,\,0.2$ for both purely compressive ($F_{\rm sol}=0$) and purely solenoidal ($F_{\rm sol}=1$) driving (see \autoref{subsec:models} for details). We find that $\tcorr$ has a strong effect on the density statistics of compressive driving but leaves solenoidal models largely unchanged. This asymmetric sensitivity introduces an additional degeneracy in the $b$-parameter: for compressive \FSD, the same driving mode can produce a wide range of $b$ values depending on $\tcorr$, further undermining its reliability as a diagnostic of the driving mode. Our results are consistent with the recent finding by \citet{2026ApJ...998..270S} that the density variance--Mach number relation in compressively driven turbulence depends explicitly on the driving correlation time. Below, we examine this dependence through density-field morphology (\autoref{subsec:tcorr_morphology}), time-averaged global statistics (\autoref{subsec:tcorr_global_stat}), one-dimensional PDFs (\autoref{subsec:tcorr_pdfs}), and joint $s$--$\mathcal{M}$ distributions.

\begin{table*}
        \caption{Same as \autoref{table:main_stats}, but for $\tcorr$-varying \FSDcomp\ models, without power spectrum slopes and with $\tcorr/\teddy$. \label{table:tcorr_comp_stats}}
        \begin{tabular}{l | c c c c c}
            \hline
            Quantity & \FSDtModel{Comp}{001} & \FSDtModel{Comp}{01} & \FSDtModel{Comp}{05} & \FSDtModel{Comp}{1} & \FSDtModel{Comp}{2}\\
            \hline
                $\mathcal{M}_V$ & $1.77\pm0.02$ & $1.84\pm0.02$ & $2.16\pm0.02$ & $2.42\pm0.04$ & $2.9\pm0.07$ \\ 
                $\mathcal{M}_M$ & $1.75\pm0.01$ & $1.71\pm0.01$ & $1.75\pm0.02$ & $1.77\pm0.03$ & $1.78\pm0.04$ \\ 
                $\sigma_\rho$ & $1.27\pm0.03$ & $1.44\pm0.04$ & $2.5\pm0.09$ & $3.9\pm0.2$ & $6.9\pm0.5$ \\ 
                $b$ & $0.71\pm0.02$ & $0.78\pm0.02$ & $1.18\pm0.04$ & $1.6\pm0.09$ & $2.4\pm0.1$ \\ 
                $b_M$ & $0.72\pm0.02$ & $0.84\pm0.03$ & $1.5\pm0.06$ & $2.2\pm0.2$ & $3.9\pm0.3$ \\ 
            \hline
                $\abrackets{s}_V$ & $-0.56$ & $-0.73$ & $-1.6$ & $-2.5$ & $-4.2$ \\ 
                $\sigma_{s,V}$ & $1.1$ & $1.3$ & $2.0$ & $2.5$ & $3.4$ \\ 
                $\mathcal{S}_{s,V}$ & $-0.20$ & $-0.33$ & $-0.33$ & $-0.26$ & $-0.18$ \\ 
                $\mathcal{K}_{s,V}$ & $3.0$ & $3.0$ & $3.0$ & $2.9$ & $2.8$ \\ 
                $\abrackets{s}_M$ & $0.52$ & $0.63$ & $1.2$ & $1.8$ & $2.7$ \\ 
                $\sigma_{s,M}$ & $0.97$ & $1.0$ & $1.4$ & $1.6$ & $1.8$ \\ 
                $\mathcal{S}_{s,M}$ & $-0.25$ & $-0.38$ & $-0.47$ & $-0.58$ & $-0.69$ \\ 
                $\mathcal{K}_{s,M}$ & $3.1$ & $3.2$ & $3.3$ & $3.4$ & $3.6$ \\ 
            \hline
                $L_{\rm in}/L$ & $0.22$ & $0.21$ & $0.19$ & $0.19$ & $0.2$ \\ 
                $\teddy$ & $0.063$ & $0.056$ & $0.045$ & $0.04$ & $0.035$ \\ 
                $\tcorr/\teddy$ & $0.016$ & $0.18$ & $1.1$ & $2.5$ & $5.8$ \\ 
                $\rcomp$ & $0.62$ & $0.63$ & $0.66$ & $0.68$ & $0.69$ \\ 
                $\abrackets{\Rcomp}_V$ & $0.62$ & $0.63$ & $0.66$ & $0.68$ & $0.71$ \\ 
                $\abrackets{\Rcomp}_M$ & $0.5$ & $0.48$ & $0.43$ & $0.41$ & $0.39$ \\ 
            \hline
        \end{tabular}
\end{table*}

\begin{table*}
        \caption{Same as \autoref{table:tcorr_comp_stats}, but for $\tcorr$-varying \FSDsol\ models. \label{table:tcorr_sol_stats}}
        \begin{tabular}{l | c c c c c}
            \hline
            Quantity & \FSDtModel{Sol}{001} & \FSDtModel{Sol}{01} & \FSDtModel{Sol}{05} & \FSDtModel{Sol}{1} & \FSDtModel{Sol}{2}\\
            \hline
                $\mathcal{M}_V$ & $2.22\pm0.01$ & $2.24\pm0.01$ & $2.34\pm0.01$ & $2.37\pm0.02$ & $2.40\pm0.02$ \\ 
                $\mathcal{M}_M$ & $2.10\pm0.01$ & $2.09\pm0.008$ & $2.15\pm0.01$ & $2.17\pm0.01$ & $2.19\pm0.01$ \\ 
                $\sigma_\rho$ & $0.90\pm0.01$ & $0.90\pm0.01$ & $0.91\pm0.01$ & $0.91\pm0.01$ & $0.90\pm0.01$ \\ 
                $b$ & $0.41\pm0.006$ & $0.40\pm0.007$ & $0.39\pm0.005$ & $0.38\pm0.006$ & $0.38\pm0.006$ \\ 
                $b_M$ & $0.43\pm0.006$ & $0.43\pm0.008$ & $0.42\pm0.006$ & $0.42\pm0.007$ & $0.41\pm0.006$ \\ 
            \hline
                $\abrackets{s}_V$ & $-0.31$ & $-0.31$ & $-0.32$ & $-0.32$ & $-0.32$ \\ 
                $\sigma_{s,V}$ & $0.79$ & $0.79$ & $0.81$ & $0.81$ & $0.81$ \\ 
                $\mathcal{S}_{s,V}$ & $-0.048$ & $-0.052$ & $-0.073$ & $-0.081$ & $-0.10$ \\ 
                $\mathcal{K}_{s,V}$ & $3.0$ & $2.9$ & $2.9$ & $2.9$ & $3.0$ \\ 
                $\abrackets{s}_M$ & $0.30$ & $0.30$ & $0.31$ & $0.31$ & $0.31$ \\ 
                $\sigma_{s,M}$ & $0.77$ & $0.77$ & $0.78$ & $0.78$ & $0.77$ \\ 
                $\mathcal{S}_{s,M}$ & $-0.073$ & $-0.088$ & $-0.11$ & $-0.13$ & $-0.14$ \\ 
                $\mathcal{K}_{s,M}$ & $3.0$ & $3.0$ & $3.0$ & $3.0$ & $3.0$ \\ 
            \hline
                $L_{\rm in}/L$ & $0.19$ & $0.19$ & $0.22$ & $0.22$ & $0.24$ \\ 
                $\teddy$ & $0.042$ & $0.043$ & $0.046$ & $0.046$ & $0.049$ \\ 
                $\tcorr/\teddy$ & $0.024$ & $0.23$ & $1.1$ & $2.2$ & $4.1$ \\ 
                $\rcomp$ & $0.19$ & $0.19$ & $0.17$ & $0.16$ & $0.15$ \\ 
                $\abrackets{\Rcomp}_V$ & $0.35$ & $0.34$ & $0.33$ & $0.33$ & $0.32$ \\ 
                $\abrackets{\Rcomp}_M$ & $0.32$ & $0.31$ & $0.3$ & $0.3$ & $0.29$ \\ 
            \hline
        \end{tabular}
\end{table*}

\begin{figure*}
    \centering
    \includegraphics[width=\textwidth]{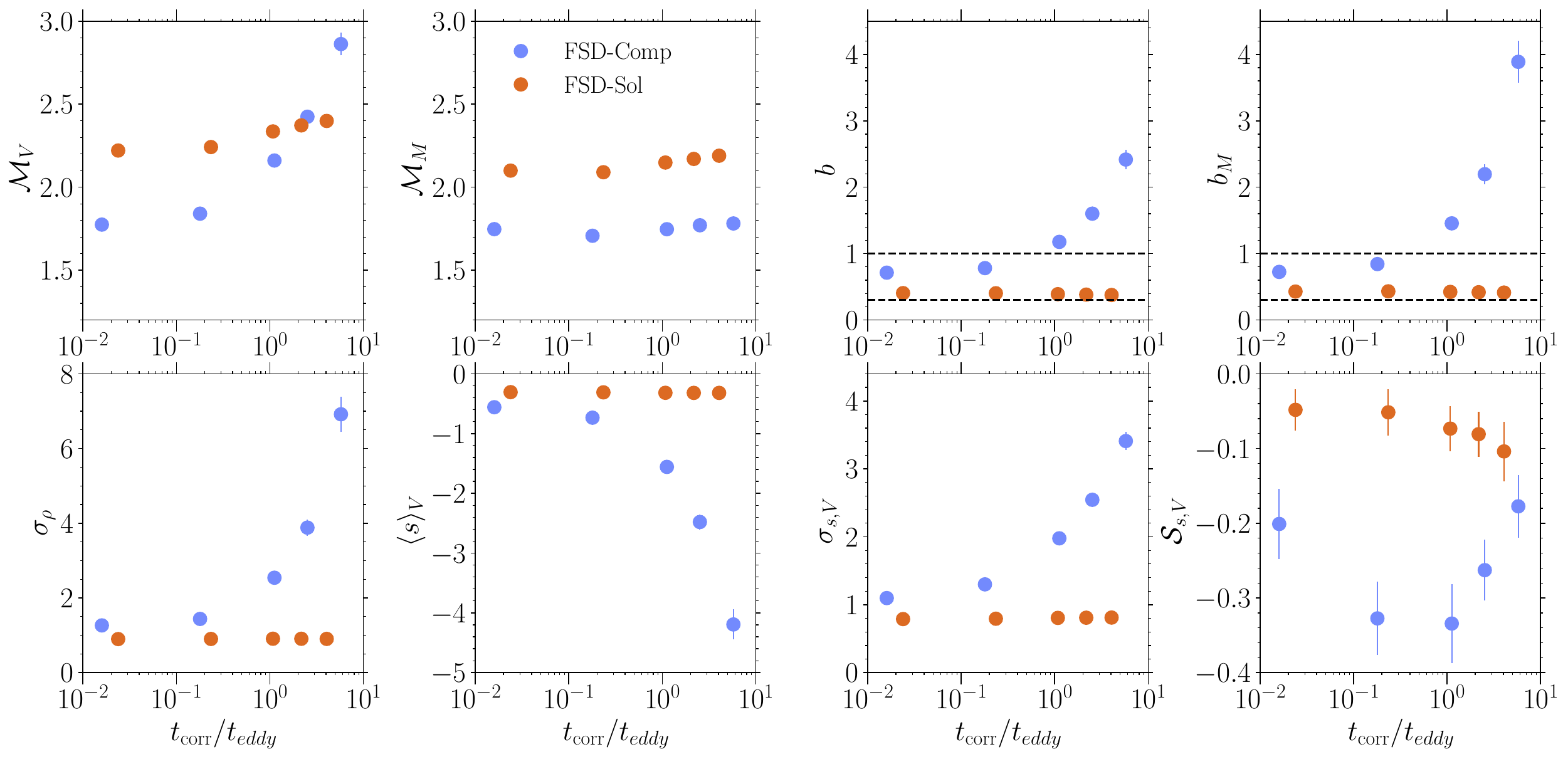}
    \caption{Dependence of time-averaged quantities on $\tcorr$ for the compressive (\FSDtModel{Comp}{ZZZ}; purple) and solenoidal (\FSDtModel{Sol}{ZZZ}; blue) model sets. {\bf Top row:} volume and mass-weighted rms Mach numbers, $\mathcal{M}_V$ (first) and $\mathcal{M}_M$ (second) and volume- and mass-weighted $b$-parameters, $b=\sigma_\rho/\mathcal{M}_V$ (third) and $b_M=\sigma_\rho/\mathcal{M}_M$ (fourth). {\bf Bottom row:} from left to right, linear density standard deviation $\sigma_\rho$ and first to third moments of logarithmic density,  $\abrackets{s}_V$,  $\sigma_{s,V}$, and $\mathcal{S}_{s,V}$. The dashed horizontal lines in the $b$-parameter panels mark $b=1$ and $b=1/3$. The compressive models show strong dependence on $\tcorr$. \label{fig:tcorr_var_dependance}}
\end{figure*}

\subsection{Morphology} \label{subsec:tcorr_morphology}

\autoref{fig:tcorr_slices} shows $z=0$ slices of $s$ for all 10 models. Longer $\tcorr$ in the compressive models (top row; left to right) produces progressively more extreme density contrasts: the low-density voids deepen and expand, while the high-density structures sharpen and steepen. The solenoidal models (bottom row), by contrast, show little visible change across the full range of $\tcorr$. This asymmetry is physically intuitive: a longer correlation time allows the compressive acceleration field to persist in a given direction, driving coherent converging and diverging flows that amplify density contrasts. Solenoidal forcing, being dominated by vortical motions, does not generate strong density fluctuations regardless of how long the pattern is maintained.

\autoref{fig:tcorr_divcurl} shows $\Rcomp$ for the same models. As $\tcorr$ increases, progressively more extended regions of high $\Rcomp$ emerge in the compressive models (top row), reflecting the buildup of large-scale, coherent compressive motions that the longer-lived forcing field can drive. This spatial organization of $\Rcomp$ mirrors the density morphology in \autoref{fig:tcorr_slices}: the low density structures visible at large $\tcorr$ are precisely where $\Rcomp$ is high. The solenoidal models (bottom row) show little variation in $\Rcomp$ with $\tcorr$, consistent with the insensitivity of their density statistics.

The quantitative measures of volume and mass-weighted mean values of $\Rcomp$ show systematic variations with $\tcorr$ (\autoref{table:tcorr_comp_stats} and \autoref{table:tcorr_sol_stats}). From $\tcorr/t_0=0.001$ to $0.2$ (left to right in \autoref{fig:tcorr_divcurl}), $\abrackets{\Rcomp}_V=0.62\to 0.71$ and $\abrackets{\Rcomp}_M=0.5\to0.39$ in the \FSDcomp\ models. The increase of the volume-weighted mean $\abrackets{\Rcomp}_V$ as $\tcorr$ gets longer is consistent with the visual impression, while the decrease of the mass-weighted mean $\abrackets{\Rcomp}_M$ reflects that the coherent compressive motions actually promote solenoidal mode generation preferentially in denser gas. In the \FSDsol\ models, a weaker decreasing trend is seen in both volume and mass-weighted mean of $\Rcomp$: $\abrackets{\Rcomp}_V=0.35\to 0.32$ and $\abrackets{\Rcomp}_M=0.32\to 0.29$. This likely reflects the fact that more coherent driving preserves the injected modal character more effectively. Solenoidal driving results in the fraction consistent with the natural mix $\Rcomp\sim1/3$ for a wide range of correlation times.

We also measure the compressive power ratio $\rcomp$ (\autoref{eq:rcomp}) from the Helmholtz-decomposed velocity power spectra following \autoref{subsec:spectra}, though we do not present the individual spectra for these lower-resolution models\footnote{In the fiducial model set, the power spectra in the $256^3$ models begin to diverge from the $512^3$ models at $kL/2\pi\sim15$.}. For the \FSDcomp\ models, $\rcomp$ increases weakly from $0.62$ to $0.69$ as $\tcorr/t_0$ increases from $0.001$ to $0.2$, mirroring the trend in $\abrackets{\Rcomp}_V$. The \FSDsol\ models show a slight decrease from $0.19$ to $0.15$, consistent with their overall insensitivity to $\tcorr$. The power ratio measurement $\rcomp$ weights more on the higher velocity region, and the compressive mode is mainly generated at shocks and colliding flows where velocity is lower, yielding $\rcomp <\abrackets{\Rcomp}_V$. The opposite trend is observed in the \PSD\ models where the high-velocity expanding motions are injected locally, yielding $\rcomp>\abrackets{\Rcomp}_V$.

\subsection{Global Statistics}\label{subsec:tcorr_global_stat}

We now examine how $\tcorr$ modulates the global statistics of the \FSD\ models. \autoref{fig:tcorr_mach_dens_hist} and \autoref{fig:tcorr_b_hist} show the time evolution of $\mathcal{M}_V$, $\mathcal{M}_M$, $\sigma_\rho$, $b$, and $b_M$; time-averaged values are reported in \autoref{table:tcorr_comp_stats} and \autoref{table:tcorr_sol_stats}, and their explicit dependence on $\tcorr$ is shown in \autoref{fig:tcorr_var_dependance}. The injection scale $L_{\rm in}$, eddy turnover time $\teddy$, and normalized correlation time $\tcorr/\teddy$ for each model are also listed in these tables.

For the \FSDcomp\ models, $\tcorr$ has a dramatic effect on all density-related statistics. As $\tcorr/t_0$ increases from $0.001$ to $0.2$, $\sigma_\rho$ rises from $1.27$ to $6.92$, while the Mach number shows only a moderate increase ($\mathcal{M}_V = 1.77 \to 2.86$). The resulting $b$ spans from $0.71$ to $2.42$, a factor of $\sim3.4$ variation driven entirely by the choice of correlation time. Since the mass-weighted Mach number remains nearly constant ($\mathcal{M}_M \approx 1.75$), $b_M$ rises even more steeply, from $0.72$ to $3.89$. The near-constancy of $\mathcal{M}_M$ in contrast to the substantial increase in $\mathcal{M}_V$ indicates that the extra kinetic energy at large $\tcorr$ resides predominantly in low-density gas. 

\begin{figure*}
    \centering
    \includegraphics[width=\textwidth]{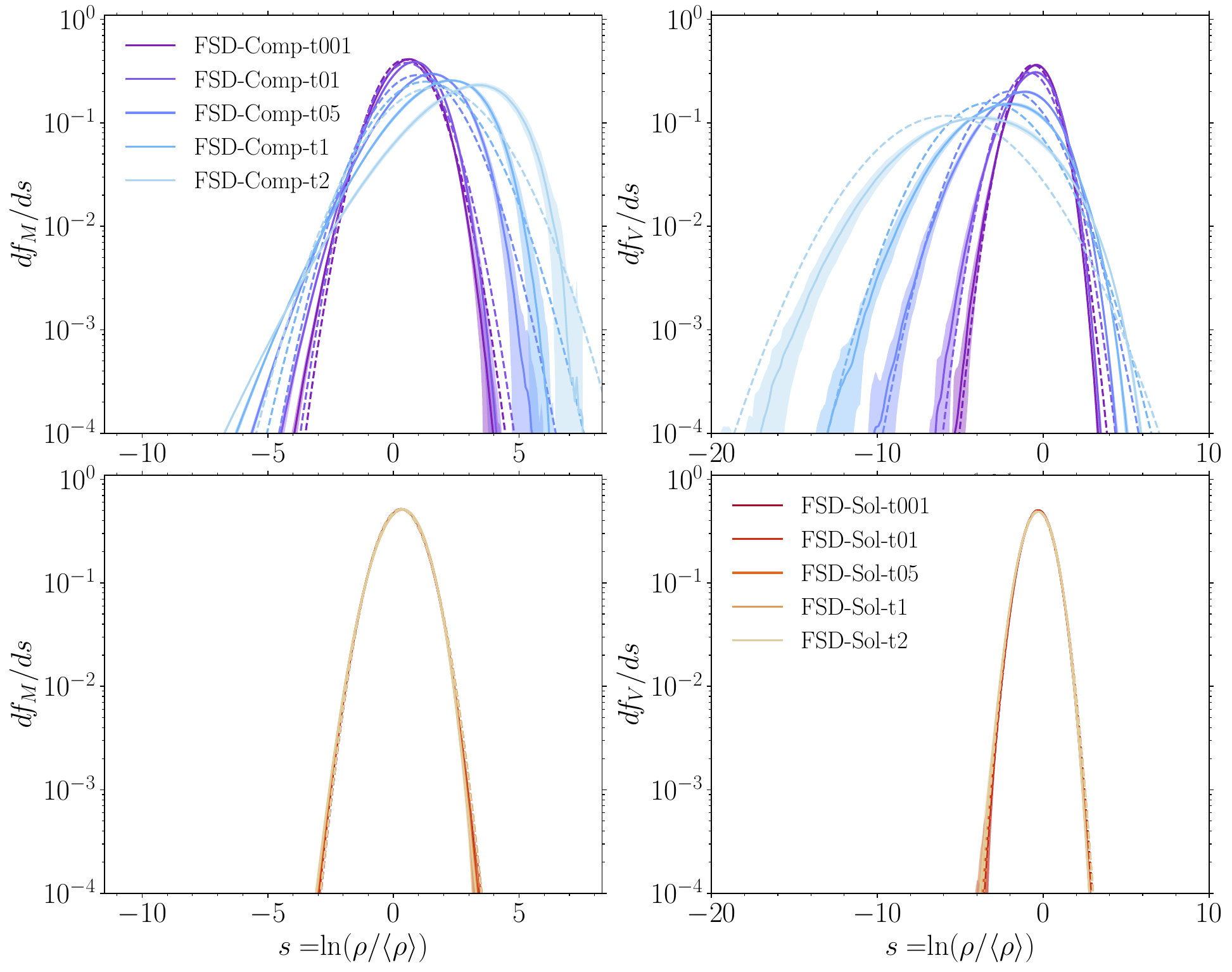}
    \caption{Probability density functions (PDFs) of the logarithmic density $s$ weighted by mass (left column) and volume (right column). The top row collects the \FSDcomp\ models and the bottom row collects the \FSDsol\ models.
    The mean values over $t/t_0=0.4-1$ are shown in solid lines, while the shaded regions depict temporal fluctuations using standard deviations.
    The predicted log-normal PDFs (\autoref{eq:lognorm}) given $\sigma_s$ and $\sigma_{s,M}$ for volume and mass weighted PDFs, respectively, in \autoref{table:tcorr_comp_stats} for the \FSDcomp\ models and \autoref{table:tcorr_sol_stats} \FSDsol\ models are shown in dashed lines. \label{fig:Tcorr_pdf_grid}
    }
\end{figure*}


The \FSDsol\ models, by contrast, are remarkably insensitive to $\tcorr$. Across the same range of $\tcorr/t_0$, $\sigma_\rho$ is essentially unchanged at $0.90$, and $b$ and $b_M$ remain nearly constant at $\sim0.4$ (\autoref{table:tcorr_sol_stats}). The volume-weighted Mach number increases modestly ($2.22\to 2.40$), reflecting the same tendency for longer correlation times to allow more sustained acceleration. Because solenoidal forcing generates vortical motions rather than convergent flows, however, this does not translate into appreciable density fluctuations, especially for the low energy injection rate that leads to moderately supersonic flows.

This strong asymmetry between compressive and solenoidal driving is summarized in \autoref{fig:tcorr_var_dependance}, including the volume- and mass-weighted statistics for the logarithmic density $s$. The \FSDcomp\ models display a substantially stronger dependence on $\tcorr$ than the \FSDsol\ models for all quantities shown. This underscores that $\tcorr$ is effectively a hidden parameter in the $\sigma_\rho$--$\mathcal{M}$ relation and hence the $b$-parameter.
The strong dependence of the density variance on $\tcorr$ for compressive driving is also consistent with the findings of \citet{2026ApJ...998..270S}. Especially, their new fitting function for a parameter $B\equiv\sigma_{s,V}^2/\mathcal{M}_V$ as a function of $\ln(\tcorr/\teddy)$ is in good agreement with our measurements. Overall, our log-density statistics (\autoref{table:tcorr_comp_stats} and \autoref{table:tcorr_sol_stats}) are consistent with their results.

\subsection{Probability Density Functions} \label{subsec:tcorr_pdfs}

The increasing $\sigma_s$ with $\tcorr$ (\autoref{fig:tcorr_var_dependance}) indicates a widening of the density PDF for the \FSDcomp\ models. \autoref{fig:Tcorr_pdf_grid} shows that the PDFs do not merely widen: the $s$-PDFs develop increasing skewness as $\tcorr/t_0$ increases, in particular for the mass-weighted PDFs. The volume-weighted PDF is most strongly affected by low-density regions, which shift the distribution toward lower $s$ values. The mass-weighted PDF, by contrast, shifts toward higher density with a sharper cutoff at the high-density end, visible skewness, and extended low-density tails. In \autoref{subsec:pdfs}, we attributed similar non-Gaussian tails in the fiducial \FSDcomp\ models to intermittent events such as strong shock collisions and rarefaction waves. The trend with $\tcorr$ demonstrates that these intermittent events grow in both frequency and severity as the forcing field persists longer, consistent with the picture that a longer-lived compressive acceleration field drives more coherent convergent and divergent flows.
Quantitatively, the volume-weighted skewness $\mathcal{S}_{s,V}$ peaks in magnitude at intermediate correlation times ($\mathcal{S}_{s,V} \approx -0.33$ for \FSDtModel{Comp}{01} and \FSDtModel{Comp}{05}) and relaxes to $-0.18$ at $\tcorr/t_0=0.2$, while the kurtosis $\mathcal{K}_{s,V}$ decreases from $\approx 3.0$ to $2.8$, indicating a platykurtic (flattened) distribution at long $\tcorr$ (\autoref{table:tcorr_comp_stats}). The mass-weighted statistics show a clearer monotonic trend: $\mathcal{S}_{s,M}$ becomes steadily more negative from $-0.25$ (\FSDtModel{Comp}{001}) to $-0.69$ (\FSDtModel{Comp}{2}), while $\mathcal{K}_{s,M}$ increases from $3.0$ to $3.6$, reflecting the growing prominence of the extended low-density tail in the mass-weighted PDF.

The \FSDsol\ models show essentially no sensitivity to $\tcorr$ in their $s$-PDFs (\autoref{fig:Tcorr_pdf_grid}), consistent with the near-constant $\sigma_\rho$ and $b$ reported in \autoref{subsec:tcorr_global_stat}. Their higher-order statistics remain close to Gaussian at all $\tcorr$: $|\mathcal{S}_{s,V}| \lesssim 0.1$ and $\mathcal{K}_{s,V} \approx 3.0$, with similarly small departures in the mass-weighted moments (\autoref{table:tcorr_sol_stats}). Because solenoidal forcing generates vortical rather than convergent flows, the compressions and expansions that produce density fluctuations are generated only indirectly through nonlinear interactions, and their level is mainly set by the overall Mach number rather than $\tcorr$.

The velocity component PDFs follow Gaussian profiles as in \autoref{fig:512_vel_pdf_grid}, which we opt not to show in the paper. The \FSDsol\ models show no appreciable change in either volume- or mass-weighting. Even for the \FSDcomp\ models, only the width of the volume-weighted velocity PDFs increases slightly with $\tcorr$, consistent with the moderate rise in $\mathcal{M}_V$ seen in \autoref{fig:tcorr_mach_dens_hist}. The mass-weighted velocity PDFs are nearly unchanged across the full range of $\tcorr$. 

\begin{figure*}
    \centering
    \includegraphics[width=0.8\linewidth]{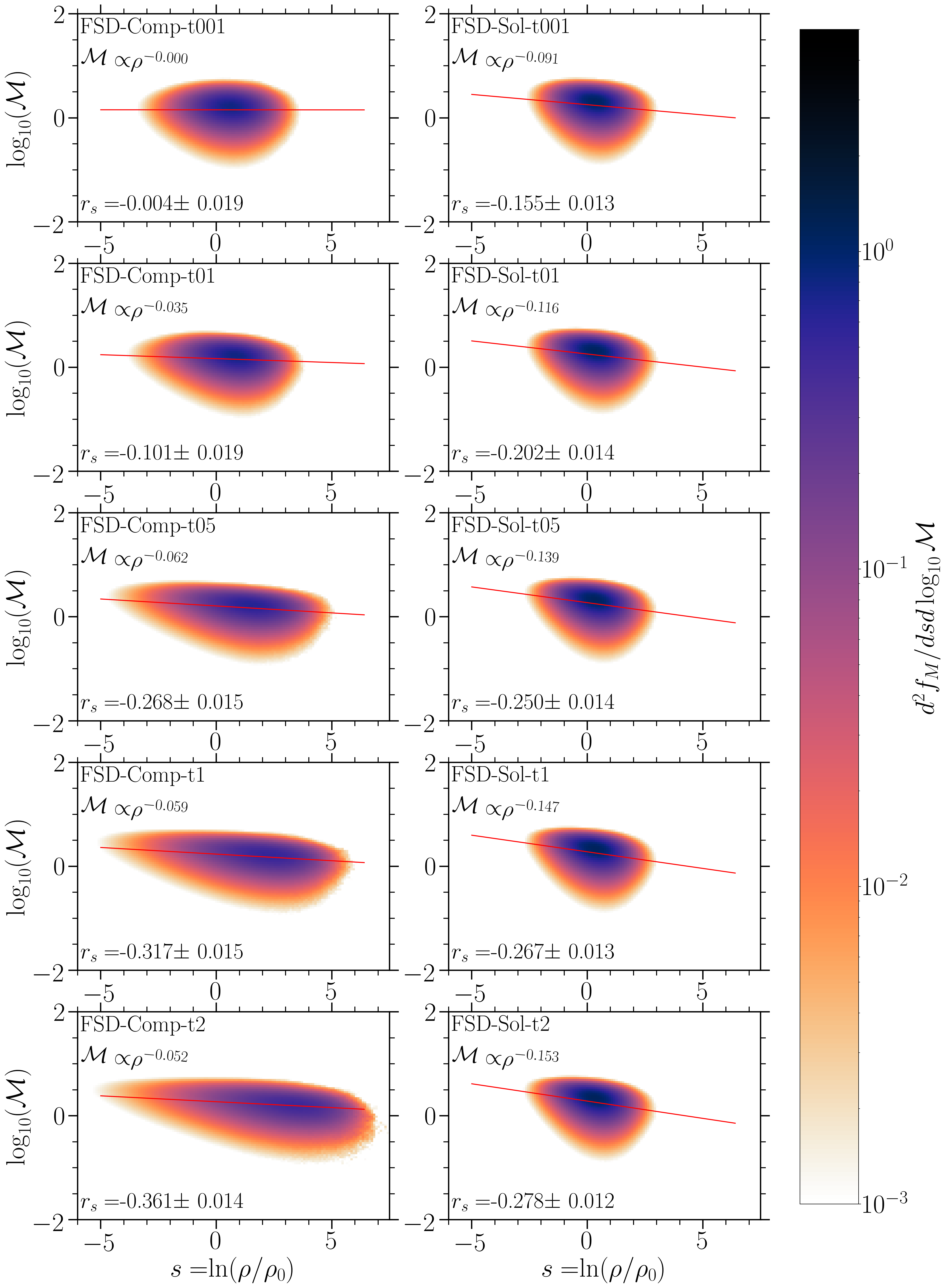}
    \caption{Mass-weighted correlation PDFs between logarithmic density $s$ and Mach number $\log_{10}\mathcal{M}$ of $\tcorr/\teddy$ varying models. Purely compressive models (\FSDcomp) on the left column, and purely solenoidal models (\FSDsol) on the right column. Rows going from top to bottom represent increasing $\tcorr$. Red lines indicate the linear regression lines. The Spearman correlation coefficient is reported as $r_s$. The magnitude of $r_s$ increases monotonically with increasing $\tcorr/\teddy$. \label{fig:tcorr_2D_pdf_grid}}
\end{figure*}

The connection between intermittent events and density--velocity coupling motivates examining the $s$--$\mathcal{M}$ joint distribution. \autoref{fig:tcorr_2D_pdf_grid} shows that the \FSDcomp\ models develop an increasingly negative $s$--$\mathcal{M}$ correlation with longer $\tcorr$. At the shortest correlation time (\FSDtModel{Comp}{001}), the Spearman correlation coefficient is essentially zero ($r_s \sim 0$), indicating that density and velocity fluctuations are nearly independent. As $\tcorr$ increases, the correlation strengthens monotonically, reaching $r_s = -0.361$ at $\tcorr/t_0 = 0.2$. The strong negative correlation is qualitatively reminiscent of the \PSD\ models ($r_s^l<-0.5$; \autoref{fig:512_2D_pdf_grid}), where low-density bubble interiors host the highest velocities. However, the slope in the \FSDcomp\ models ($\mathcal{M}\propto \rho^{-0.05}$) is not as steep as in the \PSD\ models ($\mathcal{M}\propto \rho^{-0.3}$), and the upturn at high densities seen in the \PSD\ joint PDFs is absent. Again, this upturn is a distinct density--velocity structure that \FSD\ driving cannot reproduce.

The solenoidal models also show a negative $s$--$\mathcal{M}$ correlation, but with a different character. Even at the shortest $\tcorr$, \FSDtModel{Sol}{001} already has $r_s = -0.155$. The magnitude of the correlation coefficient $|r_s|$ increases modestly with $\tcorr$, the rate of increase is weaker than for the compressive models. The anti-correlation between density and velocity implies high density structures are formed where velocity decreases by shocks and colliding flows and where compressive modes exist.


\section{Summary and Discussion} \label{sec:summary}

\subsection{Summary} \label{subsec:summary}

A simple relation, the $b$-parameter (the ratio of density fluctuations to the rms Mach number) and the driving mode of interstellar turbulence, has been widely used: $b \approx 1/3$ for purely solenoidal driving and $b \approx 1$ for purely compressive driving \citep{2010A&A...512A..81F}. This mapping has been established using Fourier Space Driving (FSD) simulations and applied broadly to infer the driving mode from observed measurements. In this paper, we challenge the universality of this mapping by comparing FSD against Point Source Driving (PSD), which injects radial momentum at stochastic locations mimicking SNe and represents a physically distinct and more realistic driving mechanism for the ISM. As a first step, we focus on isothermal, hydrodynamic turbulence in a periodic box using AthenaK, which isolates the effect of the driving mechanism from the additional complexity of thermal instability, gravitational collapse, magnetic fields, and stratification. We ask whether turbulence statistics, and in particular the $b$-parameter, depend on the driving mechanism even when global energy injection rates and Mach numbers are matched. The comparison spans morphology, one- and two-dimensional probability density functions, power spectra, and Helmholtz decomposition of the velocity field (\autoref{sec:psd_fsd}). We additionally examine whether the $b$-parameter remains a reliable diagnostic within \FSD\ itself by varying the forcing correlation time $\tcorr$ (\autoref{sec:tcorr}). Our principal findings are as follows.

\begin{itemize}

\item We design the \FSD\ models to match the \PSD\ models in Mach number and driving scale: the \FSDhigh\ and \FSDlow\ models bracket the volume- and mass-weighted Mach numbers of \PSDhigh\ and \PSDlow, while the injection scale $L_{\rm in}/L \approx 0.2$ and eddy turnover time $\teddy$ of both methods are comparable (\autoref{table:main_stats}). Despite this matching, the detailed statistics differ substantially.
Morphologically, the \PSD\ models display large, circular low-density voids surrounded by thin, dense shells that fragment into high-density filaments where they collide (\autoref{fig:slices}). This expanding bubble geometry is a characteristic of localized driving, which none of the \FSD\ models reproduce. The locally injected compressive modes in the \PSD\ models generate solenoidal modes when the flows are interacting as the bubble sweeps up the density. The expanding bubble driving imprints positive correlations between generated solenoidal mode and density (\autoref{fig:rcomp_slices}) as well as density and velocity (\autoref{fig:512_s_pdf_grid}) in the swept-up regions. These are unique feature cannot be reproduced by volume-filling driving in the \FSD\ models, which always create negative correlation between density and velocity.


\item The \PSD\ models have $b \approx 0.33$--$0.49$ (comparable to the \FSDsol\ values of $0.37$--$0.39$), yet their velocity power is predominantly compressive ($\rcomp \approx 0.76$--$0.81$; \autoref{fig:512_helm_grid}), even exceeding the \FSDcomp\ models ($\rcomp \approx 0.64$--$0.65$) and far above the \FSDsol\ models ($\rcomp \approx 0.17$--$0.20$). The \PSD\ density power spectra align with the \FSDsol\ models rather than \FSDcomp\ (\autoref{fig:512_spectra_row}), despite the predominantly compressive character of the \PSD\ velocity field. The velocity PDFs of the \PSD\ models exhibit distinct non-Gaussian tails extending to large $|\vel|$ (\autoref{fig:512_vel_pdf_grid}), reflecting a high level of intermittency driven by the localized, impulsive momentum injections. The \FSD\ models, by design, produce near-Gaussian velocity distributions. The density PDFs are highly non-Gaussian in both \PSD\ and \FSDcomp\ models. 

\item For the compressive driving (\FSDcomp), the density fluctuations are highly sensitive to the choice of correlation time $\tcorr$ between acceleration field realizations. Varying $\tcorr/\teddy$ from $0.016$ to $5.8$ drives $\sigma_\rho$ from $1.27$ to $6.9$, while the Mach number changes only modestly ($\mathcal{M}_V = 1.77$ to $2.9$; $\mathcal{M}_M \approx 1.75$, nearly constant). Because $\tcorr$ strongly amplifies density fluctuations but leaves the velocity statistics largely unchanged, with a factor of $\sim$3--5 variation in $b$ and $b_M$ driven entirely by the choice of correlation time. In contrast, the \FSDsol\ models are nearly insensitive to $\tcorr$. This asymmetry means that while $b \approx 1/3$ does reliably identify purely solenoidal driving, the mapping beyond that is non-trivial. 

\item The widely used relation of \citet{2010A&A...512A..81F} predicts $b \approx 1$ for purely compressive and $b \approx 1/3$ for purely solenoidal driving. The \PSD\ models have $b = 0.33$--$0.49$, nominally suggesting solenoidal driving. The energy injection is locally purely compressive, and the Helmholtz decomposition of the velocity field reveals predominantly compressive modes (\autoref{fig:512_helm_grid}). Using the mass-weighted $b$-parameter, $b_M \approx 0.74$--$0.79$ for \PSD, places it similar to $b_M\sim 0.7$ of the shortest correlation time \FSDcomp\ model with $\tcorr/\teddy\sim0.02$.
\end{itemize}

\subsection{$b$-parameter as a Diagnostic for Turbulence Driving Modes}\label{subsec:summary_bparam}

The $b$-parameter, defined as $b \equiv \sigma_\rho / \mathcal{M}$, is a dimensionless measure of the relative amplitude of density and velocity fluctuations in a turbulent medium.
\citet{2008ApJ...688L..79F} and \citet{2010A&A...512A..81F} showed that in \FSD\ simulations with volume-filling stochastic forcing, $b$ maps onto the fraction of compressive modes in the driving field, ranging from $b \approx 1/3$ for purely solenoidal to $b \approx 1$ for purely compressive forcing.
This calibration was established using simulations that systematically vary the solenoidal-to-compressive mode mixture \citep{2010A&A...512A..81F, 2011ApJ...727L..21P, 2012ApJ...761..156F, 2012MNRAS.423.2680M}.
Observationally, $b$ is inferred from column density maps and line width-based Mach number estimates using power-spectrum-based reconstruction methods \citep{2010MNRAS.403.1507B, 2010MNRAS.405L..56B, 2014MNRAS.442.1451B}, extended to galaxy-scale H\,{\sc i} mapping with spatially resolved kernels \citep{2023MNRAS.526..982G}.
This makes $b$ one of the few quantitative bridges between turbulence simulations and ISM observations, and it is now widely used to infer the dominant driving mode by appealing to the \FSD\ calibration.

Measured values of $b$ span a broad range of environments and tracers.
In molecular clouds, CO isotopologue emission has yielded $b \approx 0.5$ in IC\,5146 \citep{1997ApJ...474..730P}, $b \approx 0.5$ in Taurus \citep{2013A&A...549A..53K}, $b > 0.4$ in Galactic Ring Survey molecular clouds \citep{2013ApJ...779...50G}, $b \approx 0.22$ in the Brick within the Central Molecular Zone \citep{2016ApJ...832..143F}, $b \approx 0.7$--$1.0$ in the Pillars of Creation \citep{2021MNRAS.500.1721M}, and $b \approx 0.9$ in the Papillon Nebula in the LMC \citep{2022MNRAS.509.2180S}.
A similar picture emerges in the diffuse atomic ISM, where $b$ is reconstructed from H\,{\sc i} 21~cm emission.
\citet{2021ApJ...908..186M} measured $b \sim 0.7$ in the Milky Way warm neutral medium (WNM) at high latitudes, \citet{2023MNRAS.526..982G} found $b \sim 0.5$ in the SMC WNM, and \citet{2024MNRAS.530.4317G} reported $b \sim 1.0$ in extra-planar H\,{\sc i} clouds.
Taken together, many measurements across both molecular clouds and the diffuse ISM consistently point toward $b \geq 0.4$, which under the standard \FSD\ calibration would be interpreted as implying predominantly compressive driving as a near-universal feature of the ISM. Furthermore, the region-to-region variation of $b$ is interpreted as respective variations in the driving mode.

These inferences, however, are not robust outside the conditions under which the calibration was established.

Our \PSD\ models (localized compressive energy injection) produce $b = 0.33$--$0.49$ (\autoref{table:main_stats}), values that the \FSD\ calibration would associate with solenoidal or weakly mixed driving. However, this measurement is somewhat misleading as the low-density bubble interiors retain high velocities from the initial momentum injection, inflating the volume-weighted rms Mach number $\mathcal{M}_V$ well above the mass-weighted value ($\mathcal{M}_V/\mathcal{M}_M \approx 2.3$ and $1.7$ for \PSDhigh\ and \PSDlow, respectively; \autoref{table:main_stats}). Since this low-density, high-velocity bubble interior gas is not sampled by standard ISM tracers such as CO or H\,{\sc i} 21~cm emission, a more physically meaningful measure for the \PSD\ models is $b_M=0.74$--$0.79$. Still, $b_M$ falls between the solenoidal ($b \approx 1/3$) and compressive ($b \approx 1$) limits of the \FSD\ calibration. This may in part reflect subsequent turbulence generated by the interaction of bubble expansions with each other and with the inhomogeneous background. Either way, it cannot identify the mode of energy injection.

Even within the \FSD\ framework, the driving-mode interpretation of $b$ is degenerate with $\tcorr$: our \FSDcomp\ models span $b = 0.71$--$2.4$ with no change in driving mode (\autoref{table:tcorr_comp_stats}), a factor of $>3$ variation driven solely by the forcing correlation time. This finding is also consistent with recent independent simulation studies: \citet{2025ApJ...987..122G} demonstrate that density PDFs vary significantly with $\tcorr$ at fixed Mach number in compressively driven turbulence, and \citet{2026ApJ...998..270S} propose a modified $\sigma_\rho$--$\mathcal{M}$ relation that explicitly incorporates $\tcorr$ as a parameter. This complicates observational inferences drawn from comparing $b$ parameters. The region-to-region variation of $b$ may be solely due to the change in correlation rather than the mode of energy injection.
Since $\tcorr$ is not directly observable and varies with the physical driver (e.g., SN rate, cloud-cloud collision frequency), this introduces a hidden parameter that complicates any cross-environment comparison of $b$ values.

Furthermore, there exist other physical mechanisms (e.g., thermal instability, gravity, galactic shear, to name a few) that enhance density variances without causing velocity fluctuations (or vice versa) in the same way that stochastic forcing does \citep[e.g.,][]{2022MNRAS.510.3778M}. \citet{2026MNRAS.546f2277G} analyzed a solar neighborhood model of TIGRESS simulations \citep{2017ApJ...846..133K}, in which the multiphase ISM in a differentially rotating disk is modelled with star formation and feedback self-consistently. They applied the Gaussian kernel method of \citet{2023MNRAS.526..982G} to isolate the turbulent component of density and velocity fluctuations, focusing on the WNM. Over $\sim$100~Myr of evolution (approximately half an orbital period), $b$ fluctuates from $\sim 0.4$ to $\sim 1$, with a time-averaged value of $b \sim 0.5$. A brief peak of $b$ preceding star formation peak is interpreted as a signature of compressive modes enhancing star formation (i.e., likely gravity-driven; \citealt{2012ApJ...761..156F}), but the star formation peaks that inject the compressive mode of energy by SNe do not enhance but reduce $b$. This reduction is broadly reminiscent of our PSD finding that localized compressive injection can inflate $\mathcal{M}$ disproportionately relative to $\sigma_\rho$; however, in the TIGRESS context this behavior is entangled with thermal phase changes, magnetic fields, and galactic shear, so a direct mechanistic comparison is not straightforward. Taken together, these results caution against interpreting $b$ as a direct indicator of the energy injection mode: even in the simplified setting of isothermal periodic-box simulations, $b$ is degenerate with the correlation time and spatial geometry of the driving, and in realistic ISM environments additional physical processes further entangle the density-velocity statistics.

\subsection{Caveats and Future Perspectives} \label{subsec:caveats}

All simulations in this study adopt an isothermal equation of state, which precludes the formation of a multiphase medium and eliminates baroclinic vorticity generation \citep[e.g.,][]{2011PhRvL.107k4504F,2025arXiv250907354B}. Also, the \PSD\ models in this work inject momentum at random, spatially uncorrelated positions with a uniform probability, which is a simplification of how SNe occur in the real ISM. Future controlled \PSD\ simulations that systematically vary the temporal and spatial correlations of sources could provide a more robust $b$-parameter calibration for localized compressive driving. A detailed transfer function analysis of turbulence characteristics of SN-driven multi-phase turbulence simulations begins to reveal various surprises and departures from the conventional Kolmogorov theory of turbulence \citep{2025ApJ...994..193B}. Further developments of systematic analysis of even simple statistics explored here applied to realistic ISM simulations would provide valuable insights on the extent to which such additional statistics have the power to break the degeneracy found here.

The usefulness of $b$-parameter measurements can be tested from another direction. The maps like \autoref{fig:rcomp_slices} show locality of the fraction in different velocity modes. Spatially resolved measurements of $b$ may still be useful if they correlate with local $\Rcomp$ or $\rcomp$. Although they will not tell the energy injection mode from sources, they will tell what is the dominant mode of turbulent velocity field in the particular region. A systematic study of the correlation between spatially resolved $b$ and $\Rcomp$ is a natural extension of this work.

The $b$-parameter diagnostic illustrates a broader challenge in astrophysics: simplified parametrizations calibrated under idealized conditions are valuable precisely because they enable observational diagnostics.
But their application must be balanced against systematic assessment of the conditions under which they remain valid.
The $b$-parameter is one example where the calibration domain is narrower than its widespread application implies.

\addcontentsline{toc}{section}{References/Acknowledgements }

\section*{Acknowledgements}
This work was supported by NASA ATP grant No. 80NSSC22K0717.
This work used resources provided by Princeton Research Computing, a consortium that includes the Princeton Institute for Computational Science and Engineering (PICSciE) and the Office of Information Technology’s Research Computing division.

The authors would like to thank Minghao Guo for providing \texttt{AthenaK} binary data output tool kit (\texttt{AthenaKit}).
This research has made use of NASA's Astrophysics Data System.
This work has made use of \texttt{Astropy} \citep{2022ApJ...935..167A, 2018AJ....156..123A, 2013A&A...558A..33A}, \texttt{AthenaK} \citep{2024arXiv240916053S}, \texttt{Numpy} \citep{harris2020array}, \texttt{Lmfit} \citep{2016ascl.soft06014N}, and \texttt{adstex} (\url{https://github.com/yymao/adstex}).

\section*{Data Availability}
The data used in this work will be shared on reasonable request to the corresponding author.

\bibliographystyle{mnras}
\bibliography{main}

@string{june = {June}}

@article{1941DoSSR..30..301K,
 adsurl = {https://ui.adsabs.harvard.edu/abs/1941DoSSR..30..301K},
 author = {{Kolmogorov}, A.},
 journal = {Akademiia Nauk SSSR Doklady},
 month = {January},
 pages = {301-305},
 title = {{The Local Structure of Turbulence in Incompressible Viscous Fluid for Very Large Reynolds' Numbers}},
 volume = {30},
 year = {1941}
}

@inproceedings{1955IAUS....2..121L,
 adsurl = {https://ui.adsabs.harvard.edu/abs/1955IAUS....2..121L},
 author = {{Lighthill}, M.~J.},
 booktitle = {Gas Dynamics of Cosmic Clouds},
 month = {January},
 pages = {121},
 series = {IAU Symposium},
 title = {{The Effect of Compressibility on Turbulence}},
 volume = {2},
 year = {1955}
}

@book{1992pavi.book.....S,
 adsurl = {https://ui.adsabs.harvard.edu/abs/1992pavi.book.....S},
 author = {{Shu}, F.~H.},
 title = {{The physics of astrophysics. Volume II: Gas dynamics.}},
 year = {1992}
}

@article{1994ApJ...423..681V,
 adsurl = {https://ui.adsabs.harvard.edu/abs/1994ApJ...423..681V},
 author = {{Vazquez-Semadeni}, Enrique},
 doi = {10.1086/173847},
 journal = {\apj},
 month = {March},
 pages = {681},
 title = {{Hierarchical Structure in Nearly Pressureless Flows as a Consequence of Self-similar Statistics}},
 volume = {423},
 year = {1994}
}

@article{1997ApJ...474..730P,
 adsurl = {https://ui.adsabs.harvard.edu/abs/1997ApJ...474..730P},
 archiveprefix = {arXiv},
 author = {{Padoan}, Paolo and {Jones}, Bernard J.~T. and {Nordlund}, {\r{A}}ke P.},
 doi = {10.1086/303482},
 eprint = {astro-ph/9603061},
 journal = {\apj},
 month = {January},
 number = {2},
 pages = {730-734},
 primaryclass = {astro-ph},
 title = {{Supersonic Turbulence in the Interstellar Medium: Stellar Extinction Determinations as Probes of the Structure and Dynamics of Dark Clouds}},
 volume = {474},
 year = {1997}
}

@article{1997MNRAS.288..145P,
 adsurl = {https://ui.adsabs.harvard.edu/abs/1997MNRAS.288..145P},
 archiveprefix = {arXiv},
 author = {{Padoan}, Paolo and {Nordlund}, Ake and {Jones}, Bernard J.~T.},
 doi = {10.1093/mnras/288.1.145},
 eprint = {astro-ph/9703110},
 journal = {\mnras},
 month = {June},
 number = {1},
 pages = {145-152},
 primaryclass = {astro-ph},
 title = {{The universality of the stellar initial mass function}},
 volume = {288},
 year = {1997}
}

@article{1998PhRvE..58.4501P,
 adsurl = {https://ui.adsabs.harvard.edu/abs/1998PhRvE..58.4501P},
 archiveprefix = {arXiv},
 author = {{Passot}, Thierry and {V{\'a}zquez-Semadeni}, Enrique},
 doi = {10.1103/PhysRevE.58.4501},
 eprint = {physics/9802019},
 journal = {\pre},
 month = {October},
 number = {4},
 pages = {4501-4510},
 primaryclass = {physics.flu-dyn},
 title = {{Density probability distribution in one-dimensional polytropic gas dynamics}},
 volume = {58},
 year = {1998}
}

@article{1999ApJS..123....3L,
 adsurl = {https://ui.adsabs.harvard.edu/abs/1999ApJS..123....3L},
 archiveprefix = {arXiv},
 author = {{Leitherer}, Claus and {Schaerer}, Daniel and {Goldader}, Jeffrey D. and {Delgado}, Rosa M. Gonz{\'a}lez and {Robert}, Carmelle and {Kune}, Denis Foo and {de Mello}, Du{\'\i}lia F. and {Devost}, Daniel and {Heckman}, Timothy M.},
 doi = {10.1086/313233},
 eprint = {astro-ph/9902334},
 journal = {\apjs},
 month = {July},
 number = {1},
 pages = {3-40},
 primaryclass = {astro-ph},
 title = {{Starburst99: Synthesis Models for Galaxies with Active Star Formation}},
 volume = {123},
 year = {1999}
}

@article{2001ApJ...546..980O,
 adsurl = {https://ui.adsabs.harvard.edu/abs/2001ApJ...546..980O},
 archiveprefix = {arXiv},
 author = {{Ostriker}, Eve C. and {Stone}, James M. and {Gammie}, Charles F.},
 doi = {10.1086/318290},
 eprint = {astro-ph/0008454},
 journal = {\apj},
 month = {January},
 number = {2},
 pages = {980-1005},
 primaryclass = {astro-ph},
 title = {{Density, Velocity, and Magnetic Field Structure in Turbulent Molecular Cloud Models}},
 volume = {546},
 year = {2001}
}

@article{2001MNRAS.322..231K,
 adsurl = {https://ui.adsabs.harvard.edu/abs/2001MNRAS.322..231K},
 archiveprefix = {arXiv},
 author = {{Kroupa}, Pavel},
 doi = {10.1046/j.1365-8711.2001.04022.x},
 eprint = {astro-ph/0009005},
 journal = {\mnras},
 month = {April},
 number = {2},
 pages = {231-246},
 primaryclass = {astro-ph},
 title = {{On the variation of the initial mass function}},
 volume = {322},
 year = {2001}
}

@article{2002ApJ...576..870P,
 adsurl = {https://ui.adsabs.harvard.edu/abs/2002ApJ...576..870P},
 archiveprefix = {arXiv},
 author = {{Padoan}, Paolo and {Nordlund}, {\r{A}}ke},
 doi = {10.1086/341790},
 eprint = {astro-ph/0011465},
 journal = {\apj},
 month = {September},
 number = {2},
 pages = {870-879},
 primaryclass = {astro-ph},
 title = {{The Stellar Initial Mass Function from Turbulent Fragmentation}},
 volume = {576},
 year = {2002}
}

@article{2004ARA&A..42..211E,
 adsurl = {https://ui.adsabs.harvard.edu/abs/2004ARA&A..42..211E},
 archiveprefix = {arXiv},
 author = {{Elmegreen}, Bruce G. and {Scalo}, John},
 doi = {10.1146/annurev.astro.41.011802.094859},
 eprint = {astro-ph/0404451},
 journal = {\araa},
 month = {September},
 number = {1},
 pages = {211-273},
 primaryclass = {astro-ph},
 title = {{Interstellar Turbulence I: Observations and Processes}},
 volume = {42},
 year = {2004}
}

@article{2004RvMP...76..125M,
 adsurl = {https://ui.adsabs.harvard.edu/abs/2004RvMP...76..125M},
 archiveprefix = {arXiv},
 author = {{Mac Low}, Mordecai-Mark and {Klessen}, Ralf S.},
 doi = {10.1103/RevModPhys.76.125},
 eprint = {astro-ph/0301093},
 journal = {Reviews of Modern Physics},
 month = {January},
 number = {1},
 pages = {125-194},
 primaryclass = {astro-ph},
 title = {{Control of star formation by supersonic turbulence}},
 volume = {76},
 year = {2004}
}

@article{2005ApJ...630L..45K,
 adsurl = {https://ui.adsabs.harvard.edu/abs/2005ApJ...630L..45K},
 author = {{Kim}, Jongsoo and {Ryu}, Dongsu},
 doi = {10.1086/491600},
 journal = {\apjl},
 pages = {L45--L48},
 title = {{Density Power Spectrum of Compressible Hydrodynamic Turbulent Flows}},
 volume = {630},
 year = {2005}
}

@article{2007ApJ...665..416K,
 adsurl = {https://ui.adsabs.harvard.edu/abs/2007ApJ...665..416K},
 archiveprefix = {arXiv},
 author = {{Kritsuk}, Alexei G. and {Norman}, Michael L. and {Padoan}, Paolo and {Wagner}, Rick},
 doi = {10.1086/519443},
 eprint = {0704.3851},
 journal = {\apj},
 month = {August},
 number = {1},
 pages = {416-431},
 primaryclass = {astro-ph},
 title = {{The Statistics of Supersonic Isothermal Turbulence}},
 volume = {665},
 year = {2007}
}

@article{2007ARA&A..45..565M,
 adsurl = {https://ui.adsabs.harvard.edu/abs/2007ARA&A..45..565M},
 archiveprefix = {arXiv},
 author = {{McKee}, Christopher F. and {Ostriker}, Eve C.},
 doi = {10.1146/annurev.astro.45.051806.110602},
 eprint = {0707.3514},
 journal = {\araa},
 month = {September},
 number = {1},
 pages = {565-687},
 primaryclass = {astro-ph},
 title = {{Theory of Star Formation}},
 volume = {45},
 year = {2007}
}

@article{2008ApJ...682L..97L,
 adsurl = {https://ui.adsabs.harvard.edu/abs/2008ApJ...682L..97L},
 archiveprefix = {arXiv},
 author = {{Lemaster}, M. Nicole and {Stone}, James M.},
 doi = {10.1086/590929},
 eprint = {0806.1525},
 journal = {\apjl},
 month = {August},
 number = {2},
 pages = {L97},
 primaryclass = {astro-ph},
 title = {{Density Probability Distribution Functions in Supersonic Hydrodynamic and MHD Turbulence}},
 volume = {682},
 year = {2008}
}

@article{2008ApJ...688L..79F,
 adsurl = {https://ui.adsabs.harvard.edu/abs/2008ApJ...688L..79F},
 archiveprefix = {arXiv},
 author = {{Federrath}, Christoph and {Klessen}, Ralf S. and {Schmidt}, Wolfram},
 doi = {10.1086/595280},
 eprint = {0808.0605},
 journal = {\apjl},
 month = {December},
 number = {2},
 pages = {L79},
 primaryclass = {astro-ph},
 title = {{The Density Probability Distribution in Compressible Isothermal Turbulence: Solenoidal versus Compressive Forcing}},
 volume = {688},
 year = {2008}
}

@article{2009ApJ...691.1092L,
 adsurl = {https://ui.adsabs.harvard.edu/abs/2009ApJ...691.1092L},
 archiveprefix = {arXiv},
 author = {{Lemaster}, M. Nicole and {Stone}, James M.},
 doi = {10.1088/0004-637X/691/2/1092},
 eprint = {0809.4005},
 journal = {\apj},
 month = {February},
 number = {2},
 pages = {1092-1108},
 primaryclass = {astro-ph},
 title = {{Dissipation and Heating in Supersonic Hydrodynamic and MHD Turbulence}},
 volume = {691},
 year = {2009}
}

@article{2010A&A...512A..81F,
 adsurl = {https://ui.adsabs.harvard.edu/abs/2010A&A...512A..81F},
 archiveprefix = {arXiv},
 author = {{Federrath}, C. and {Roman-Duval}, J. and {Klessen}, R.~S. and {Schmidt}, W. and {Mac Low}, M. -M.},
 doi = {10.1051/0004-6361/200912437},
 eid = {A81},
 eprint = {0905.1060},
 journal = {\aap},
 month = {March},
 pages = {A81},
 primaryclass = {astro-ph.SR},
 title = {{Comparing the statistics of interstellar turbulence in simulations and observations. Solenoidal versus compressive turbulence forcing}},
 volume = {512},
 year = {2010}
}

@article{2010A&A...513A..67B,
 adsurl = {https://ui.adsabs.harvard.edu/abs/2010A&A...513A..67B},
 archiveprefix = {arXiv},
 author = {{Brunt}, C.~M.},
 doi = {10.1051/0004-6361/200913506},
 eid = {A67},
 eprint = {1002.1239},
 journal = {\aap},
 month = {April},
 pages = {A67},
 primaryclass = {astro-ph.GA},
 title = {{The density variance - Mach number relation in the Taurus molecular cloud}},
 volume = {513},
 year = {2010}
}

@article{2010MNRAS.403.1507B,
 adsurl = {https://ui.adsabs.harvard.edu/abs/2010MNRAS.403.1507B},
 archiveprefix = {arXiv},
 author = {{Brunt}, C.~M. and {Federrath}, C. and {Price}, D.~J.},
 doi = {10.1111/j.1365-2966.2009.16215.x},
 eprint = {1001.1046},
 journal = {\mnras},
 month = {April},
 number = {3},
 pages = {1507-1515},
 primaryclass = {astro-ph.GA},
 title = {{A method for reconstructing the variance of a 3D physical field from 2D observations: application to turbulence in the interstellar medium}},
 volume = {403},
 year = {2010}
}

@article{2010MNRAS.405L..56B,
 adsurl = {https://ui.adsabs.harvard.edu/abs/2010MNRAS.405L..56B},
 archiveprefix = {arXiv},
 author = {{Brunt}, Christopher M. and {Federrath}, Christoph and {Price}, Daniel J.},
 doi = {10.1111/j.1745-3933.2010.00858.x},
 eprint = {1003.4151},
 journal = {\mnras},
 month = {June},
 number = {1},
 pages = {L56-L60},
 primaryclass = {astro-ph.GA},
 title = {{A method for reconstructing the PDF of a 3D turbulent density field from 2D observations}},
 volume = {405},
 year = {2010}
}

@article{2011ApJ...727L..21P,
 adsurl = {https://ui.adsabs.harvard.edu/abs/2011ApJ...727L..21P},
 archiveprefix = {arXiv},
 author = {{Price}, Daniel J. and {Federrath}, Christoph and {Brunt}, Christopher M.},
 doi = {10.1088/2041-8205/727/1/L21},
 eid = {L21},
 eprint = {1010.3754},
 journal = {\apjl},
 month = {January},
 number = {1},
 pages = {L21},
 primaryclass = {astro-ph.GA},
 title = {{The Density Variance-Mach Number Relation in Supersonic, Isothermal Turbulence}},
 volume = {727},
 year = {2011}
}

@article{2011PhRvL.107k4504F,
 adsurl = {https://ui.adsabs.harvard.edu/abs/2011PhRvL.107k4504F},
 archiveprefix = {arXiv},
 author = {{Federrath}, C. and {Chabrier}, G. and {Schober}, J. and {Banerjee}, R. and {Klessen}, R.~S. and {Schleicher}, D.~R.~G.},
 doi = {10.1103/PhysRevLett.107.114504},
 eid = {114504},
 eprint = {1109.1760},
 journal = {\prl},
 month = {September},
 number = {11},
 pages = {114504},
 primaryclass = {physics.flu-dyn},
 title = {{Mach Number Dependence of Turbulent Magnetic Field Amplification: Solenoidal versus Compressive Flows}},
 volume = {107},
 year = {2011}
}

@book{2011piim.book.....D,
 adsurl = {https://ui.adsabs.harvard.edu/abs/2011piim.book.....D},
 author = {{Draine}, Bruce T.},
 title = {{Physics of the Interstellar and Intergalactic Medium}},
 year = {2011}
}

@article{2012ApJ...755L..19B,
 adsurl = {https://ui.adsabs.harvard.edu/abs/2012ApJ...755L..19B},
 archiveprefix = {arXiv},
 author = {{Burkhart}, Blakesley and {Lazarian}, A.},
 doi = {10.1088/2041-8205/755/1/L19},
 eid = {L19},
 eprint = {1205.3792},
 journal = {\apjl},
 month = {August},
 number = {1},
 pages = {L19},
 primaryclass = {astro-ph.GA},
 title = {{The Column Density Variance-\{\textbackslashcal M\}\_s Relationship}},
 volume = {755},
 year = {2012}
}

@article{2012ApJ...761..156F,
 adsurl = {https://ui.adsabs.harvard.edu/abs/2012ApJ...761..156F},
 archiveprefix = {arXiv},
 author = {{Federrath}, Christoph and {Klessen}, Ralf S.},
 doi = {10.1088/0004-637X/761/2/156},
 eid = {156},
 eprint = {1209.2856},
 journal = {\apj},
 month = {December},
 number = {2},
 pages = {156},
 primaryclass = {astro-ph.SR},
 title = {{The Star Formation Rate of Turbulent Magnetized Clouds: Comparing Theory, Simulations, and Observations}},
 volume = {761},
 year = {2012}
}

@article{2012MNRAS.423.2680M,
 adsurl = {https://ui.adsabs.harvard.edu/abs/2012MNRAS.423.2680M},
 archiveprefix = {arXiv},
 author = {{Molina}, F.~Z. and {Glover}, S.~C.~O. and {Federrath}, C. and {Klessen}, R.~S.},
 doi = {10.1111/j.1365-2966.2012.21075.x},
 eprint = {1203.2117},
 journal = {\mnras},
 month = {July},
 number = {3},
 pages = {2680-2689},
 primaryclass = {astro-ph.GA},
 title = {{The density variance-Mach number relation in supersonic turbulence - I. Isothermal, magnetized gas}},
 volume = {423},
 year = {2012}
}

@article{2013A&A...549A..53K,
 adsurl = {https://ui.adsabs.harvard.edu/abs/2013A&A...549A..53K},
 archiveprefix = {arXiv},
 author = {{Kainulainen}, J. and {Tan}, J.~C.},
 doi = {10.1051/0004-6361/201219526},
 eid = {A53},
 eprint = {1210.8130},
 journal = {\aap},
 month = {January},
 pages = {A53},
 primaryclass = {astro-ph.GA},
 title = {{High-dynamic-range extinction mapping of infrared dark clouds. Dependence of density variance with sonic Mach number in molecular clouds}},
 volume = {549},
 year = {2013}
}

@article{2013A&A...558A..33A,
 adsurl = {https://ui.adsabs.harvard.edu/abs/2013A&A...558A..33A},
 archiveprefix = {arXiv},
 author = {{Astropy Collaboration} and {Robitaille}, Thomas P. and {Tollerud}, Erik J. and {Greenfield}, Perry and {Droettboom}, Michael and {Bray}, Erik and {Aldcroft}, Tom and {Davis}, Matt and {Ginsburg}, Adam and {Price-Whelan}, Adrian M. and {Kerzendorf}, Wolfgang E. and {Conley}, Alexander and {Crighton}, Neil and {Barbary}, Kyle and {Muna}, Demitri and {Ferguson}, Henry and {Grollier}, Fr{\'e}d{\'e}ric and {Parikh}, Madhura M. and {Nair}, Prasanth H. and {Unther}, Hans M. and {Deil}, Christoph and {Woillez}, Julien and {Conseil}, Simon and {Kramer}, Roban and {Turner}, James E.~H. and {Singer}, Leo and {Fox}, Ryan and {Weaver}, Benjamin A. and {Zabalza}, Victor and {Edwards}, Zachary I. and {Azalee Bostroem}, K. and {Burke}, D.~J. and {Casey}, Andrew R. and {Crawford}, Steven M. and {Dencheva}, Nadia and {Ely}, Justin and {Jenness}, Tim and {Labrie}, Kathleen and {Lim}, Pey Lian and {Pierfederici}, Francesco and {Pontzen}, Andrew and {Ptak}, Andy and {Refsdal}, Brian and {Servillat}, Mathieu and {Streicher}, Ole},
 doi = {10.1051/0004-6361/201322068},
 eid = {A33},
 eprint = {1307.6212},
 journal = {\aap},
 month = {October},
 pages = {A33},
 primaryclass = {astro-ph.IM},
 title = {{Astropy: A community Python package for astronomy}},
 volume = {558},
 year = {2013}
}

@article{2013ApJ...779...50G,
 adsurl = {https://ui.adsabs.harvard.edu/abs/2013ApJ...779...50G},
 archiveprefix = {arXiv},
 author = {{Ginsburg}, Adam and {Federrath}, Christoph and {Darling}, Jeremy},
 doi = {10.1088/0004-637X/779/1/50},
 eid = {50},
 eprint = {1310.0809},
 journal = {\apj},
 month = {December},
 number = {1},
 pages = {50},
 primaryclass = {astro-ph.GA},
 title = {{A Measurement of the Turbulence-driven Density Distribution in a Non-star-forming Molecular Cloud}},
 volume = {779},
 year = {2013}
}

@article{2013SSRv..178..163B,
 adsurl = {https://ui.adsabs.harvard.edu/abs/2013SSRv..178..163B},
 archiveprefix = {arXiv},
 author = {{Brandenburg}, A. and {Lazarian}, A.},
 doi = {10.1007/s11214-013-0009-3},
 eprint = {1307.5496},
 journal = {\ssr},
 month = {October},
 number = {2-4},
 pages = {163-200},
 primaryclass = {astro-ph.SR},
 title = {{Astrophysical Hydromagnetic Turbulence}},
 volume = {178},
 year = {2013}
}

@article{2014MNRAS.442.1451B,
 adsurl = {https://ui.adsabs.harvard.edu/abs/2014MNRAS.442.1451B},
 archiveprefix = {arXiv},
 author = {{Brunt}, C.~M. and {Federrath}, C.},
 doi = {10.1093/mnras/stu888},
 eprint = {1405.1285},
 journal = {\mnras},
 month = {August},
 number = {2},
 pages = {1451-1469},
 primaryclass = {astro-ph.GA},
 title = {{An observational method to measure the relative fractions of solenoidal and compressible modes in interstellar clouds}},
 volume = {442},
 year = {2014}
}

@inproceedings{2014prpl.conf...77P,
 adsurl = {https://ui.adsabs.harvard.edu/abs/2014prpl.conf...77P},
 archiveprefix = {arXiv},
 author = {{Padoan}, P. and {Federrath}, C. and {Chabrier}, G. and {Evans}, II, N.~J. and {Johnstone}, D. and {J{\o}rgensen}, J.~K. and {McKee}, C.~F. and {Nordlund}, {\r{A}}.},
 booktitle = {Protostars and Planets VI},
 doi = {10.2458/azu_uapress_9780816531240-ch004},
 editor = {{Beuther}, Henrik and {Klessen}, Ralf S. and {Dullemond}, Cornelis P. and {Henning}, Thomas},
 eprint = {1312.5365},
 month = {January},
 pages = {77-100},
 primaryclass = {astro-ph.GA},
 title = {{The Star Formation Rate of Molecular Clouds}},
 year = {2014}
}

@article{2015ApJ...802...99K,
 adsurl = {https://ui.adsabs.harvard.edu/abs/2015ApJ...802...99K},
 archiveprefix = {arXiv},
 author = {{Kim}, Chang-Goo and {Ostriker}, Eve C.},
 doi = {10.1088/0004-637X/802/2/99},
 eid = {99},
 eprint = {1410.1537},
 journal = {\apj},
 month = {April},
 number = {2},
 pages = {99},
 primaryclass = {astro-ph.GA},
 title = {{Momentum Injection by Supernovae in the Interstellar Medium}},
 volume = {802},
 year = {2015}
}

@article{2016ApJ...822...11P,
 adsurl = {https://ui.adsabs.harvard.edu/abs/2016ApJ...822...11P},
 archiveprefix = {arXiv},
 author = {{Padoan}, Paolo and {Pan}, Liubin and {Haugb{\o}lle}, Troels and {Nordlund}, {\r{A}}ke},
 doi = {10.3847/0004-637X/822/1/11},
 eid = {11},
 eprint = {1509.04663},
 journal = {\apj},
 month = {May},
 number = {1},
 pages = {11},
 primaryclass = {astro-ph.GA},
 title = {{Supernova Driving. I. The Origin of Molecular Cloud Turbulence}},
 volume = {822},
 year = {2016}
}

@article{2016ApJ...832..143F,
 adsurl = {https://ui.adsabs.harvard.edu/abs/2016ApJ...832..143F},
 archiveprefix = {arXiv},
 author = {{Federrath}, C. and {Rathborne}, J.~M. and {Longmore}, S.~N. and {Kruijssen}, J.~M.~D. and {Bally}, J. and {Contreras}, Y. and {Crocker}, R.~M. and {Garay}, G. and {Jackson}, J.~M. and {Testi}, L. and {Walsh}, A.~J.},
 doi = {10.3847/0004-637X/832/2/143},
 eid = {143},
 eprint = {1609.05911},
 journal = {\apj},
 month = {December},
 number = {2},
 pages = {143},
 primaryclass = {astro-ph.GA},
 title = {{The Link between Turbulence, Magnetic Fields, Filaments, and Star Formation in the Central Molecular Zone Cloud G0.253+0.016}},
 volume = {832},
 year = {2016}
}

@software{2016ascl.soft06014N,
 adsurl = {https://ui.adsabs.harvard.edu/abs/2016ascl.soft06014N},
 author = {{Newville}, Matthew and {Stensitzki}, Till and {Allen}, Daniel B. and {Rawlik}, Michal and {Ingargiola}, Antonino and {Nelson}, Andrew},
 eid = {ascl:1606.014},
 howpublished = {Astrophysics Source Code Library, record ascl:1606.014},
 month = {June},
 title = {{Lmfit: Non-Linear Least-Square Minimization and Curve-Fitting for Python}},
 year = {2016}
}

@article{2016PhRvF...1h2403I,
 adsurl = {https://ui.adsabs.harvard.edu/abs/2016PhRvF...1h2403I},
 author = {{Ishihara}, Takashi and {Morishita}, Koji and {Yokokawa}, Mitsuo and {Uno}, Atsuya and {Kaneda}, Yukio},
 doi = {10.1103/PhysRevFluids.1.082403},
 eid = {082403},
 journal = {Physical Review Fluids},
 month = {December},
 number = {8},
 pages = {082403},
 title = {{Energy spectrum in high-resolution direct numerical simulations of turbulence}},
 volume = {1},
 year = {2016}
}

@article{2017ApJ...846..133K,
 adsurl = {https://ui.adsabs.harvard.edu/abs/2017ApJ...846..133K},
 archiveprefix = {arXiv},
 author = {{Kim}, Chang-Goo and {Ostriker}, Eve C.},
 doi = {10.3847/1538-4357/aa8599},
 eid = {133},
 eprint = {1612.03918},
 journal = {\apj},
 month = {September},
 number = {2},
 pages = {133},
 primaryclass = {astro-ph.GA},
 title = {{Three-phase Interstellar Medium in Galaxies Resolving Evolution with Star Formation and Supernova Feedback (TIGRESS): Algorithms, Fiducial Model, and Convergence}},
 volume = {846},
 year = {2017}
}

@article{2018AJ....156..123A,
 adsurl = {https://ui.adsabs.harvard.edu/abs/2018AJ....156..123A},
 archiveprefix = {arXiv},
 author = {{Astropy Collaboration} and {Price-Whelan}, A.~M. and {Sip{\H{o}}cz}, B.~M. and {G{\"u}nther}, H.~M. and {Lim}, P.~L. and {Crawford}, S.~M. and {Conseil}, S. and {Shupe}, D.~L. and {Craig}, M.~W. and {Dencheva}, N. and {Ginsburg}, A. and {VanderPlas}, J.~T. and {Bradley}, L.~D. and {P{\'e}rez-Su{\'a}rez}, D. and {de Val-Borro}, M. and {Aldcroft}, T.~L. and {Cruz}, K.~L. and {Robitaille}, T.~P. and {Tollerud}, E.~J. and {Ardelean}, C. and {Babej}, T. and {Bach}, Y.~P. and {Bachetti}, M. and {Bakanov}, A.~V. and {Bamford}, S.~P. and {Barentsen}, G. and {Barmby}, P. and {Baumbach}, A. and {Berry}, K.~L. and {Biscani}, F. and {Boquien}, M. and {Bostroem}, K.~A. and {Bouma}, L.~G. and {Brammer}, G.~B. and {Bray}, E.~M. and {Breytenbach}, H. and {Buddelmeijer}, H. and {Burke}, D.~J. and {Calderone}, G. and {Cano Rodr{\'\i}guez}, J.~L. and {Cara}, M. and {Cardoso}, J.~V.~M. and {Cheedella}, S. and {Copin}, Y. and {Corrales}, L. and {Crichton}, D. and {D'Avella}, D. and {Deil}, C. and {Depagne}, {\'E}. and {Dietrich}, J.~P. and {Donath}, A. and {Droettboom}, M. and {Earl}, N. and {Erben}, T. and {Fabbro}, S. and {Ferreira}, L.~A. and {Finethy}, T. and {Fox}, R.~T. and {Garrison}, L.~H. and {Gibbons}, S.~L.~J. and {Goldstein}, D.~A. and {Gommers}, R. and {Greco}, J.~P. and {Greenfield}, P. and {Groener}, A.~M. and {Grollier}, F. and {Hagen}, A. and {Hirst}, P. and {Homeier}, D. and {Horton}, A.~J. and {Hosseinzadeh}, G. and {Hu}, L. and {Hunkeler}, J.~S. and {Ivezi{\'c}}, {\v{Z}}. and {Jain}, A. and {Jenness}, T. and {Kanarek}, G. and {Kendrew}, S. and {Kern}, N.~S. and {Kerzendorf}, W.~E. and {Khvalko}, A. and {King}, J. and {Kirkby}, D. and {Kulkarni}, A.~M. and {Kumar}, A. and {Lee}, A. and {Lenz}, D. and {Littlefair}, S.~P. and {Ma}, Z. and {Macleod}, D.~M. and {Mastropietro}, M. and {McCully}, C. and {Montagnac}, S. and {Morris}, B.~M. and {Mueller}, M. and {Mumford}, S.~J. and {Muna}, D. and {Murphy}, N.~A. and {Nelson}, S. and {Nguyen}, G.~H. and {Ninan}, J.~P. and {N{\"o}the}, M. and {Ogaz}, S. and {Oh}, S. and {Parejko}, J.~K. and {Parley}, N. and {Pascual}, S. and {Patil}, R. and {Patil}, A.~A. and {Plunkett}, A.~L. and {Prochaska}, J.~X. and {Rastogi}, T. and {Reddy Janga}, V. and {Sabater}, J. and {Sakurikar}, P. and {Seifert}, M. and {Sherbert}, L.~E. and {Sherwood-Taylor}, H. and {Shih}, A.~Y. and {Sick}, J. and {Silbiger}, M.~T. and {Singanamalla}, S. and {Singer}, L.~P. and {Sladen}, P.~H. and {Sooley}, K.~A. and {Sornarajah}, S. and {Streicher}, O. and {Teuben}, P. and {Thomas}, S.~W. and {Tremblay}, G.~R. and {Turner}, J.~E.~H. and {Terr{\'o}n}, V. and {van Kerkwijk}, M.~H. and {de la Vega}, A. and {Watkins}, L.~L. and {Weaver}, B.~A. and {Whitmore}, J.~B. and {Woillez}, J. and {Zabalza}, V. and {Astropy Contributors}},
 doi = {10.3847/1538-3881/aabc4f},
 eid = {123},
 eprint = {1801.02634},
 journal = {\aj},
 month = {September},
 number = {3},
 pages = {123},
 primaryclass = {astro-ph.IM},
 title = {{The Astropy Project: Building an Open-science Project and Status of the v2.0 Core Package}},
 volume = {156},
 year = {2018}
}

@article{2018ApJ...858L..19G,
 adsurl = {https://ui.adsabs.harvard.edu/abs/2018ApJ...858L..19G},
 archiveprefix = {arXiv},
 author = {{Grete}, Philipp and {O'Shea}, Brian W. and {Beckwith}, Kris},
 doi = {10.3847/2041-8213/aac0f5},
 eid = {L19},
 eprint = {1803.05481},
 journal = {\apjl},
 month = {May},
 number = {2},
 pages = {L19},
 primaryclass = {physics.flu-dyn},
 title = {{As a Matter of Force{\textemdash}Systematic Biases in Idealized Turbulence Simulations}},
 volume = {858},
 year = {2018}
}

@article{2020ApJ...904..160F,
 adsurl = {https://ui.adsabs.harvard.edu/abs/2020ApJ...904..160F},
 archiveprefix = {arXiv},
 author = {{Ferrand}, R. and {Galtier}, S. and {Sahraoui}, F. and {Federrath}, C.},
 doi = {10.3847/1538-4357/abb76e},
 eid = {160},
 eprint = {2303.06960},
 journal = {\apj},
 month = {December},
 number = {2},
 pages = {160},
 primaryclass = {astro-ph.GA},
 title = {{Compressible Turbulence in the Interstellar Medium: New Insights from a High-resolution Supersonic Turbulence Simulation}},
 volume = {904},
 year = {2020}
}

@article{2020ApJ...905...35K,
 adsurl = {https://ui.adsabs.harvard.edu/abs/2020ApJ...905...35K},
 archiveprefix = {arXiv},
 author = {{Koo}, Bon-Chul and {Kim}, Chang-Goo and {Park}, Sangwook and {Ostriker}, Eve C.},
 doi = {10.3847/1538-4357/abc1e7},
 eid = {35},
 eprint = {2011.06322},
 journal = {\apj},
 month = {December},
 number = {1},
 pages = {35},
 primaryclass = {astro-ph.GA},
 title = {{Radiative Supernova Remnants and Supernova Feedback}},
 volume = {905},
 year = {2020}
}

@article{2020ApJS..249....4S,
 adsurl = {https://ui.adsabs.harvard.edu/abs/2020ApJS..249....4S},
 archiveprefix = {arXiv},
 author = {{Stone}, James M. and {Tomida}, Kengo and {White}, Christopher J. and {Felker}, Kyle G.},
 doi = {10.3847/1538-4365/ab929b},
 eid = {4},
 eprint = {2005.06651},
 journal = {\apjs},
 month = {July},
 number = {1},
 pages = {4},
 primaryclass = {astro-ph.IM},
 title = {{The Athena++ Adaptive Mesh Refinement Framework: Design and Magnetohydrodynamic Solvers}},
 volume = {249},
 year = {2020}
}

@article{2021ApJ...908..186M,
 adsurl = {https://ui.adsabs.harvard.edu/abs/2021ApJ...908..186M},
 archiveprefix = {arXiv},
 author = {{Marchal}, Antoine and {Miville-Desch{\^e}nes}, Marc-Antoine},
 doi = {10.3847/1538-4357/abd108},
 eid = {186},
 eprint = {2012.03160},
 journal = {\apj},
 month = {February},
 number = {2},
 pages = {186},
 primaryclass = {astro-ph.GA},
 title = {{Thermal and Turbulent Properties of the Warm Neutral Medium in the Solar Neighborhood}},
 volume = {908},
 year = {2021}
}

@article{2021MNRAS.500.1721M,
 adsurl = {https://ui.adsabs.harvard.edu/abs/2021MNRAS.500.1721M},
 archiveprefix = {arXiv},
 author = {{Menon}, Shyam H. and {Federrath}, Christoph and {Klaassen}, Pamela and {Kuiper}, Rolf and {Reiter}, Megan},
 doi = {10.1093/mnras/staa3271},
 eprint = {2010.09861},
 journal = {\mnras},
 month = {January},
 number = {2},
 pages = {1721-1740},
 primaryclass = {astro-ph.GA},
 title = {{On the compressive nature of turbulence driven by ionizing feedback in the pillars of the Carina Nebula}},
 volume = {500},
 year = {2021}
}

@article{2022ApJ...935..167A,
 adsurl = {https://ui.adsabs.harvard.edu/abs/2022ApJ...935..167A},
 archiveprefix = {arXiv},
 author = {{Astropy Collaboration} and {Price-Whelan}, Adrian M. and {Lim}, Pey Lian and {Earl}, Nicholas and {Starkman}, Nathaniel and {Bradley}, Larry and {Shupe}, David L. and {Patil}, Aarya A. and {Corrales}, Lia and {Brasseur}, C.~E. and {N{\"o}the}, Maximilian and {Donath}, Axel and {Tollerud}, Erik and {Morris}, Brett M. and {Ginsburg}, Adam and {Vaher}, Eero and {Weaver}, Benjamin A. and {Tocknell}, James and {Jamieson}, William and {van Kerkwijk}, Marten H. and {Robitaille}, Thomas P. and {Merry}, Bruce and {Bachetti}, Matteo and {G{\"u}nther}, H. Moritz and {Aldcroft}, Thomas L. and {Alvarado-Montes}, Jaime A. and {Archibald}, Anne M. and {B{\'o}di}, Attila and {Bapat}, Shreyas and {Barentsen}, Geert and {Baz{\'a}n}, Juanjo and {Biswas}, Manish and {Boquien}, M{\'e}d{\'e}ric and {Burke}, D.~J. and {Cara}, Daria and {Cara}, Mihai and {Conroy}, Kyle E. and {Conseil}, Simon and {Craig}, Matthew W. and {Cross}, Robert M. and {Cruz}, Kelle L. and {D'Eugenio}, Francesco and {Dencheva}, Nadia and {Devillepoix}, Hadrien A.~R. and {Dietrich}, J{\"o}rg P. and {Eigenbrot}, Arthur Davis and {Erben}, Thomas and {Ferreira}, Leonardo and {Foreman-Mackey}, Daniel and {Fox}, Ryan and {Freij}, Nabil and {Garg}, Suyog and {Geda}, Robel and {Glattly}, Lauren and {Gondhalekar}, Yash and {Gordon}, Karl D. and {Grant}, David and {Greenfield}, Perry and {Groener}, Austen M. and {Guest}, Steve and {Gurovich}, Sebastian and {Handberg}, Rasmus and {Hart}, Akeem and {Hatfield-Dodds}, Zac and {Homeier}, Derek and {Hosseinzadeh}, Griffin and {Jenness}, Tim and {Jones}, Craig K. and {Joseph}, Prajwel and {Kalmbach}, J. Bryce and {Karamehmetoglu}, Emir and {Ka{\l}uszy{\'n}ski}, Miko{\l}aj and {Kelley}, Michael S.~P. and {Kern}, Nicholas and {Kerzendorf}, Wolfgang E. and {Koch}, Eric W. and {Kulumani}, Shankar and {Lee}, Antony and {Ly}, Chun and {Ma}, Zhiyuan and {MacBride}, Conor and {Maljaars}, Jakob M. and {Muna}, Demitri and {Murphy}, N.~A. and {Norman}, Henrik and {O'Steen}, Richard and {Oman}, Kyle A. and {Pacifici}, Camilla and {Pascual}, Sergio and {Pascual-Granado}, J. and {Patil}, Rohit R. and {Perren}, Gabriel I. and {Pickering}, Timothy E. and {Rastogi}, Tanuj and {Roulston}, Benjamin R. and {Ryan}, Daniel F. and {Rykoff}, Eli S. and {Sabater}, Jose and {Sakurikar}, Parikshit and {Salgado}, Jes{\'u}s and {Sanghi}, Aniket and {Saunders}, Nicholas and {Savchenko}, Volodymyr and {Schwardt}, Ludwig and {Seifert-Eckert}, Michael and {Shih}, Albert Y. and {Jain}, Anany Shrey and {Shukla}, Gyanendra and {Sick}, Jonathan and {Simpson}, Chris and {Singanamalla}, Sudheesh and {Singer}, Leo P. and {Singhal}, Jaladh and {Sinha}, Manodeep and {Sip{\H{o}}cz}, Brigitta M. and {Spitler}, Lee R. and {Stansby}, David and {Streicher}, Ole and {{\v{S}}umak}, Jani and {Swinbank}, John D. and {Taranu}, Dan S. and {Tewary}, Nikita and {Tremblay}, Grant R. and {de Val-Borro}, Miguel and {Van Kooten}, Samuel J. and {Vasovi{\'c}}, Zlatan and {Verma}, Shresth and {de Miranda Cardoso}, Jos{\'e} Vin{\'\i}cius and {Williams}, Peter K.~G. and {Wilson}, Tom J. and {Winkel}, Benjamin and {Wood-Vasey}, W.~M. and {Xue}, Rui and {Yoachim}, Peter and {Zhang}, Chen and {Zonca}, Andrea and {Astropy Project Contributors}},
 doi = {10.3847/1538-4357/ac7c74},
 eid = {167},
 eprint = {2206.14220},
 journal = {\apj},
 month = {August},
 number = {2},
 pages = {167},
 primaryclass = {astro-ph.IM},
 title = {{The Astropy Project: Sustaining and Growing a Community-oriented Open-source Project and the Latest Major Release (v5.0) of the Core Package}},
 volume = {935},
 year = {2022}
}

@software{2022ascl.soft04001F,
 adsurl = {https://ui.adsabs.harvard.edu/abs/2022ascl.soft04001F},
 archiveprefix = {ascl},
 author = {{Federrath}, C. and {Roman-Duval}, J. and {Klessen}, R.~S. and {Schmidt}, W. and {Mac Low}, M.-M.},
 eid = {ascl:2204.001},
 eprint = {2204.001},
 howpublished = {Astrophysics Source Code Library, record ascl:2204.001},
 month = {April},
 title = {{TG: Turbulence Generator}},
 year = {2022}
}

@article{2022MNRAS.509.2180S,
 adsurl = {https://ui.adsabs.harvard.edu/abs/2022MNRAS.509.2180S},
 archiveprefix = {arXiv},
 author = {{Sharda}, Piyush and {Menon}, Shyam H. and {Federrath}, Christoph and {Krumholz}, Mark R. and {Beattie}, James R. and {Jameson}, Katherine E. and {Tokuda}, Kazuki and {Burkhart}, Blakesley and {Crocker}, Roland M. and {Law}, Charles J. and {Seta}, Amit and {Gaetz}, Terrance J. and {Pingel}, Nickolas M. and {Seitenzahl}, Ivo R. and {Sano}, Hidetoshi and {Fukui}, Yasuo},
 doi = {10.1093/mnras/stab3048},
 eprint = {2109.03983},
 journal = {\mnras},
 month = {January},
 number = {2},
 pages = {2180-2193},
 primaryclass = {astro-ph.GA},
 title = {{First extragalactic measurement of the turbulence driving parameter: ALMA observations of the star-forming region N159E in the Large Magellanic Cloud}},
 volume = {509},
 year = {2022}
}

@article{2022MNRAS.510.3778M,
 adsurl = {https://ui.adsabs.harvard.edu/abs/2022MNRAS.510.3778M},
 archiveprefix = {arXiv},
 author = {{Mohapatra}, Rajsekhar and {Jetti}, Mrinal and {Sharma}, Prateek and {Federrath}, Christoph},
 doi = {10.1093/mnras/stab3603},
 eprint = {2107.07722},
 journal = {\mnras},
 month = {March},
 number = {3},
 pages = {3778-3793},
 primaryclass = {astro-ph.GA},
 title = {{Characterizing the turbulent multiphase haloes with periodic box simulations}},
 volume = {510},
 year = {2022}
}

@article{2022MNRAS.514.3139M,
 adsurl = {https://ui.adsabs.harvard.edu/abs/2022MNRAS.514.3139M},
 archiveprefix = {arXiv},
 author = {{Mohapatra}, Rajsekhar and {Federrath}, Christoph and {Sharma}, Prateek},
 doi = {10.1093/mnras/stac1610},
 eprint = {2206.03602},
 journal = {\mnras},
 month = {August},
 number = {3},
 pages = {3139-3159},
 primaryclass = {astro-ph.GA},
 title = {{Multiphase turbulence in galactic haloes: effect of the driving}},
 volume = {514},
 year = {2022}
}

@article{2023A&A...669A..10Z,
 adsurl = {https://ui.adsabs.harvard.edu/abs/2023A&A...669A..10Z},
 archiveprefix = {arXiv},
 author = {{Zari}, Eleonora and {Frankel}, Neige and {Rix}, Hans-Walter},
 doi = {10.1051/0004-6361/202244194},
 eid = {A10},
 eprint = {2206.02616},
 journal = {\aap},
 month = {January},
 pages = {A10},
 primaryclass = {astro-ph.GA},
 title = {{Did the Milky Way just light up? The recent star formation history of the Galactic disc}},
 volume = {669},
 year = {2023}
}

@article{2023MNRAS.524.2379H,
 adsurl = {https://ui.adsabs.harvard.edu/abs/2023MNRAS.524.2379H},
 archiveprefix = {arXiv},
 author = {{Hu}, Yue and {Lazarian}, A.},
 doi = {10.1093/mnras/stad1996},
 eprint = {2302.05047},
 journal = {\mnras},
 month = {September},
 number = {2},
 pages = {2379-2394},
 primaryclass = {astro-ph.GA},
 title = {{Mapping the Galactic magnetic field orientation and strength in three dimensions}},
 volume = {524},
 year = {2023}
}

@article{2023MNRAS.526..982G,
 adsurl = {https://ui.adsabs.harvard.edu/abs/2023MNRAS.526..982G},
 archiveprefix = {arXiv},
 author = {{Gerrard}, Isabella A. and {Federrath}, Christoph and {Pingel}, Nickolas M. and {McClure-Griffiths}, Naomi M. and {Marchal}, Antoine and {Joncas}, Gilles and {Clark}, Susan E. and {Stanimirovi{\'c}}, Sne{\v{z}}ana and {Lee}, Min-Young and {van Loon}, Jacco Th and {Dickey}, John and {D{\'e}nes}, Helga and {Ma}, Yik Ki and {Dempsey}, James and {Lynn}, Callum},
 doi = {10.1093/mnras/stad2718},
 eprint = {2309.10755},
 journal = {\mnras},
 month = {November},
 number = {1},
 pages = {982-999},
 primaryclass = {astro-ph.GA},
 title = {{A new method for spatially resolving the turbulence-driving mixture in the ISM with application to the Small Magellanic Cloud}},
 volume = {526},
 year = {2023}
}

@article{2024arXiv240916053S,
 adsurl = {https://ui.adsabs.harvard.edu/abs/2026ApJS..283...27S},
 archiveprefix = {arXiv},
 author = {{Stone}, James M. and {Mullen}, Patrick D. and {Fielding}, Drummond and {Grete}, Philipp and {Guo}, Minghao and {Kempski}, Philipp and {Most}, Elias R. and {White}, Christopher J. and {Wong}, George N.},
 doi = {10.3847/1538-4365/ae3717},
 eid = {27},
 eprint = {2409.16053},
 journal = {\apjs},
 month = {March},
 number = {1},
 pages = {27},
 primaryclass = {astro-ph.IM},
 title = {{AthenaK: A Performance-portable Version of the Athena++ Adaptive Mesh Refinement Framework}},
 volume = {283},
 year = {2026}
}

@article{2024MNRAS.530.4317G,
 adsurl = {https://ui.adsabs.harvard.edu/abs/2024MNRAS.530.4317G},
 archiveprefix = {arXiv},
 author = {{Gerrard}, Isabella A. and {Noon}, Karlie A. and {Federrath}, Christoph and {Di Teodoro}, Enrico M. and {Marchal}, Antoine and {McClure-Griffiths}, N.~M.},
 doi = {10.1093/mnras/stae1144},
 eprint = {2404.18349},
 journal = {\mnras},
 month = {June},
 number = {4},
 pages = {4317-4330},
 primaryclass = {astro-ph.GA},
 title = {{Turbulence statistics of H I clouds entrained in the Milky Way's nuclear wind}},
 volume = {530},
 year = {2024}
}

@article{2025ApJ...987..122G,
 adsurl = {https://ui.adsabs.harvard.edu/abs/2025ApJ...987..122G},
 archiveprefix = {arXiv},
 author = {{Grete}, Philipp and {Scannapieco}, Evan and {Br{\"u}ggen}, Marcus and {Pan}, Liubin},
 doi = {10.3847/1538-4357/add936},
 eid = {122},
 eprint = {2505.23898},
 journal = {\apj},
 month = {July},
 number = {2},
 pages = {122},
 primaryclass = {astro-ph.GA},
 title = {{The Density Distribution of Compressively Forced, Supersonic Turbulence Depends on the Driving Correlation Time}},
 volume = {987},
 year = {2025}
}

@article{2025ApJ...990...49G,
 adsurl = {https://ui.adsabs.harvard.edu/abs/2025ApJ...990...49G},
 archiveprefix = {arXiv},
 author = {{Guo}, Minghao and {Kim}, Chang-Goo and {Stone}, James M.},
 doi = {10.3847/1538-4357/adeb85},
 eid = {49},
 eprint = {2411.12809},
 journal = {\apj},
 month = {September},
 number = {1},
 pages = {49},
 primaryclass = {astro-ph.GA},
 title = {{Evolution of Supernova Remnants in a Cloudy Multiphase Interstellar Medium}},
 volume = {990},
 year = {2025}
}

@article{2025ApJ...994...80L,
 adsurl = {https://ui.adsabs.harvard.edu/abs/2025ApJ...994...80L},
 author = {{Lee}, Bumhyun and {Lee}, Min-Young and {Cho}, Jungyeon and {Pingel}, Nickolas M. and {Ma}, Yik Ki and {Jameson}, Katie and {Dempsey}, James and {D{\'e}nes}, Helga and {Dickey}, John M. and {Federrath}, Christoph and {Gibson}, Steven and {Joncas}, Gilles and {Kemp}, Ian and {Kim}, Shin-Jeong and {Lynn}, Callum and {Marchal}, Antoine and {McClure-Griffiths}, N.~M. and {Nguyen}, Hiep and {Seta}, Amit and {Soler}, Juan D. and {Stanimirovi{\'c}}, Sne{\v{z}}ana and {van Loon}, Jacco Th.},
 doi = {10.3847/1538-4357/ae1008},
 eid = {80},
 journal = {\apj},
 month = {November},
 number = {1},
 pages = {80},
 title = {{Study of H I Turbulence in the SMC Using Multipoint Structure Functions}},
 volume = {994},
 year = {2025}
}

@article{2025ApJ...994..193B,
 adsurl = {https://ui.adsabs.harvard.edu/abs/2025ApJ...994..193B},
 archiveprefix = {arXiv},
 author = {{Beattie}, James R. and {Kolborg}, Anne Noer and {Ramirez-Ruiz}, Enrico and {Federrath}, Christoph},
 doi = {10.3847/1538-4357/ae07cd},
 eid = {193},
 eprint = {2501.09855},
 journal = {\apj},
 month = {December},
 number = {2},
 pages = {193},
 primaryclass = {astro-ph.GA},
 title = {{So Long Kolmogorov: The Forward and Backward Turbulence Cascades in a Supernovae-driven, Multiphase Interstellar Medium}},
 volume = {994},
 year = {2025}
}

@article{2025arXiv250907354B,
 adsurl = {https://ui.adsabs.harvard.edu/abs/2025arXiv250907354B},
 archiveprefix = {arXiv},
 author = {{Beattie}, James R.},
 doi = {10.48550/arXiv.2509.07354},
 eid = {arXiv:2509.07354},
 eprint = {2509.07354},
 journal = {arXiv e-prints},
 month = {September},
 pages = {arXiv:2509.07354},
 primaryclass = {astro-ph.GA},
 title = {{Supernovae drive large-scale, incompressible turbulence through small-scale instabilities}},
 year = {2025}
}

@article{2025arXiv251100229M,
 adsurl = {https://ui.adsabs.harvard.edu/abs/2025arXiv251100229M},
 archiveprefix = {arXiv},
 author = {{Mohapatra}, Rajsekhar and {Dutta}, Alankar and {Sharma}, Prateek},
 doi = {10.48550/arXiv.2511.00229},
 eid = {arXiv:2511.00229},
 eprint = {2511.00229},
 journal = {arXiv e-prints},
 month = {October},
 pages = {arXiv:2511.00229},
 primaryclass = {astro-ph.GA},
 title = {{Tracing Multiphase Structure in the Circumgalactic Medium: Insights from Magnetohydrodynamic Turbulence Simulations}},
 year = {2025}
}

@article{2025JFM..1019R...2Y,
 adsurl = {https://ui.adsabs.harvard.edu/abs/2025JFM..1019R...2Y},
 author = {{Yeung}, P.~K. and {Ravikumar}, Kiran and {Uma-Vaideswaran}, Rohini and {Dotson}, Daniel L. and {Sreenivasan}, Katepalli R. and {Pope}, Stephen B. and {Meneveau}, Charles and {Nichols}, Stephen},
 doi = {10.1017/jfm.2025.10493},
 eid = {R2},
 journal = {Journal of Fluid Mechanics},
 month = {September},
 pages = {R2},
 title = {{Small-scale properties from exascale computations of turbulence on a \textbackslashmathbf\{32 768(3\})  periodic cube}},
 volume = {1019},
 year = {2025}
}

@article{2025NatAs...9.1195B,
 adsurl = {https://ui.adsabs.harvard.edu/abs/2025NatAs...9.1195B},
 archiveprefix = {arXiv},
 author = {{Beattie}, James R. and {Federrath}, Christoph and {Klessen}, Ralf S. and {Cielo}, Salvatore and {Bhattacharjee}, Amitava},
 doi = {10.1038/s41550-025-02551-5},
 eprint = {2504.07136},
 journal = {Nature Astronomy},
 month = {August},
 pages = {1195-1205},
 primaryclass = {astro-ph.GA},
 title = {{The spectrum of magnetized turbulence in the interstellar medium}},
 volume = {9},
 year = {2025}
}

@article{2025OJAp....8E..44H,
 adsurl = {https://ui.adsabs.harvard.edu/abs/2025OJAp....8E..44H},
 archiveprefix = {arXiv},
 author = {{Hopkins}, Philip F.},
 doi = {10.33232/001c.132375},
 eid = {44},
 eprint = {2404.16987},
 journal = {The Open Journal of Astrophysics},
 month = {April},
 pages = {44},
 primaryclass = {astro-ph.GA},
 title = {{The Importance of Subtleties in the Scaling of the 'Terminal Momentum' For Galaxy Formation Simulations}},
 volume = {8},
 year = {2025}
}

@article{2026ApJ...997...33C,
 adsurl = {https://ui.adsabs.harvard.edu/abs/2026ApJ...997...33C},
 archiveprefix = {arXiv},
 author = {{Connor}, Isabelle and {Beattie}, James R. and {Kolborg}, Anne Noer and {Ramirez-Ruiz}, Enrico},
 doi = {10.3847/1538-4357/ae17b1},
 eid = {33},
 eprint = {2509.01653},
 journal = {\apj},
 month = {January},
 number = {1},
 pages = {33},
 primaryclass = {astro-ph.GA},
 title = {{Cascading from the Winds to the Disk: The Universality of Supernovae-driven Turbulence in Different Galactic Interstellar Media}},
 volume = {997},
 year = {2026}
}

@article{2026ApJ...998..270S,
 adsurl = {https://ui.adsabs.harvard.edu/abs/2026ApJ...998..270S},
 archiveprefix = {arXiv},
 author = {{Scannapieco}, Evan and {Br{\"u}ggen}, Marcus and {Grete}, Philipp and {Pan}, Liubin},
 doi = {10.3847/1538-4357/ae394d},
 eid = {270},
 eprint = {2509.09811},
 journal = {\apj},
 month = {February},
 number = {2},
 pages = {270},
 primaryclass = {astro-ph.GA},
 title = {{An Improved Fit to the Density Fluctuations in Supersonic Isothermal Turbulence}},
 volume = {998},
 year = {2026}
}

@article{2026MNRAS.546f2277G,
 adsurl = {https://ui.adsabs.harvard.edu/abs/2026MNRAS.546f2277G},
 archiveprefix = {arXiv},
 author = {{Gerrard}, Isabella A. and {Federrath}, Christoph},
 doi = {10.1093/mnras/staf2277},
 eid = {staf2277},
 eprint = {2601.01427},
 journal = {\mnras},
 month = {February},
 number = {2},
 pages = {staf2277},
 primaryclass = {astro-ph.GA},
 title = {{Turbulence driving in a star-forming Milky-Way-type galaxy}},
 volume = {546},
 year = {2026}
}

@article{Fuchs2009,
 adsurl = {https://ui.adsabs.harvard.edu/abs/2009AJ....137..266F},
 archiveprefix = {arXiv},
 author = {{Fuchs}, B. and {Jahrei{\ss}}, H. and {Flynn}, C.},
 doi = {10.1088/0004-6256/137/1/266},
 eprint = {0810.1656},
 journal = {\aj},
 month = {January},
 number = {1},
 pages = {266-271},
 primaryclass = {astro-ph},
 title = {{A Schmidt-Kennicutt Law for Star Formation in the Milky Way Disk}},
 volume = {137},
 year = {2009}
}

@article{harris2020array,
 adsurl = {https://ui.adsabs.harvard.edu/abs/2020Natur.585..357H},
 archiveprefix = {arXiv},
 author = {{Harris}, Charles R. and {Millman}, K. Jarrod and {van der Walt}, St{\'e}fan J. and {Gommers}, Ralf and {Virtanen}, Pauli and {Cournapeau}, David and {Wieser}, Eric and {Taylor}, Julian and {Berg}, Sebastian and {Smith}, Nathaniel J. and {Kern}, Robert and {Picus}, Matti and {Hoyer}, Stephan and {van Kerkwijk}, Marten H. and {Brett}, Matthew and {Haldane}, Allan and {del R{\'\i}o}, Jaime Fern{\'a}ndez and {Wiebe}, Mark and {Peterson}, Pearu and {G{\'e}rard-Marchant}, Pierre and {Sheppard}, Kevin and {Reddy}, Tyler and {Weckesser}, Warren and {Abbasi}, Hameer and {Gohlke}, Christoph and {Oliphant}, Travis E.},
 doi = {10.1038/s41586-020-2649-2},
 eprint = {2006.10256},
 journal = {\nat},
 month = {September},
 number = {7825},
 pages = {357-362},
 primaryclass = {cs.MS},
 title = {{Array programming with NumPy}},
 volume = {585},
 year = {2020}
}

@article{Jenkins2011,
 adsurl = {https://ui.adsabs.harvard.edu/abs/2011ApJ...734...65J},
 archiveprefix = {arXiv},
 author = {{Jenkins}, Edward B. and {Tripp}, Todd M.},
 doi = {10.1088/0004-637X/734/1/65},
 eid = {65},
 eprint = {1104.2323},
 journal = {\apj},
 month = {June},
 number = {1},
 pages = {65},
 primaryclass = {astro-ph.GA},
 title = {{The Distribution of Thermal Pressures in the Diffuse, Cold Neutral Medium of Our Galaxy. II. An Expanded Survey of Interstellar C I Fine-structure Excitations}},
 volume = {734},
 year = {2011}
}

\appendix

\section{Resolution Convergence} \label{sec:rez_dep}
\begin{figure*}
    \includegraphics[width=\linewidth]{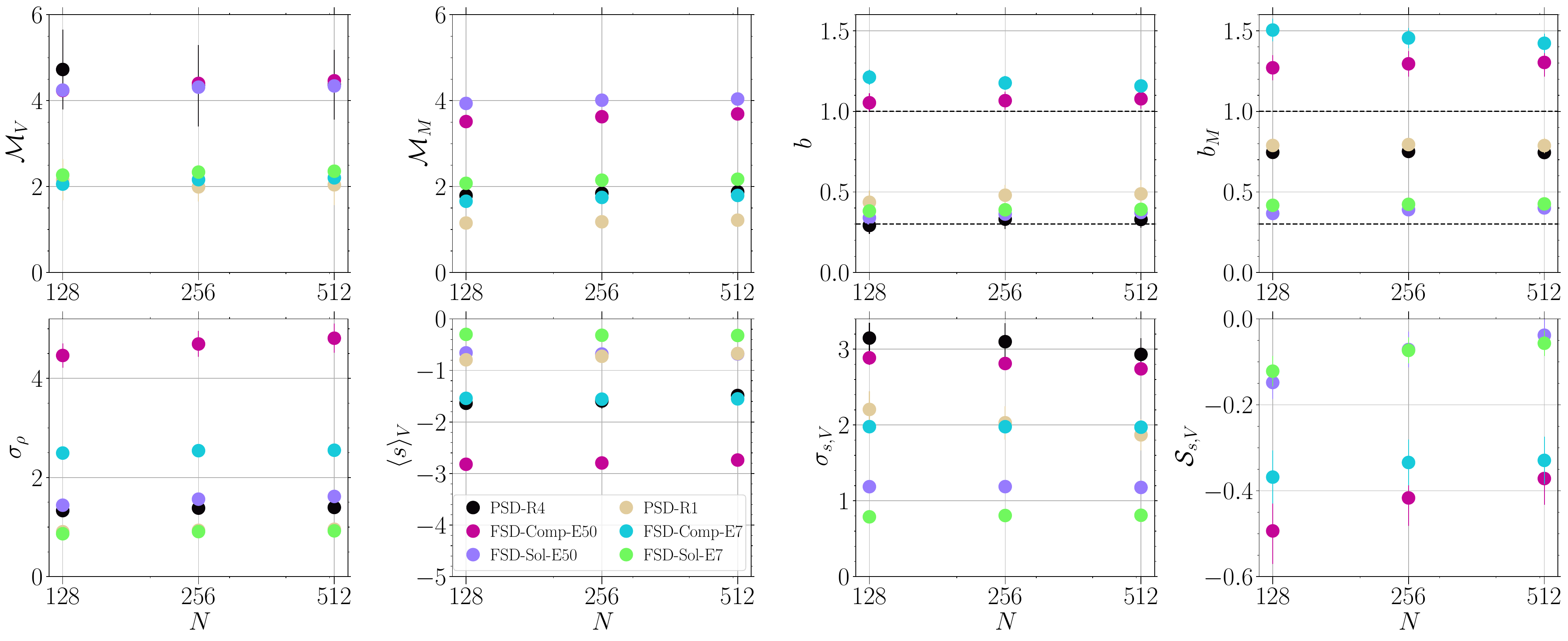}
    \caption{\label{fig:rez_dep} Same quantities as \autoref{fig:tcorr_var_dependance}, but as a function of resolution for the fiducial simulation suite.}
\end{figure*}

We run the fiducial model suite (\autoref{sec:psd_fsd}) with resolution varying from $N=128$ to $512$, as seen in \autoref{fig:rez_dep}. We find the global statistics are generally converged, especially between $N=256$ and $N=512$. Some quantities (e.g., high order moments of the logarithmic density) show non-negligible changes between the two resolutions. $b$-parameters also slightly change, but we do not expect them to change dramatically from the current measured values and converge to either $b=1/3$ or $b=1$. 

In addition to the quantities in \autoref{fig:rez_dep}, we calculate the resolution dependency of $\rcomp$,  $\abrackets{\Rcomp}_V$, and $\abrackets{\Rcomp}_M$. We find that $\rcomp$ is very well converged, with the largest difference in value between $N=128\rightarrow512$ being $<3\%$ for our fiducial models. $\abrackets{\Rcomp}_V$ and $\abrackets{\Rcomp}_M$ vary more, but better converged between $N=256\to512$ within $<5\%$, except the \PSDlow\ and \FSDsolhigh\ models show larger variations at a level of $\sim 10\%$. 

\section{PSD Parameter Sensitivity} \label{sec:appendix_psd_param}

To model turbulence driving by expanding bubbles, we inject radial momentum (\autoref{subsec:psd}). This mimics physical process like SN explosions, but given the isothermal equation of state adopted in the simulations, we make several parameter choices that may affect the results. In addition, we apply velocity ceiling and density floor to circumvent numerical difficulties. Here, we explore the sensitivity of our results to the chosen parameters.

\subsection{Velocity Ceiling and Density Floor \label{subsec:vel_ciel}}
As discussed in \autoref{subsec:psd}, our \PSD\ models have the potential to generate exceedingly high velocities. The $v_{\rm max}=500\ \rm km\ s^{-1}$ parameter, or $\mathcal{M}=51$, is applied for both \PSD\ models such that $v_{\rm inj} \le v_{\rm max}$. A total of $8.4\%$ and $3.7\%$ of momentum injection events in the \PSDhigh\ and \PSDlow\ models, respectively, have $v_{\rm inj}$ set to $v_{\rm max}$. We find that the average total volume percentage of grid cells with $\mathcal{M} \ge50$ to be $0.1\%$ for \PSDhigh\ and $0.007\%$ for \PSDlow. The average fraction of total mass within those grid cells is found to be $6\times10^{-6}$ and $1.9\times 10^{-7}$ for each respective \PSD\ model.

To see the effect of the accumulation at $v_{\rm max}$, we calculate the average $\mathcal{M}_V$, $\mathcal{M}_M$, $\sigma_{\rho}$, $b$, and $b_M$ values while ignoring any cells with high local Mach number $\mathcal{M}\ge50$. Comparing between the values presented in \autoref{table:main_stats} and the masked values, we find that $\mathcal{M}_V$ varies from $4.37\to4.19$ for \PSDhigh and $2.04\to2.02$ for \PSDlow. The mass-weighted values $\mathcal{M}_M$ as well as $\sigma_\rho$ are nearly unchanged. As a result, $b=0.33\to0.34$ for \PSDhigh, and $b=0.488\to0.492$ for \PSDlow. Again, the mass-weighted values $b_M$ are nearly unchanged. All changes are small and within the 1-$\sigma$ temporal fluctuations presented in \autoref{table:main_stats}.

When lifting $v_{\rm max}$ constraint from  $v_{\rm max}=500\kms$ to $3000\kms$ for a lower resolution version ($256^3$) of \PSDhigh, we find that $\mathcal{M}_V$ increases from $4.3$ to $5.5$, while $\mathcal{M}_M$ only varies from $1.84$ to $1.89$. With $\sigma_\rho= 1.38\to1.40$, we get $b=0.33\to 0.29$ and $b_M=0.75\to 0.74$. 

A similar assessment was made for the density floor. The volume affected by the density floor is not negligible, forming a visible peak in the density PDFs (\autoref{fig:512_s_pdf_grid}). As this low-density gas is associated with the high velocity bubble interior gas, the volume-weighted statistics (mainly Mach number) are contaminated.
In general, our conclusions drawn from the mass-weighted statistics are robust. Note that the interior of bubbles cannot be fully self-consistent with the isothermal equation of state anyway and is irrelevant for turbulence traced by CO or H{\sc I} observations.

\subsection{Injection Size Dependence} \label{sec:InjSize}

The injection radius chosen for the \PSD\ models gives the typical injection velocity $P_{\rm inj}/(\rho_0V_{\rm inj})\sim 112 (r_{\rm inj}/40\pc)^{-3}\kms$ for the mean density of the simulation domain $n_{H,0}=0.3\pcc$. This value is consistent with the shell formation radius and velocity at which the forward shock begins to cool, marking the beginning of momentum conserving stage of the radiative SN remnant evolution \citep{2015ApJ...802...99K}. Although this choice is physically motivated, the size of bubble makes up a significant portion of the simulation box as seen in \autoref{fig:slices}, causing a potential concern about the sensitivity to the choice as the locality of injection is the key difference between \PSD\ and \FSD.

We run simulations of varying $r_{\rm inj}$ for \PSDhigh\ and \PSDlow\ with the resolution $N=256$ and velocity ceiling $v_{\rm max}=3000\ \rm km\ s^{-1}$. As the typical injection velocity scales with $r_{\rm inj}^{-3}$, a larger ceiling is necessary to ensure consistent momentum injection without limiting velocity too frequently.

Varying $r_{\rm inj}$ from $40$ to $20$ for \PSDhigh\ varies $\mathcal{M}_V=5.5\to7.9$. With only a marginal change for $\sigma_\rho=1.4\to1.38$, the corresponding $b=0.29\to0.20$. The mass weighted version has a smaller change, with $\mathcal{M}_M=1.9\to2.3$, and $b_M=0.74\to0.61$. The fraction of compressive modes in the velocity fields increases with $\abrackets{\Rcomp}_V=0.66\to0.53$ and $\abrackets{\Rcomp}_M=0.53\to0.43$. The injection scale decreases for smaller $r_{\rm inj}$, but not linearly with $L_{\rm in}/L=0.25\to0.22$. The dependence to the injection size implies that there is a degeneracy for the same momentum injection per event. It is of great interest to perform similar comparative study for simulations with localized energy injection including cooling and heating and resolving energy conserving stage \citep[e.g.,][]{2015ApJ...802...99K,2025ApJ...990...49G}.


\label{lastpage}
\end{document}